\documentclass[showpacs,amsmath,amssymb,aps,twocolumn,superscriptaddress,prx]{revtex4-2}
\usepackage{mathtools}
\usepackage{algorithm}
\usepackage{algpseudocode}
\usepackage[pdftex]{graphicx} \graphicspath{{}}% Include figure files
\usepackage{grffile} % change algorithm for searching thru graphics extensions
\usepackage{comment}
\usepackage{braket}

\usepackage{float}
\usepackage{subfigure}
\usepackage{amsmath}
\usepackage{graphicx}
\usepackage{dcolumn}% Align table columns on decimal point
\usepackage{bm}% bold math
\usepackage[pdftex,colorlinks=true]{hyperref}
\usepackage{soul}
\hypersetup{
	colorlinks=true,
	linkcolor=blue,
	urlcolor=cyan,
}

\usepackage{color}
\definecolor{ForestGreen}{RGB}{34, 139, 34}

\usepackage[normalem]{ulem}

\newcommand{\affA}{State Key Laboratory of Artificial Microstructure and Mesoscopic Physics, School of Physics, Peking University, Beijing 100871, China}
\newcommand{\affC}{Frontiers Science Center for Nano-Optoelectronics, Peking University, Beijing 100871, China}
\newcommand{\affD}{Collaborative Innovation Center of Extreme Optics, Shanxi University, Taiyuan 030006, China}
\newcommand{\affB}{Weierstrass Institute for Applied Analysis and Stochastics, Mohrenstr. 39, 10117, Berlin, Germany}

\begin{document}

\title{{Stable time rondeau crystals in dissipative many-body systems}}

\author{Zhuocheng Ma}
 \affiliation{\affA}
 \author{Jin Yan}
 \affiliation{\affB}
 	\author{Hongzheng Zhao}
	\email{hzhao@pku.edu.cn}
 \affiliation{\affA}
 \author{Liang-You Peng}
 \email{liangyou.peng@pku.edu.cn}
 \affiliation{\affA }
 \affiliation{ \affC }
 \affiliation{ \affD}
	\date{\today}

\begin{abstract}
Driven systems offer the potential to realize a wide range of non-equilibrium phenomena that are inaccessible in static systems, such as discrete time crystals. 
Time rondeau crystals with a partial temporal order have been proposed as a distinctive prethermal phase of matter in systems driven by structured random protocols. Yet, heating is inevitable in closed systems and time rondeau crystals eventually melt. We introduce dissipation to counteract heating and demonstrate stable time rondeau crystals, which persist indefinitely, in a many-body interacting system. A key ingredient is synchronization in the non-interacting limit, which allows for stable time rondeau order without generating excessive heating. The presence of many-body interaction competes with synchronization and a de-synchronization phase transition occurs at a finite interaction strength. This transition is well captured via a linear stability analysis of the underlying stochastic processes.
\end{abstract}
	\maketitle
\let\oldaddcontentsline\addcontentsline
\renewcommand{\addcontentsline}[3]{}

% 这个tex要用pdflatex编译
% 这个tex要用pdflatex编译
% 这个tex要用pdflatex编译

\section{Introduction}
Identifying different forms of order and disorder is an everlasting research subject in science.
In thermal equilibrium, the spontaneous symmetry breaking leads to a long-range spatial order. In periodically driven (Floquet) systems, a discrete time crystalline (DTC) order can be similarly defined~\cite{khemani2016phase,else2016floquet,dtc_prl2017,PhysRevLett.122.015701}, where the discrete time translational symmetry is broken. 
By using quasi-periodic drives, one goes beyond this Floquet lore~\cite{verdeny2016quasi,nandy2017aperiodically,mori2021rigorous,wen2021periodically,long2022many,he2023quasi,gallone2024prethermalization,ghosh2024slow,schmid2024self}, establishing the deterministic, yet non-periodic, quasicrystalline temporal order ~\cite{dumitrescu2018logarithmically,zhao2019floquet,else2020long}.

Interestingly, the organization principle of the natural world is far richer than merely being deterministic. In fact, order and disorder can naturally coexist in our daily life: the oxygen ions in ice establish long-range spatial order, while the location of protons bonding the oxygen ions is highly disordered. Despite the ubiquity of such partial order in space, one of its temporal cousins - the time rondeau crystal (TRC) - has not been discovered until very recently~\cite{PhysRevB.108.L100203,moon2024experimental}. {The rondeau is a pattern built around a recurring main theme that alternates with one or more contrasting variations.
Time rondeau order, induced by structured random drives, exhibits both stroboscopic long-time order and short-time disorder at all other times, notably enriching the classification of non-equilibrium temporal orders.} It is different from the conventional Floquet DTC, where the micromotion within a drive cycle exhibits the same temporal order as the stroboscopic evolution.

Yet, randomness in the driving typically opens additional heating channels~\cite{zhao2021random,wen2022periodically2,yan2024prethermalization,tiwari2024dynamical},  which even many-body localization can not prevent~\cite{levi2016robustness,gopalakrishnan2017noise,rieder2018localization,zhao2022localization}. Indeed, a TRC in a closed system appears to be a transient meta-stable (prethermal) phenomenon~\cite{lazarides2014equilibrium,kim2014testing,Isolated2014,bukov2015prethermal,kuwahara2016floquet,else2017prethermal,mori2018floquet,Josephson-PhysRevB,howell2019asymptotic,luitz2020prethermalization,rubio2020floquet,pizzi2021classical,ye2021floquet,peng2021floquet,fleckenstein2021thermalization,ikeda2021fermi,munoz2022floquet,beatrez2023critical,jin2023fractionalized,ho2023quantum,yue2023prethermal,hou2024floquet,PhysRevX.14.041070,qi2024topological}: although an increasing driving frequency parametrically prolongs its lifetime, eventual heat death seems inevitable for any fixed frequency~\cite{PhysRevB.108.L100203,moon2024experimental}. One fundamental question thus arises:  Are there types of TRCs that are absolutely stable to arbitrary perturbations of both the initial state and the equations of motion (EOM), s.t. they are infinitely long lived in the thermodynamic limit?

One potential way to avoid heating is by introducing dissipation, a strategy applicable to both quantum and classical systems~\cite{piazza2015self,buvca2019non,mori2023floquet,kawabata2023symmetry,russomanno2023spatiotemporally,yi2024theory}. Dissipation can generate contractive dynamics around target attractors with desired features, like the DTC order~\cite{DTC_qed2018,iemini2018boundary,kessler2021observation,wu2024dissipative,wu2024dissipative,Solanki2024Exotic}, and perturbations can be strongly damped. Yet, applying this idea to stabilize TRC requires addressing two formidable
challenges: (i) Stroboscopic temporal order can be fragile when the drive involves randomness. For instance, as illustrated in Ref.~\cite{CDTC2020}, temporal fluctuations akin to a finite-temperature bath can lead to DTC order with a finite lifetime.
(ii) The complex interplay between dissipation and many-body interactions makes it unclear how to ensure the stability of TRC in the thermodynamic limit.

\begin{figure}[t]
    \centering   \includegraphics[width=0.9\linewidth]{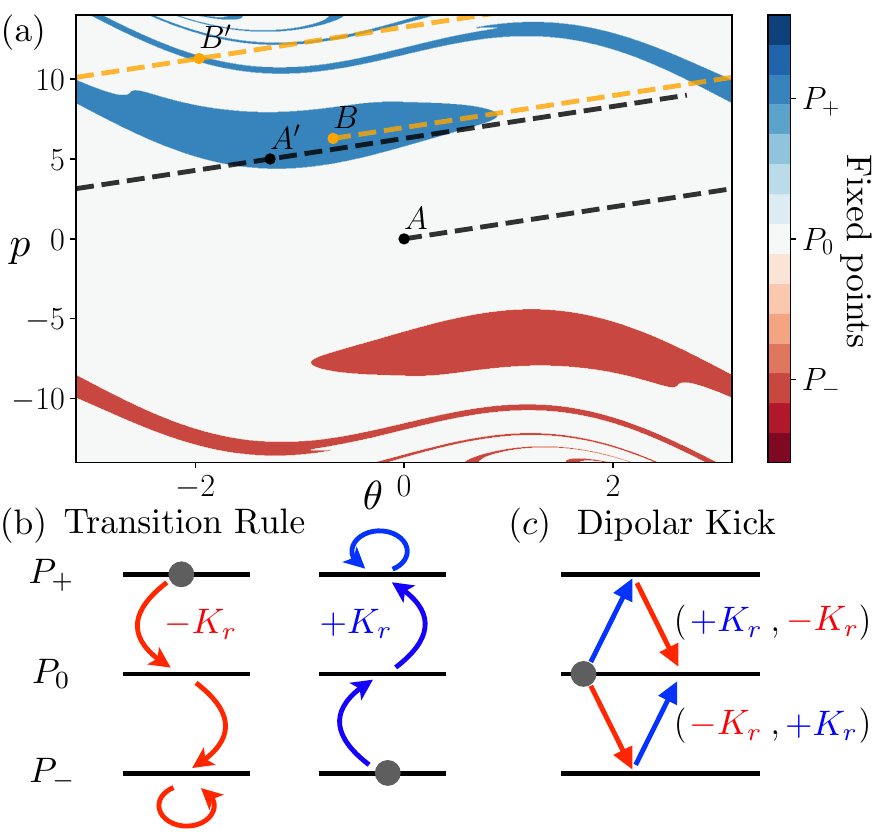}
\caption{(a) Basins of fixed points in the dissipative kicked rotor model. Transition rules at stroboscopic times can be determined by tuning the random kick strength $K_r$. For instance, after a positive kick $+K_r$, the rotor can suddenly jump from point $A$ to $A'$, which is located in the basin of the fixed point $P_+$ (blue), realizing $P_0{\to} P_+$.  (b) Exact transition rules. (c) A dipolar kick induces a time rondeau order without generating unwanted heating.}
    \label{fig:transitionrule}
\end{figure}

We provide an affirmative answer by studying systems with multistability, i.e., multiple attractors coexist s.t. the system evolves to one of them after a sufficiently long waiting time $T$, a ubiquitous feature in dissipative systems~\cite{boccaletti2002synchronization,pisarchik2014control,landa2020multistability,alaeian2022exact}. Perhaps one of the simplest physical examples is the periodically kicked classical rotor system in the presence of damping~\cite{zaslavsky1978simplest}, exhibiting a tunable number of fixed points in the phase space. We introduce a random protocol with a minimal ``dipolar" correlation that enables precise transition rules between these fixed points. Therefore, the rotor randomly traverses multiple pathways between states and returns to the same initial state after a complete dipolar kick, cf. Fig.~\ref{fig:transitionrule}, thereby establishing the rondeau order without inducing excessive heating. 
Remarkably, a simple analysis of the concomitant Markovian chain reveals that synchronization occurs~\cite{matsumoto1983noise,van1994noise,zhou2002noise,teramae2004robustness,goldobin2005synchronization,goychuk2006quantum,huang2020synchronization,schmolke2022noise}: even for a randomly distributed initial ensemble of non-interacting rotors, the system inevitably exhibits synchronized motion with stable rondeau order. {Note that in Ref.~\cite{PhysRevB.108.L100203}, the Ising symmetry is crucial for the realization of TRCs. However, symmetry is not essential in our construction since multistability can also exist in the absence of manifest symmetries.}

The presence of many-body interactions and a finite waiting time $T$ introduces uncertainties in the transition rules, which compete with perfect synchronization. Our central finding is that the synchronized phase remains robust against perturbations, with a de-synchronization phase transition occurring at a finite critical interaction strength in the thermodynamic limit. Finally, to investigate the asymptotic behavior of spatial fluctuations, we perform a linear stability analysis (LSA), which quantitatively captures the phase boundary obtained by many-body simulations.

\section{Random kicked one-rotor model}
\par We start from a periodically kicked rotor described by the Hamiltonian $H(t){=}p^2/2 {-}K_0\cos\theta \sum_n \delta(t{-}n)$. By introducing damping, we obtain the discretized equations of motion
\begin{equation}
    \begin{array}{l}
p(t+1)=\gamma p(t)-K_{0} \sin \theta(t), \\
\theta(t+1)=\theta(t)+p(t+1)(\bmod \, 2\pi),
\end{array}
\label{EOMa}
\end{equation}
where $(p,\theta)$ denote the angular momentum and the angle, {$1-\gamma$} is the dissipation rate, and $K_0$ is the strength of periodic kicks~\cite{zaslavsky1978simplest}. The long-time behavior of the system depends on specific parameter values, which we fix as $\gamma{=}0.8,K_0{=}2$, such that the system exhibits three stationary states, or fixed points: {$(\Theta_-{=}\arcsin[2\pi(1-\gamma)/K_0],P_-{=}{-}2\pi),(\Theta_0{=}0,P_0{=}0),(\Theta_+{=}-\arcsin[2\pi(1-\gamma)/K_0],P_+{=}+2\pi)$}. {Other parameter regimes can also lead to three fixed points, see discussions in Ref.~\cite{yan2025bifurcationsintermittencycoupleddissipative}.}

Their corresponding basins are depicted in Fig.~\ref{fig:transitionrule}(a). After a sufficiently long waiting time $T$, any initial condition in a colored region will asymptotically evolve towards the corresponding fixed point.

Supposing initially the system is located at one of the fixed points, we now introduce an additional random kick with the Hamiltonian
$V(t){=}{-}\sum_m K(t)\theta\delta(t{-}mT)$ where $K(t)$ takes a binary value $\pm K_r$, and we dub $mT$ ($m$ takes integer values) as stroboscopic times. Consequently,
a sudden change occurs at $t=mT$, $p{\to} p {\pm} K_r$ and $\theta {\to} \theta {\pm} K_r\ \text{mod} \  2\pi$. For simplicity, we consider $T{\to}\infty$ such that before the next random kick, the rotor can equilibrate again at one of the three fixed points. Therefore, $V(t)$ induces precise transitions between different fixed points without generating excessive heating.

Clearly, such a transition strongly depends on the specific value of $K_r$. 
For example, if we choose  $K_r \in(4.93,5.06)$~\footnote{We discuss different parameter regimes and the stability of the transition rule in the Sec.~SM~1 in the Supplementary Material} and consider a rotor starting from the fixed point $(\Theta_0, P_0)$, point $A$ in  Fig.~\ref{fig:transitionrule}(a). After a positive kick $+K_r$, the rotor suddenly jumps from point $A$ to $A'$, which is located in the basin of the fixed point $P_+$, the blue region in Fig.~\ref{fig:transitionrule}(a). 
Therefore, between two neighboring stroboscopic times, the transition $P_0\to P_+$ is realized. On the other hand, if the initial point starts from point $B$, $(\Theta_+,P_+)$, it jumps to $B'$ after $+K_r$, which also sits in the basin of $P_+$. Hence, stroboscopically, the rotor actually remains unchanged. 
Consequently, the following transition rules are achieved:
$P_-{\to} P_0{\to}P_+{\to} P_+,$ cf. blue lines in Fig.~\ref{fig:transitionrule}(b).
In other words, the system absorbs the input momentum from the drive, unless the rotor already has a maximum momentum, in which case the excess momentum is damped. A similar effect occurs for a negative kick ${-}K_r$:
     $P_+{\to} P_0{\to} P_-{\to}P_-.$

\begin{figure}[t]
    \centering
\includegraphics[width=0.9\linewidth]{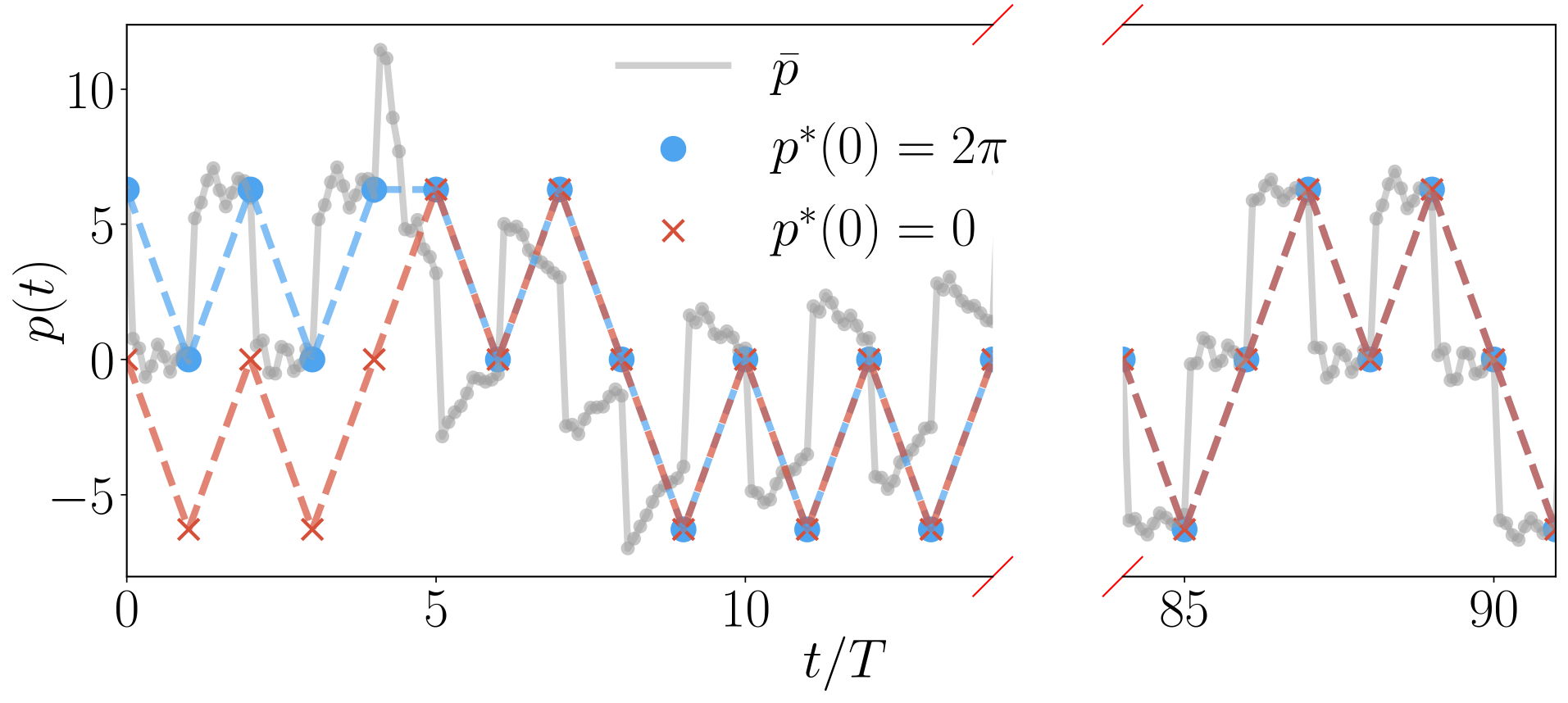}
    \caption{
    Red and blue trajectories of a single rotor quickly synchronize, exhibiting the time rondeau order where the long-time order at $t{=}2mT$ coexists with the short-time disorder. 
    The gray line depicts the average momentum in a many-rotor system, which follows the single-rotor trajectory.   
    We use $J{=}0.25,$ $K_r{=}5.5$ for the numerical simulation.
    The initial $\theta_i$ is randomly sampled within
$[0, 2\pi]$ and $p_i$ is sampled around $P_+$ according to a Gaussian distribution of a standard deviation $0.1$.
    }
    \label{fig:P_tra}
\end{figure}

We employ this set of transition rules to realize the time rondeau order. 
 The key is to introduce a dipolar structure to the random kick, i.e., at two consecutive stroboscopic times we randomly select one of the two kick sequences, $(+K_r,-K_r)$ or $(-K_r,+K_r)$. 
 To see its dynamical consequences, we consider a simple yet insightful scenario where a single rotor starts from the initial condition $p(0){=}0, \theta(0){=}0$. As illustrated in Fig.~\ref{fig:transitionrule}(c), after this dipolar kick the rotor always returns to its origin, establishing a long-time order, while the rotor can traverse multiple pathways, either
$P_0 {\to} P_+ {\to} P_0$ or $P_0 {\to} P_- {\to} P_0,$
demonstrating a short-time disorder. A typical trajectory is plotted in Fig.~\ref{fig:P_tra} in red. {The coexistence of the long-time order and the short-time disorder is the defining feature of a time rondeau.
 }

Remarkably, regardless of the initial condition of the system, the rondeau order always appears at long times. This can be revealed by considering an ensemble of uncoupled rotors, starting from an arbitrary distribution of three fixed points, $\boldsymbol{W}(0)$. Its stroboscopic evolution can be obtained by $\boldsymbol{W}(2mT{+}2T){=}\boldsymbol{A}\boldsymbol{W}(2mT)$ with a stochastic matrix of the Markovian chain~\cite{meyn2012markov}:
\begin{eqnarray}
    \boldsymbol{A}{=}
\begin{pmatrix}
  1/2 &  & \\
 1/2 & 1 & 1/2\\
  &  & 1/2
\end{pmatrix},
\end{eqnarray}
where the matrix elements $A_{i,j}$ denote the probability of updating the momentum from the $i$-th to $j$-th fixed points after one dipolar kick. 
Note, this dynamical system has a unique feature that if the momentum at time $2m
T$ deviates from $P_0$, the system always has the probability of $1/2$ to correct it back to $P_0$ after one dipolar kick. Hence, the system exponentially converges to the stationary solution within four stroboscopic periods on average where the entire ensemble synchronizes and occupies $P_0$ at stroboscopic times~\cite{huang2020synchronization}. As shown in Fig.~\ref{fig:P_tra}, the blue trajectory quickly converges to the red one if their random kick sequences are the same, {despite different initial conditions}. 
Therefore, the transition rule (Fig.~\ref{fig:transitionrule}) provides exceptional stability for the rondeau order, which is robust against any initial state imperfections.

\section{Many-body system with a finite waiting time}
The analysis above is exact only for uncoupled systems in the limit $T{\to}\infty$. Many-body interactions and a finite $T$ inevitably introduce perturbations to the exact transition rule, potentially destabilizing the rondeau order. Yet, we will show that the TRC is indeed robust, and a de-synchronization phase transition occurs at a critical interaction strength in the thermodynamic limit.

To show this, we consider a many-rotor chain of size $L$ with nearest-neighbor interactions, $H_I{=}{-}J\sum_i[\cos(\theta_i-\theta_{i+1}){+}\cos(\theta_i{-}\theta_{i-1})]\sum_n \delta(t{-}n)$ with $J{>}0$, resulting in
\begin{equation}
    \begin{aligned}
p_{i}(t&+1)=\gamma p_{i}(t)-K_{0} \sin \theta_{i}(t)\pm K_{r}\delta_{t,mT}  \\
& +J[\sin(\theta_i(t)-\theta_{i+1}(t))+\sin(\theta_i(t)-\theta_{i-1}(t))], \\
\theta_{i}(t&+1) =\theta_{i}(t)+p_{i}(t+1)(\bmod \, 2\pi),
\end{aligned}
\label{EOMs}
\end{equation}
where $i$ denotes the site number and $\pm$ depends on the driving sequence being applied. In the following discussion, we fix $T{=}10$, which is far from the $T{\to}\infty$ limit even for the non-interacting limit; see details in Sec.~SM~1 in the Supplementary Material (SM)~\cite{supplement}.

\begin{figure}[t]
    \centering    \includegraphics[width=\linewidth]{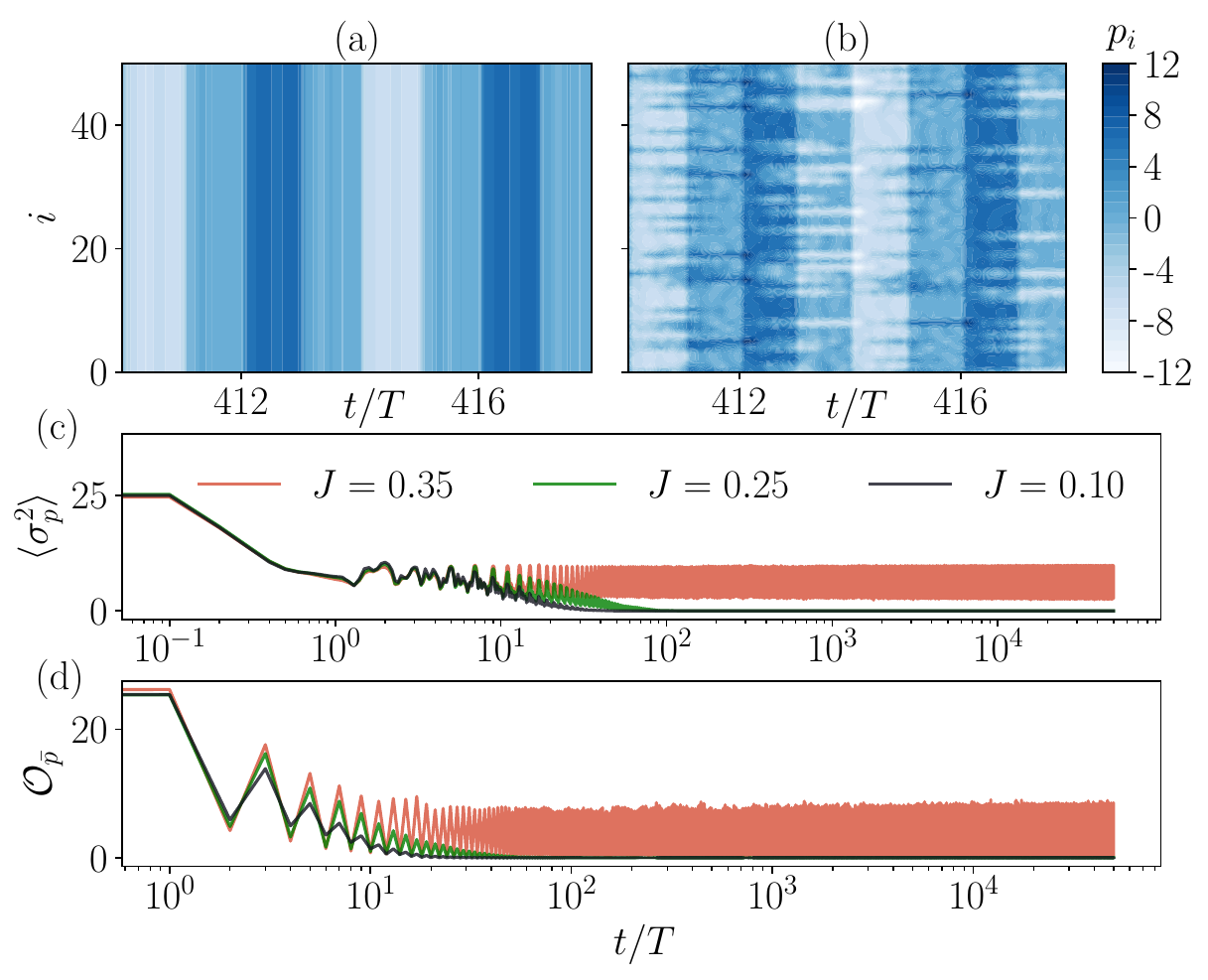}
    \caption{(a) Spatial-temporal distribution of momentum in the synchronized phase for weak $J{=}0.25$. (b) Defects persist for stronger interaction $J{=}0.35$. Both (a) and (b)
    use the same random kick sequence. (c) and (d) Momentum variance and time rondeau order parameter for different $J$. Ensemble averages over different initial states and random drive realizations are performed. 
    The system demonstrates synchronized TRC for a weak interaction (black and green) while a larger interaction induces de-synchronization (red). 
    The initial momentum is sampled from a Gaussian distribution with a standard deviation $6$ and zero average.
    We use parameters $K_r=5.5, L=256$ and a periodic boundary condition for numerical simulations.}
    \label{fig:avar-t}
\end{figure}

We first consider weak interaction strength and a spatially inhomogeneous initial state, where $\theta_i$ is randomly sampled within $[0,2\pi]$ and $p_i$ is sampled around a fixed point according to a Gaussian distribution. As shown in Fig.~\ref{fig:P_tra}, at early times ($t/T{\in}[4,11]$) the mean angular momentum $\bar{p}{=}\sum_ip_i/L$ (grey) exhibits a notable deviation from the exact single-rotor trajectory $p^*$ at stroboscopic times (blue), while $\bar{p}$ starts to follow $p^*$ at longer times. We define 
$
\mathcal{O}_{\bar{p}}(m)=[\bar{p}(mT)-p^*(m)]^2,$
which quantifies their deviation as the TRC order parameter.
As shown in Fig.~\ref{fig:avar-t}(d), indeed, $\mathcal{O}_{\bar{p}}$ decays to zero (green and black), confirming the appearance of TRC in the presence of many-body interactions. 

At long times, all rotors synchronize {and become spatially ordered} just as in the non-interacting case, where the system develops a homogeneous distribution of angular momentum, cf. Fig.~\ref{fig:avar-t}(a).
This can be confirmed via the spatial variance $\sigma^2_p{=}\sum_{i=1}^L (p_i{-}\bar{p})^2/L$. As shown in Fig.~\ref{fig:avar-t}(c), after a short transient regime, the ensemble-averaged value $\langle\sigma^2_p\rangle$ eventually drops to zero (black and green). Such a decay 
 occurs exponentially fast in time, and the corresponding time scale $\tau$ converges to a finite value as long as $L$ is sufficiently large, as detailed in Sec.~SM~2.1~\cite{supplement}, confirming that such a synchronized TRC remains stable in the thermodynamic limit.

Crucially, we note that for larger $J$ this time scale also increases, cf. Fig.~\ref{fig:avar-t}(c). This happens because a stronger interaction can maintain and even enhance the spatial inhomogeneity, via generating defects on top of the synchronized background, Fig.~\ref{fig:avar-t}(b). Hence, in general, it takes a longer time for dissipation to stabilize the system. 
Perfect synchronization breaks down for large $J$ and a finite-size system may exhibit intermittent synchronization~\cite{berger2021intermittent}, where the full synchronization and non-synchronized dynamics alternate irregularly in time; see Sec.~SM~2.1 for details~\cite{supplement}. Yet, when $L{\to}\infty$, intermittent synchronization is unstable and the de-synchronization phase transition occurs at a critical value $J_c$. A finite density of defects survives indefinitely in the de-synchronization phase with non-vanishing spatial fluctuations, while the rondeau order diminishes with non-zero $\mathcal{O}_{\bar{p}}$ at long times, see red curves in Fig.~\ref{fig:avar-t}(c) and (d).

\begin{figure}[t]
    \centering
\includegraphics[width=\linewidth]{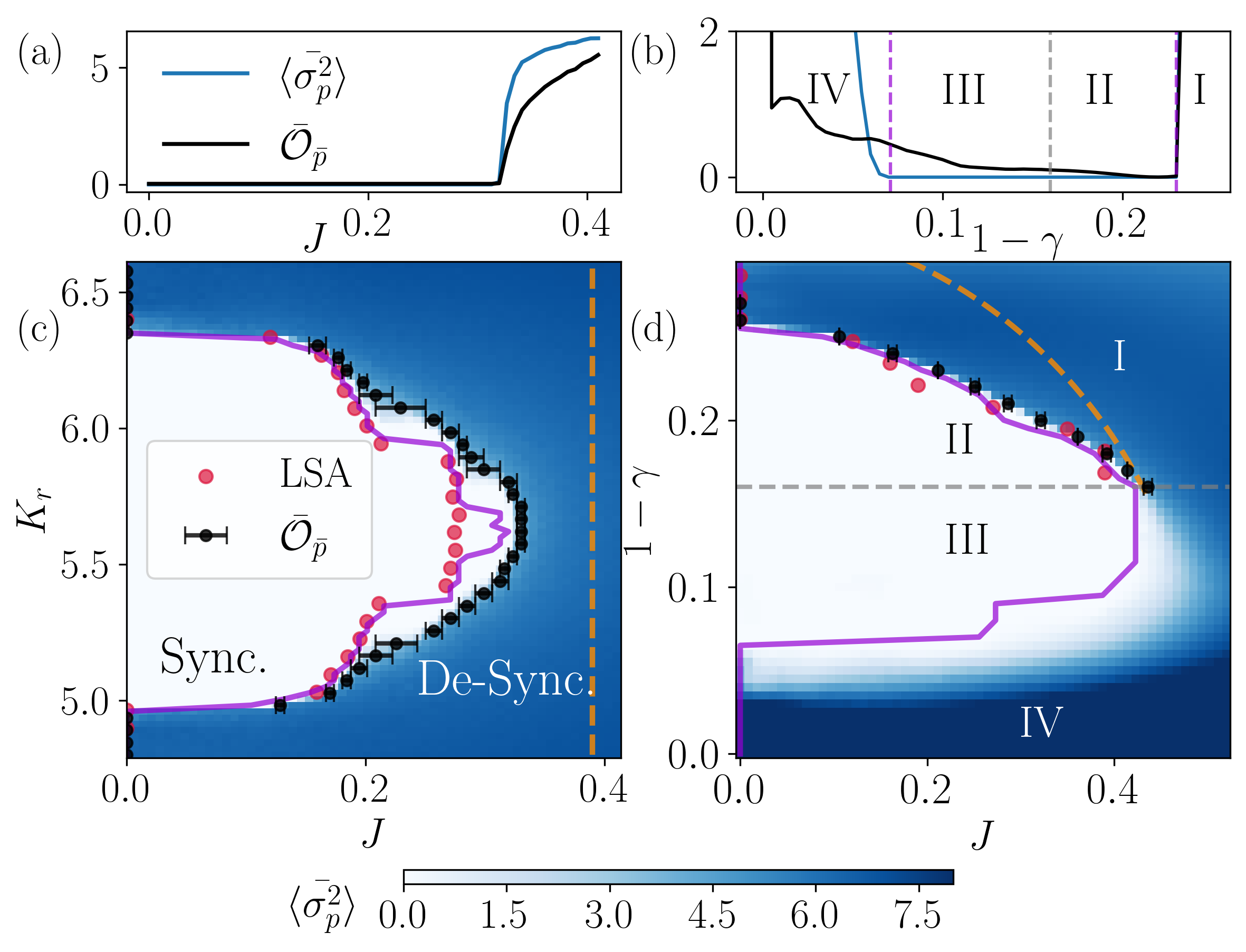}
    \caption{(a) Dependence of order parameters on $J$ for $K_r{=}5.5,\gamma{=}0.8$. The phase transition occurs at $J_c{\approx}0.32$. {(b) Dependence of order parameters on the dissipation rate $1{-}\gamma$ for $K_r{=}5.5,J{=}0.2$.  Regions I and IV correspond to the de-synchronization phase. In Regions II and III, the system is synchronized, $\langle\bar{\sigma}_p^2\rangle{=}0$.
    However, only for $1{-}\gamma{>}0.16$ (gray dashed line in panels (b) and (d)), the single kicked rotor has three fixed points 
and the many-body system exhibits the desired synchronized TRC.}
    (c) Phase diagram for different $K_r$ and $J$. {(d) Phase diagram for different $1{-}\gamma$ and $J$.  The color in both (c) and (d) denotes the value of $\langle\bar{\sigma}_p^2\rangle$ where we truncate its maximum value to $8$.} White and blue regions correspond to the synchronization and de-synchronization phases, respectively. The purple curve corresponds to $\langle\bar{\sigma}_p^2\rangle{=}10^{-5}$. A linear stability analysis (LSA) captures the synchronization phase transition (red). The $T{\to}\infty$ assumption predicts the critical interaction strength, orange dashed lines in (c) and (d), which notably overestimate the phase boundary. The black dots denote the phase boundary obtained by $\bar{\mathcal{O}}_{\bar{p}}$. Initial state distribution is the same as in Fig.~\ref{fig:avar-t}.  
    }
    \label{fig:phase}
\end{figure}

In Fig.~\ref{fig:phase}(a), we plot the long-time saturation value $\langle\bar{\sigma}_p^2\rangle$ of the momentum variance and $\bar{\mathcal{O}}_{\bar{p}}$ for different $J$ and the phase transition occurs around $J_c{\approx} 0.32$ when $K_r{=}5.5$~\footnote{Long-time saturation values are obtained by averaging order parameters(system size $L=500$) at 100 stroboscopic times, starting from $t=10^5$.}. The phase transition point shows no dependence on the initial states, as synchronization occurs at early times (see Sec.~SM~2.3 for the numerical verification~\cite{supplement}). We further scan over a wide range of the random kick strengths $K_r$ and $J$ and map out the phase diagram in Fig.~\ref{fig:phase}(c): white and blue regions correspond to the synchronization ($\langle\bar{\sigma}_p^2\rangle{=}0$) and de-synchronization phases ($\langle\bar{\sigma}_p^2\rangle{\neq} 0$), respectively, and the purple line corresponds to $\langle\bar{\sigma}_p^2\rangle{=}10^{-5}$. Black dots correspond to the average of three $J$ values such that the order parameter
$\bar{\mathcal{O}}_{\bar{p}}$ equals 0.2, 1.1 and 2, which match with the purple line with good accuracy. 

{We further analyze the phase diagram for different dissipation rate $1{-}\gamma$. We first fix $J{=}0.2$ and plot the order parameters in Fig.~\ref{fig:phase}(b). For a large dissipation rate, the system is de-synchronized (Region I) where both $\langle\bar{\sigma}_p^2\rangle$ and $\bar{\mathcal{O}}_{\bar{p}}$ have sizable values. In Regions II and III,
the system is synchronized with vanishing $\langle\bar{\sigma}_p^2\rangle$. However, only for $1{-}\gamma{>}0.16$ (gray dashed line in panels (b) and (d)), the single kicked rotor has three fixed points 
and hence the many-body system exhibits the desired synchronized TRC phase. Instead, for $1{-}\gamma{<}0.16$, an increasing number of fixed points appears for a smaller dissipation rate and eventually the single rotor exhibits chaotic motions. Hence, the driving protocol designed for the rondeau order already fails at the single rotor level. Similarly, as shown in Region IV, heating cannot be properly suppressed and
the system enters the de-synchronization phase. In particular, when dissipation is absent ($1{-}\gamma{=}0$), the system heats up to an infinite temperature ensemble where both order parameters diverge.  The phase diagram for different values of $1{-}\gamma$ and $J$ is depicted in Fig.~\ref{fig:phase}(d), where the synchronized TRC exists in the white region above the gray dashed line. These observations highlight the importance of a suitably chosen range of dissipation rate in realizing the stable TRC phase, which cannot persist indefinitely in closed systems due to the heating effect.} 

\section{Linear Stability Analysis} 
The de-synchronization phase transition can be well captured by a linear stability analysis (red dots in Fig.~\ref{fig:phase}(c) {and (d)}). 
The synchronized evolution can be captured by a mean-field solution $(\bar{p}(t),\bar{\theta}(t))$, i.e., the spatial average of momentum and angle which follow the single-rotor EOM, Eq.~\eqref{EOMa}. Many-body interactions generate spatial fluctuations, $\Delta \theta_i(t){=}\theta_i(t){-}\bar{\theta}(t)$ and $\Delta p_i(t){=}p_i(t){-}\bar{p}(t)$. We assume $\Delta \theta_i(t){\ll} 1$, $\Delta p_i(t){\ll} 1$ and only keep its leading order contributions to Eq.~\eqref{EOMs}, $\Delta p_{i}(t{+}1){=}\gamma \Delta p_{i}(t)-K_{0} \cos \bar{\theta}(t)\Delta \theta_{i}(t){+}J[2\Delta \theta_i(t){-}\Delta \theta_{i+1}(t){-}\Delta \theta_{i-1}(t)], \Delta \theta_{i}(t+1){=}\Delta \theta_{i}(t){+}\Delta p_{i}(t{+}1) (\text {mod} 2\pi).$ 
A Fourier transformation, $\Delta p_k(t){=}\sum_j \Delta p_j e^{-ijk}/\sqrt{N}$ and $\Delta \theta_k(t)=\sum_j \Delta \theta_j e^{-ijk}/\sqrt{N}$, leads to the decoupled EOMs 
\begin{eqnarray}
\begin{aligned}
\begin{pmatrix}
 \Delta p_k(t+1)\\
\Delta \theta_k(t+1)
\end{pmatrix} 
&=\boldsymbol{A}_k(t)
\begin{pmatrix}
 \Delta p_k(t)\\
\Delta \theta_k(t)
\end{pmatrix},\  
\end{aligned}
\end{eqnarray}
for each quasi-momentum $k$ mode with the Jacobian matrix
\begin{eqnarray}
\begin{aligned}
\boldsymbol{A}_k(\bar{\theta}(t)) &=
\begin{pmatrix}
 \gamma  & 2J(1-\cos(k))-K_0\cos\bar{\theta}(t)\\
 \gamma  & 1+2J(1-\cos(k))-K_0\cos\bar{\theta}(t)
\end{pmatrix}.
\end{aligned}
\end{eqnarray}
The dependence on the random kick sequence and $K_r$ are entirely contained in $\bar{\theta}(t)$ and $\boldsymbol{A}_k$ varies in time. 

For simplicity, we first consider the limit $T{\to} \infty$, where the transition rules are exact, and $(\bar{p}(t),\bar{\theta}(t))$ only has three possible choices - three fixed points. The stability of this mean-field solution can be determined by all the eigenvalues $\lambda_k$ of $\boldsymbol{A}_k$ for each fixed point: 
If $\text{max}|\lambda_k|{<}1$ for all $k$, the trajectory is stable and fluctuations eventually vanish. {Larger $J$ may generate unstable $k$ modes with $|\lambda_k|{>}1$} and the critical value is determined when $\text{max}|\lambda_k(J_c)|{=}1$. This leads to the orange line in Fig.~\ref{fig:phase} {(c) and (d)}; however, it notably overestimates $J_c$. Crucially, the dependence of $J_c$ on the random kick strength $K_r$ cannot be captured.

\begin{figure}
    \centering
\includegraphics[width=\linewidth]{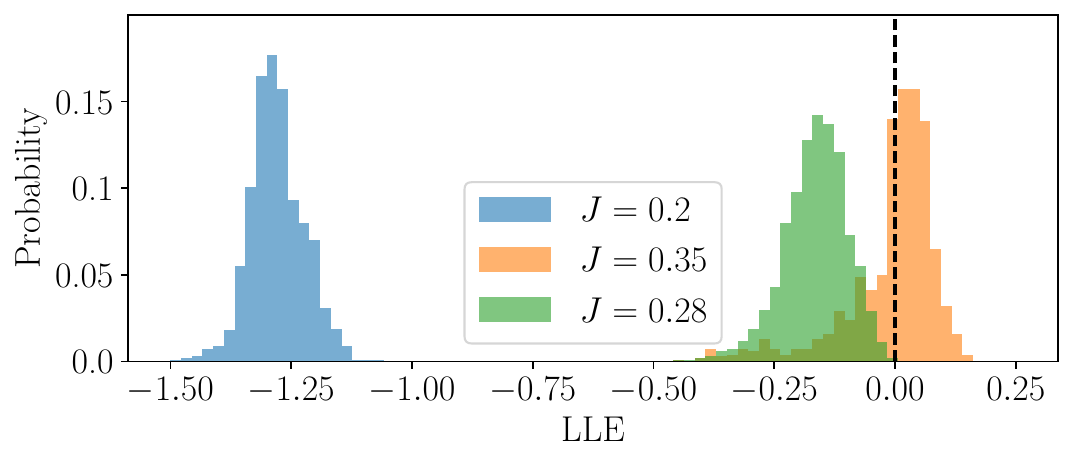}
    \caption{Distribution of largest Lyapunov exponents (LLEs). We consider a de-synchronization phase transition occurs when 5\% of LLEs become positive. Numerically we use $K_r{=}5.5, m{=}300$ and 1000 different mean-field trajectories. Convergence of the distribution is discussed in Sec.~SM~2.4~\cite{supplement}.}
    \label{fig:lle}
\end{figure}
A finite $T$ plays a crucial role in determining the TRC stability and the deviation between $(\bar{p}(t),\bar{\theta}(t))$ and the fixed points cannot be neglected. We define $\boldsymbol{D}_k$ as the product of $\boldsymbol{A}_k$ calculated along a mean-field trajectory $\bar{\theta}(t)$~\cite{wolf1985determining} during $m$ dipolar drives
$
\boldsymbol{D}_{k}=\prod_{i=0}^{2mT-1} \boldsymbol{A}_k(\bar{\theta}(i))
$.
For a given trajectory, we obtain the eigenvalues $\lambda_k$ of $\boldsymbol{D}_{k}$ and the corresponding largest Lyapunov exponents (LLEs), $\lim_{m\to\infty}\log(\max|\lambda_k|)/m$, can be obtained. In Fig.~\ref{fig:lle}, we plot the distribution of LLEs for different mean-field trajectories. In the synchronization phase (blue) LLEs are negative, while for large $J$ a notable fraction of LLEs become positive. We estimate the critical $J_c$ when 5\% of LLEs become positive, and as shown in Fig.~\ref{fig:phase}(b) and (d), the result matches well with the phase boundary obtained by the many-body simulation. However, below the gray dashed in  panel (d), this method is no longer applicable since the fixed points are unstable even for a single-rotor.

\section{Discussion} 
Our Letter opens a promising avenue for stabilizing the partial temporal order via dissipation, demonstrating the existence of TRCs that are absolutely
stable against perturbations in both initial states
and many-body interactions. One important ingredient is multistability, such that the system jumps among different fixed points without generating unwanted heating. A dipolar structure encodes the rondeau order and synchronization notably strengthens its stability. 

Going beyond, dissipative systems can exhibit a versatile structure of stationary states, such as limit cycles~\cite{piazza2015self} and (quasi-)periodic orbits~\cite{russomanno2023spatiotemporally} in addition to the fixed points considered here. We expect that the partial temporal order can be significantly enriched in these settings. It is worth noting that the dipolar structure can be straightforwardly extended to higher-order multipoles and the aperiodic Thue-Morse sequence~\cite{zhao2021random}. We anticipate enhanced stability of the synchronized TRCs for higher multipolar orders and a systematic study would be worth pursuing.

The presence of many-body interactions sustains spatially inhomogeneous defects. Its competition with the synchronized dynamics leads to the de-synchronization phase transition. We perform large-scale and long-time numerical simulations of the classical many-body dynamics. It allows us to map out the entire phase diagram and explore the robustness of TRCs against perturbations. {Crucially, the phase boundary of the synchronized TRC cannot be captured by the standard linear stability analysis (LSA) around the fixed points, as is normally done in periodically driven systems. Instead, properly incorporating the deviation between the transient mean-field trajectory and the fixed points notably improves the performance of the LSA. Our method thus paves the way for further investigations into the stability of non-equilibrium phases beyond the conventional Floquet phases of matter.}

While we have demonstrated stable TRCs in classical many-body systems for numerical efficiency, our construction is readily generalizable to quantum systems, where both multistability~\cite{landa2020multistability,xu2024phase,Solanki2024Exotic} and synchronization have been reported~\cite{goychuk2006quantum,xu2014synchronization,schmolke2022noise}. {In Sec.~SM~3~\cite{supplement}, we illustrate one concrete example using dissipative Rydberg atoms with bistability~\cite{PhysRevLett.122.015701}.} It remains an interesting question to investigate the competition between quantum fluctuations and synchronization, which may induce non-equilibrium quantum phase transitions that are fundamentally different. 
As detailed in Sec.~SM~4~\cite{supplement}, our rotor system can be experimentally implemented on the superconducting quantum simulation platform~\cite{cataliottiJosephsonJunctionArrays2001,Josephson-PhysRevB}. Such a system may serve as a natural testbed to reveal distinct behaviors in the non-equilibrium temporal order between classical and quantum systems.

\section{Acknowledgments}
This work is supported by the National Natural Science Foundation of China (Grants No. 12234002, 12474214, 12474486, and 92250303), by the National Key Research and Development Program of China (Grant No. 2024YFA1612101), by Innovation Program for Quantum Science and Technology
(Grant No. 2024ZD0301800), and by “The Fundamental Research Funds for the Central Universities, Peking University” and "High-performance Computing Platform of Peking University". We thank Johannes Knolle for initiating this work and stimulating discussions. We also thank Marin Bukov for many useful discussions.

\bibliography{references}

\begin{thebibliography}{88}%
\makeatletter
\providecommand \@ifxundefined [1]{%
 \@ifx{#1\undefined}
}%
\providecommand \@ifnum [1]{%
 \ifnum #1\expandafter \@firstoftwo
 \else \expandafter \@secondoftwo
 \fi
}%
\providecommand \@ifx [1]{%
 \ifx #1\expandafter \@firstoftwo
 \else \expandafter \@secondoftwo
 \fi
}%
\providecommand \natexlab [1]{#1}%
\providecommand \enquote  [1]{``#1''}%
\providecommand \bibnamefont  [1]{#1}%
\providecommand \bibfnamefont [1]{#1}%
\providecommand \citenamefont [1]{#1}%
\providecommand \href@noop [0]{\@secondoftwo}%
\providecommand \href [0]{\begingroup \@sanitize@url \@href}%
\providecommand \@href[1]{\@@startlink{#1}\@@href}%
\providecommand \@@href[1]{\endgroup#1\@@endlink}%
\providecommand \@sanitize@url [0]{\catcode `\\12\catcode `\$12\catcode `\&12\catcode `\#12\catcode `\^12\catcode `\_12\catcode `\%12\relax}%
\providecommand \@@startlink[1]{}%
\providecommand \@@endlink[0]{}%
\providecommand \url  [0]{\begingroup\@sanitize@url \@url }%
\providecommand \@url [1]{\endgroup\@href {#1}{\urlprefix }}%
\providecommand \urlprefix  [0]{URL }%
\providecommand \Eprint [0]{\href }%
\providecommand \doibase [0]{https://doi.org/}%
\providecommand \selectlanguage [0]{\@gobble}%
\providecommand \bibinfo  [0]{\@secondoftwo}%
\providecommand \bibfield  [0]{\@secondoftwo}%
\providecommand \translation [1]{[#1]}%
\providecommand \BibitemOpen [0]{}%
\providecommand \bibitemStop [0]{}%
\providecommand \bibitemNoStop [0]{.\EOS\space}%
\providecommand \EOS [0]{\spacefactor3000\relax}%
\providecommand \BibitemShut  [1]{\csname bibitem#1\endcsname}%
\let\auto@bib@innerbib\@empty
%</preamble>
\bibitem [{\citenamefont {Khemani}\ \emph {et~al.}(2016)\citenamefont {Khemani}, \citenamefont {Lazarides}, \citenamefont {Moessner},\ and\ \citenamefont {Sondhi}}]{khemani2016phase}%
  \BibitemOpen
  \bibfield  {author} {\bibinfo {author} {\bibfnamefont {V.}~\bibnamefont {Khemani}}, \bibinfo {author} {\bibfnamefont {A.}~\bibnamefont {Lazarides}}, \bibinfo {author} {\bibfnamefont {R.}~\bibnamefont {Moessner}},\ and\ \bibinfo {author} {\bibfnamefont {S.~L.}\ \bibnamefont {Sondhi}},\ }\bibfield  {title} {\bibinfo {title} {Phase structure of driven quantum systems},\ }\href {https://doi.org/10.1103/PhysRevLett.116.250401} {\bibfield  {journal} {\bibinfo  {journal} {Phys. Rev. Lett.}\ }\textbf {\bibinfo {volume} {116}},\ \bibinfo {pages} {250401} (\bibinfo {year} {2016})}\BibitemShut {NoStop}%
\bibitem [{\citenamefont {Else}\ \emph {et~al.}(2016)\citenamefont {Else}, \citenamefont {Bauer},\ and\ \citenamefont {Nayak}}]{else2016floquet}%
  \BibitemOpen
  \bibfield  {author} {\bibinfo {author} {\bibfnamefont {D.~V.}\ \bibnamefont {Else}}, \bibinfo {author} {\bibfnamefont {B.}~\bibnamefont {Bauer}},\ and\ \bibinfo {author} {\bibfnamefont {C.}~\bibnamefont {Nayak}},\ }\bibfield  {title} {\bibinfo {title} {Floquet time crystals},\ }\href {https://doi.org/10.1103/PhysRevLett.117.090402} {\bibfield  {journal} {\bibinfo  {journal} {Phys. Rev. Lett.}\ }\textbf {\bibinfo {volume} {117}},\ \bibinfo {pages} {090402} (\bibinfo {year} {2016})}\BibitemShut {NoStop}%
\bibitem [{\citenamefont {Yao}\ \emph {et~al.}(2017)\citenamefont {Yao}, \citenamefont {Potter}, \citenamefont {Potirniche},\ and\ \citenamefont {Vishwanath}}]{dtc_prl2017}%
  \BibitemOpen
  \bibfield  {author} {\bibinfo {author} {\bibfnamefont {N.~Y.}\ \bibnamefont {Yao}}, \bibinfo {author} {\bibfnamefont {A.~C.}\ \bibnamefont {Potter}}, \bibinfo {author} {\bibfnamefont {I.-D.}\ \bibnamefont {Potirniche}},\ and\ \bibinfo {author} {\bibfnamefont {A.}~\bibnamefont {Vishwanath}},\ }\bibfield  {title} {\bibinfo {title} {Discrete time crystals: Rigidity, criticality, and realizations},\ }\href {https://doi.org/10.1103/PhysRevLett.118.030401} {\bibfield  {journal} {\bibinfo  {journal} {Phys. Rev. Lett.}\ }\textbf {\bibinfo {volume} {118}},\ \bibinfo {pages} {030401} (\bibinfo {year} {2017})}\BibitemShut {NoStop}%
\bibitem [{\citenamefont {Gambetta}\ \emph {et~al.}(2019)\citenamefont {Gambetta}, \citenamefont {Carollo}, \citenamefont {Marcuzzi}, \citenamefont {Garrahan},\ and\ \citenamefont {Lesanovsky}}]{PhysRevLett.122.015701}%
  \BibitemOpen
  \bibfield  {author} {\bibinfo {author} {\bibfnamefont {F.~M.}\ \bibnamefont {Gambetta}}, \bibinfo {author} {\bibfnamefont {F.}~\bibnamefont {Carollo}}, \bibinfo {author} {\bibfnamefont {M.}~\bibnamefont {Marcuzzi}}, \bibinfo {author} {\bibfnamefont {J.~P.}\ \bibnamefont {Garrahan}},\ and\ \bibinfo {author} {\bibfnamefont {I.}~\bibnamefont {Lesanovsky}},\ }\bibfield  {title} {\bibinfo {title} {Discrete time crystals in the absence of manifest symmetries or disorder in open quantum systems},\ }\href {https://doi.org/10.1103/PhysRevLett.122.015701} {\bibfield  {journal} {\bibinfo  {journal} {Phys. Rev. Lett.}\ }\textbf {\bibinfo {volume} {122}},\ \bibinfo {pages} {015701} (\bibinfo {year} {2019})}\BibitemShut {NoStop}%
\bibitem [{\citenamefont {Verdeny}\ \emph {et~al.}(2016)\citenamefont {Verdeny}, \citenamefont {Puig},\ and\ \citenamefont {Mintert}}]{verdeny2016quasi}%
  \BibitemOpen
  \bibfield  {author} {\bibinfo {author} {\bibfnamefont {A.}~\bibnamefont {Verdeny}}, \bibinfo {author} {\bibfnamefont {J.}~\bibnamefont {Puig}},\ and\ \bibinfo {author} {\bibfnamefont {F.}~\bibnamefont {Mintert}},\ }\bibfield  {title} {\bibinfo {title} {Quasi-periodically driven quantum systems},\ }\href {https://doi.org/10.1515/zna-2016-0079} {\bibfield  {journal} {\bibinfo  {journal} {Zeitschrift f{\"u}r Naturforschung A}\ }\textbf {\bibinfo {volume} {71}},\ \bibinfo {pages} {897} (\bibinfo {year} {2016})}\BibitemShut {NoStop}%
\bibitem [{\citenamefont {Nandy}\ \emph {et~al.}(2017)\citenamefont {Nandy}, \citenamefont {Sen},\ and\ \citenamefont {Sen}}]{nandy2017aperiodically}%
  \BibitemOpen
  \bibfield  {author} {\bibinfo {author} {\bibfnamefont {S.}~\bibnamefont {Nandy}}, \bibinfo {author} {\bibfnamefont {A.}~\bibnamefont {Sen}},\ and\ \bibinfo {author} {\bibfnamefont {D.}~\bibnamefont {Sen}},\ }\bibfield  {title} {\bibinfo {title} {Aperiodically driven integrable systems and their emergent steady states},\ }\href {https://doi.org/10.1103/PhysRevX.7.031034} {\bibfield  {journal} {\bibinfo  {journal} {Phys. Rev. X}\ }\textbf {\bibinfo {volume} {7}},\ \bibinfo {pages} {031034} (\bibinfo {year} {2017})}\BibitemShut {NoStop}%
\bibitem [{\citenamefont {Mori}\ \emph {et~al.}(2021)\citenamefont {Mori}, \citenamefont {Zhao}, \citenamefont {Mintert}, \citenamefont {Knolle},\ and\ \citenamefont {Moessner}}]{mori2021rigorous}%
  \BibitemOpen
  \bibfield  {author} {\bibinfo {author} {\bibfnamefont {T.}~\bibnamefont {Mori}}, \bibinfo {author} {\bibfnamefont {H.}~\bibnamefont {Zhao}}, \bibinfo {author} {\bibfnamefont {F.}~\bibnamefont {Mintert}}, \bibinfo {author} {\bibfnamefont {J.}~\bibnamefont {Knolle}},\ and\ \bibinfo {author} {\bibfnamefont {R.}~\bibnamefont {Moessner}},\ }\bibfield  {title} {\bibinfo {title} {Rigorous bounds on the heating rate in thue-morse quasiperiodically and randomly driven quantum many-body systems},\ }\href {https://doi.org/10.1103/PhysRevLett.127.050602} {\bibfield  {journal} {\bibinfo  {journal} {Phys. Rev. Lett.}\ }\textbf {\bibinfo {volume} {127}},\ \bibinfo {pages} {050602} (\bibinfo {year} {2021})}\BibitemShut {NoStop}%
\bibitem [{\citenamefont {Wen}\ \emph {et~al.}(2021)\citenamefont {Wen}, \citenamefont {Fan}, \citenamefont {Vishwanath},\ and\ \citenamefont {Gu}}]{wen2021periodically}%
  \BibitemOpen
  \bibfield  {author} {\bibinfo {author} {\bibfnamefont {X.}~\bibnamefont {Wen}}, \bibinfo {author} {\bibfnamefont {R.}~\bibnamefont {Fan}}, \bibinfo {author} {\bibfnamefont {A.}~\bibnamefont {Vishwanath}},\ and\ \bibinfo {author} {\bibfnamefont {Y.}~\bibnamefont {Gu}},\ }\bibfield  {title} {\bibinfo {title} {Periodically, quasiperiodically, and randomly driven conformal field theories},\ }\href {https://doi.org/10.1103/PhysRevResearch.3.023044} {\bibfield  {journal} {\bibinfo  {journal} {Phys. Rev. Res.}\ }\textbf {\bibinfo {volume} {3}},\ \bibinfo {pages} {023044} (\bibinfo {year} {2021})}\BibitemShut {NoStop}%
\bibitem [{\citenamefont {Long}\ \emph {et~al.}(2022)\citenamefont {Long}, \citenamefont {Crowley},\ and\ \citenamefont {Chandran}}]{long2022many}%
  \BibitemOpen
  \bibfield  {author} {\bibinfo {author} {\bibfnamefont {D.~M.}\ \bibnamefont {Long}}, \bibinfo {author} {\bibfnamefont {P.~J.}\ \bibnamefont {Crowley}},\ and\ \bibinfo {author} {\bibfnamefont {A.}~\bibnamefont {Chandran}},\ }\bibfield  {title} {\bibinfo {title} {Many-body localization with quasiperiodic driving},\ }\href {https://doi.org/10.1103/PhysRevB.105.144204} {\bibfield  {journal} {\bibinfo  {journal} {Phys. Rev. B}\ }\textbf {\bibinfo {volume} {105}},\ \bibinfo {pages} {144204} (\bibinfo {year} {2022})}\BibitemShut {NoStop}%
\bibitem [{\citenamefont {He}\ \emph {et~al.}(2023)\citenamefont {He}, \citenamefont {Ye}, \citenamefont {Gong}, \citenamefont {Liu}, \citenamefont {Murch}, \citenamefont {Yao},\ and\ \citenamefont {Zu}}]{he2023quasi}%
  \BibitemOpen
  \bibfield  {author} {\bibinfo {author} {\bibfnamefont {G.}~\bibnamefont {He}}, \bibinfo {author} {\bibfnamefont {B.}~\bibnamefont {Ye}}, \bibinfo {author} {\bibfnamefont {R.}~\bibnamefont {Gong}}, \bibinfo {author} {\bibfnamefont {Z.}~\bibnamefont {Liu}}, \bibinfo {author} {\bibfnamefont {K.~W.}\ \bibnamefont {Murch}}, \bibinfo {author} {\bibfnamefont {N.~Y.}\ \bibnamefont {Yao}},\ and\ \bibinfo {author} {\bibfnamefont {C.}~\bibnamefont {Zu}},\ }\bibfield  {title} {\bibinfo {title} {Quasi-floquet prethermalization in a disordered dipolar spin ensemble in diamond},\ }\href {https://doi.org/10.1103/PhysRevLett.131.130401} {\bibfield  {journal} {\bibinfo  {journal} {Phys. Rev. Lett.}\ }\textbf {\bibinfo {volume} {131}},\ \bibinfo {pages} {130401} (\bibinfo {year} {2023})}\BibitemShut {NoStop}%
\bibitem [{\citenamefont {Gallone}\ and\ \citenamefont {Langella}(2024)}]{gallone2024prethermalization}%
  \BibitemOpen
  \bibfield  {author} {\bibinfo {author} {\bibfnamefont {M.}~\bibnamefont {Gallone}}\ and\ \bibinfo {author} {\bibfnamefont {B.}~\bibnamefont {Langella}},\ }\bibfield  {title} {\bibinfo {title} {Prethermalization and conservation laws in quasi-periodically driven quantum systems},\ }\href {https://link.springer.com/article/10.1007/s10955-024-03313-9} {\bibfield  {journal} {\bibinfo  {journal} {J. Stat. Phys.}\ }\textbf {\bibinfo {volume} {191}},\ \bibinfo {pages} {100} (\bibinfo {year} {2024})}\BibitemShut {NoStop}%
\bibitem [{\citenamefont {Ghosh}\ \emph {et~al.}(2024)\citenamefont {Ghosh}, \citenamefont {Bhattacharjee},\ and\ \citenamefont {Bandyopadhyay}}]{ghosh2024slow}%
  \BibitemOpen
  \bibfield  {author} {\bibinfo {author} {\bibfnamefont {S.}~\bibnamefont {Ghosh}}, \bibinfo {author} {\bibfnamefont {S.}~\bibnamefont {Bhattacharjee}},\ and\ \bibinfo {author} {\bibfnamefont {S.}~\bibnamefont {Bandyopadhyay}},\ }\bibfield  {title} {\bibinfo {title} {Slow relaxation of quasi-periodically driven integrable quantum many-body systems},\ }\href {https://arxiv.org/abs/2404.06667} {\bibfield  {journal} {\bibinfo  {journal} {arXiv preprint arXiv:2404.06667}\ } (\bibinfo {year} {2024})}\BibitemShut {NoStop}%
\bibitem [{\citenamefont {Schmid}\ \emph {et~al.}(2025)\citenamefont {Schmid}, \citenamefont {Peng}, \citenamefont {Refael},\ and\ \citenamefont {von Oppen}}]{schmid2024self}%
  \BibitemOpen
  \bibfield  {author} {\bibinfo {author} {\bibfnamefont {H.}~\bibnamefont {Schmid}}, \bibinfo {author} {\bibfnamefont {Y.}~\bibnamefont {Peng}}, \bibinfo {author} {\bibfnamefont {G.}~\bibnamefont {Refael}},\ and\ \bibinfo {author} {\bibfnamefont {F.}~\bibnamefont {von Oppen}},\ }\bibfield  {title} {\bibinfo {title} {Self-similar phase diagram of the fibonacci-driven quantum ising model},\ }\href {https://doi.org/10.1103/hn66-j8pt} {\bibfield  {journal} {\bibinfo  {journal} {Phys. Rev. Lett.}\ }\textbf {\bibinfo {volume} {134}},\ \bibinfo {pages} {240404} (\bibinfo {year} {2025})}\BibitemShut {NoStop}%
\bibitem [{\citenamefont {Dumitrescu}\ \emph {et~al.}(2018)\citenamefont {Dumitrescu}, \citenamefont {Vasseur},\ and\ \citenamefont {Potter}}]{dumitrescu2018logarithmically}%
  \BibitemOpen
  \bibfield  {author} {\bibinfo {author} {\bibfnamefont {P.~T.}\ \bibnamefont {Dumitrescu}}, \bibinfo {author} {\bibfnamefont {R.}~\bibnamefont {Vasseur}},\ and\ \bibinfo {author} {\bibfnamefont {A.~C.}\ \bibnamefont {Potter}},\ }\bibfield  {title} {\bibinfo {title} {Logarithmically slow relaxation in quasiperiodically driven random spin chains},\ }\href {https://doi.org/10.1103/PhysRevLett.120.070602} {\bibfield  {journal} {\bibinfo  {journal} {Phys. Rev. Lett.}\ }\textbf {\bibinfo {volume} {120}},\ \bibinfo {pages} {070602} (\bibinfo {year} {2018})}\BibitemShut {NoStop}%
\bibitem [{\citenamefont {Zhao}\ \emph {et~al.}(2019)\citenamefont {Zhao}, \citenamefont {Mintert},\ and\ \citenamefont {Knolle}}]{zhao2019floquet}%
  \BibitemOpen
  \bibfield  {author} {\bibinfo {author} {\bibfnamefont {H.}~\bibnamefont {Zhao}}, \bibinfo {author} {\bibfnamefont {F.}~\bibnamefont {Mintert}},\ and\ \bibinfo {author} {\bibfnamefont {J.}~\bibnamefont {Knolle}},\ }\bibfield  {title} {\bibinfo {title} {Floquet time spirals and stable discrete-time quasicrystals in quasiperiodically driven quantum many-body systems},\ }\href {https://doi.org/10.1103/PhysRevB.100.134302} {\bibfield  {journal} {\bibinfo  {journal} {Phys. Rev. B}\ }\textbf {\bibinfo {volume} {100}},\ \bibinfo {pages} {134302} (\bibinfo {year} {2019})}\BibitemShut {NoStop}%
\bibitem [{\citenamefont {Else}\ \emph {et~al.}(2020)\citenamefont {Else}, \citenamefont {Ho},\ and\ \citenamefont {Dumitrescu}}]{else2020long}%
  \BibitemOpen
  \bibfield  {author} {\bibinfo {author} {\bibfnamefont {D.~V.}\ \bibnamefont {Else}}, \bibinfo {author} {\bibfnamefont {W.~W.}\ \bibnamefont {Ho}},\ and\ \bibinfo {author} {\bibfnamefont {P.~T.}\ \bibnamefont {Dumitrescu}},\ }\bibfield  {title} {\bibinfo {title} {Long-lived interacting phases of matter protected by multiple time-translation symmetries in quasiperiodically driven systems},\ }\href {https://doi.org/10.1103/PhysRevX.10.021032} {\bibfield  {journal} {\bibinfo  {journal} {Phys. Rev. X}\ }\textbf {\bibinfo {volume} {10}},\ \bibinfo {pages} {021032} (\bibinfo {year} {2020})}\BibitemShut {NoStop}%
\bibitem [{\citenamefont {Zhao}\ \emph {et~al.}(2023)\citenamefont {Zhao}, \citenamefont {Knolle},\ and\ \citenamefont {Moessner}}]{PhysRevB.108.L100203}%
  \BibitemOpen
  \bibfield  {author} {\bibinfo {author} {\bibfnamefont {H.}~\bibnamefont {Zhao}}, \bibinfo {author} {\bibfnamefont {J.}~\bibnamefont {Knolle}},\ and\ \bibinfo {author} {\bibfnamefont {R.}~\bibnamefont {Moessner}},\ }\bibfield  {title} {\bibinfo {title} {Temporal disorder in spatiotemporal order},\ }\href {https://doi.org/10.1103/PhysRevB.108.L100203} {\bibfield  {journal} {\bibinfo  {journal} {Phys. Rev. B}\ }\textbf {\bibinfo {volume} {108}},\ \bibinfo {pages} {L100203} (\bibinfo {year} {2023})}\BibitemShut {NoStop}%
\bibitem [{\citenamefont {Moon}\ \emph {et~al.}(2025)\citenamefont {Moon}, \citenamefont {Schindler}, \citenamefont {Sun}, \citenamefont {Druga}, \citenamefont {Knolle}, \citenamefont {Moessner}, \citenamefont {Zhao}, \citenamefont {Bukov},\ and\ \citenamefont {Ajoy}}]{moon2024experimental}%
  \BibitemOpen
  \bibfield  {author} {\bibinfo {author} {\bibfnamefont {L.}~\bibnamefont {Moon}}, \bibinfo {author} {\bibfnamefont {P.}~\bibnamefont {Schindler}}, \bibinfo {author} {\bibfnamefont {Y.}~\bibnamefont {Sun}}, \bibinfo {author} {\bibfnamefont {E.}~\bibnamefont {Druga}}, \bibinfo {author} {\bibfnamefont {J.}~\bibnamefont {Knolle}}, \bibinfo {author} {\bibfnamefont {R.}~\bibnamefont {Moessner}}, \bibinfo {author} {\bibfnamefont {H.}~\bibnamefont {Zhao}}, \bibinfo {author} {\bibfnamefont {M.}~\bibnamefont {Bukov}},\ and\ \bibinfo {author} {\bibfnamefont {A.}~\bibnamefont {Ajoy}},\ }\bibfield  {title} {\bibinfo {title} {Experimental observation of a time rondeau crystal},\ }\href {https://doi.org/10.1038/s41567-025-03028-y} {\bibfield  {journal} {\bibinfo  {journal} {Nature Physics}\ ,\ \bibinfo {pages} {1}} (\bibinfo {year} {2025})}\BibitemShut {NoStop}%
\bibitem [{\citenamefont {Zhao}\ \emph {et~al.}(2021)\citenamefont {Zhao}, \citenamefont {Mintert}, \citenamefont {Moessner},\ and\ \citenamefont {Knolle}}]{zhao2021random}%
  \BibitemOpen
  \bibfield  {author} {\bibinfo {author} {\bibfnamefont {H.}~\bibnamefont {Zhao}}, \bibinfo {author} {\bibfnamefont {F.}~\bibnamefont {Mintert}}, \bibinfo {author} {\bibfnamefont {R.}~\bibnamefont {Moessner}},\ and\ \bibinfo {author} {\bibfnamefont {J.}~\bibnamefont {Knolle}},\ }\bibfield  {title} {\bibinfo {title} {Random multipolar driving: Tunably slow heating through spectral engineering},\ }\href {https://doi.org/10.1103/PhysRevLett.126.040601} {\bibfield  {journal} {\bibinfo  {journal} {Phys. Rev. Lett.}\ }\textbf {\bibinfo {volume} {126}},\ \bibinfo {pages} {040601} (\bibinfo {year} {2021})}\BibitemShut {NoStop}%
\bibitem [{\citenamefont {Wen}\ \emph {et~al.}(2022)\citenamefont {Wen}, \citenamefont {Gu}, \citenamefont {Vishwanath},\ and\ \citenamefont {Fan}}]{wen2022periodically2}%
  \BibitemOpen
  \bibfield  {author} {\bibinfo {author} {\bibfnamefont {X.}~\bibnamefont {Wen}}, \bibinfo {author} {\bibfnamefont {Y.}~\bibnamefont {Gu}}, \bibinfo {author} {\bibfnamefont {A.}~\bibnamefont {Vishwanath}},\ and\ \bibinfo {author} {\bibfnamefont {R.}~\bibnamefont {Fan}},\ }\bibfield  {title} {\bibinfo {title} {Periodically, quasi-periodically, and randomly driven conformal field theories (ii): Furstenberg's theorem and exceptions to heating phases},\ }\href {https://api.semanticscholar.org/CorpusID:237605361} {\bibfield  {journal} {\bibinfo  {journal} {SciPost Phys.}\ }\textbf {\bibinfo {volume} {13}},\ \bibinfo {pages} {082} (\bibinfo {year} {2022})}\BibitemShut {NoStop}%
\bibitem [{\citenamefont {Yan}\ \emph {et~al.}(2024)\citenamefont {Yan}, \citenamefont {Moessner},\ and\ \citenamefont {Zhao}}]{yan2024prethermalization}%
  \BibitemOpen
  \bibfield  {author} {\bibinfo {author} {\bibfnamefont {J.}~\bibnamefont {Yan}}, \bibinfo {author} {\bibfnamefont {R.}~\bibnamefont {Moessner}},\ and\ \bibinfo {author} {\bibfnamefont {H.}~\bibnamefont {Zhao}},\ }\bibfield  {title} {\bibinfo {title} {Prethermalization in aperiodically kicked many-body dynamics},\ }\href {https://doi.org/10.1103/PhysRevB.109.064305} {\bibfield  {journal} {\bibinfo  {journal} {Phys. Rev. B}\ }\textbf {\bibinfo {volume} {109}},\ \bibinfo {pages} {064305} (\bibinfo {year} {2024})}\BibitemShut {NoStop}%
\bibitem [{\citenamefont {Tiwari}\ \emph {et~al.}(2024)\citenamefont {Tiwari}, \citenamefont {Bhakuni},\ and\ \citenamefont {Sharma}}]{tiwari2024dynamical}%
  \BibitemOpen
  \bibfield  {author} {\bibinfo {author} {\bibfnamefont {V.}~\bibnamefont {Tiwari}}, \bibinfo {author} {\bibfnamefont {D.~S.}\ \bibnamefont {Bhakuni}},\ and\ \bibinfo {author} {\bibfnamefont {A.}~\bibnamefont {Sharma}},\ }\bibfield  {title} {\bibinfo {title} {Dynamical localization and slow dynamics in quasiperiodically driven quantum systems},\ }\href {https://doi.org/10.1103/PhysRevB.109.L161104} {\bibfield  {journal} {\bibinfo  {journal} {Phys. Rev. B}\ }\textbf {\bibinfo {volume} {109}},\ \bibinfo {pages} {L161104} (\bibinfo {year} {2024})}\BibitemShut {NoStop}%
\bibitem [{\citenamefont {Levi}\ \emph {et~al.}(2016)\citenamefont {Levi}, \citenamefont {Heyl}, \citenamefont {Lesanovsky},\ and\ \citenamefont {Garrahan}}]{levi2016robustness}%
  \BibitemOpen
  \bibfield  {author} {\bibinfo {author} {\bibfnamefont {E.}~\bibnamefont {Levi}}, \bibinfo {author} {\bibfnamefont {M.}~\bibnamefont {Heyl}}, \bibinfo {author} {\bibfnamefont {I.}~\bibnamefont {Lesanovsky}},\ and\ \bibinfo {author} {\bibfnamefont {J.~P.}\ \bibnamefont {Garrahan}},\ }\bibfield  {title} {\bibinfo {title} {Robustness of many-body localization in the presence of dissipation},\ }\href {https://doi.org/10.1103/PhysRevLett.116.237203} {\bibfield  {journal} {\bibinfo  {journal} {Phys. Rev. Lett.}\ }\textbf {\bibinfo {volume} {116}},\ \bibinfo {pages} {237203} (\bibinfo {year} {2016})}\BibitemShut {NoStop}%
\bibitem [{\citenamefont {Gopalakrishnan}\ \emph {et~al.}(2017)\citenamefont {Gopalakrishnan}, \citenamefont {Islam},\ and\ \citenamefont {Knap}}]{gopalakrishnan2017noise}%
  \BibitemOpen
  \bibfield  {author} {\bibinfo {author} {\bibfnamefont {S.}~\bibnamefont {Gopalakrishnan}}, \bibinfo {author} {\bibfnamefont {K.~R.}\ \bibnamefont {Islam}},\ and\ \bibinfo {author} {\bibfnamefont {M.}~\bibnamefont {Knap}},\ }\bibfield  {title} {\bibinfo {title} {Noise-induced subdiffusion in strongly localized quantum systems},\ }\href {https://doi.org/10.1103/PhysRevLett.119.046601} {\bibfield  {journal} {\bibinfo  {journal} {Phys. Rev. Lett.}\ }\textbf {\bibinfo {volume} {119}},\ \bibinfo {pages} {046601} (\bibinfo {year} {2017})}\BibitemShut {NoStop}%
\bibitem [{\citenamefont {Rieder}\ \emph {et~al.}(2018)\citenamefont {Rieder}, \citenamefont {Sieberer}, \citenamefont {Fischer},\ and\ \citenamefont {Fulga}}]{rieder2018localization}%
  \BibitemOpen
  \bibfield  {author} {\bibinfo {author} {\bibfnamefont {M.-T.}\ \bibnamefont {Rieder}}, \bibinfo {author} {\bibfnamefont {L.~M.}\ \bibnamefont {Sieberer}}, \bibinfo {author} {\bibfnamefont {M.~H.}\ \bibnamefont {Fischer}},\ and\ \bibinfo {author} {\bibfnamefont {I.~C.}\ \bibnamefont {Fulga}},\ }\bibfield  {title} {\bibinfo {title} {Localization counteracts decoherence in noisy floquet topological chains},\ }\href {https://doi.org/10.1103/PhysRevLett.120.216801} {\bibfield  {journal} {\bibinfo  {journal} {Phys. Rev. Lett.}\ }\textbf {\bibinfo {volume} {120}},\ \bibinfo {pages} {216801} (\bibinfo {year} {2018})}\BibitemShut {NoStop}%
\bibitem [{\citenamefont {Zhao}\ \emph {et~al.}(2022)\citenamefont {Zhao}, \citenamefont {Mintert}, \citenamefont {Knolle},\ and\ \citenamefont {Moessner}}]{zhao2022localization}%
  \BibitemOpen
  \bibfield  {author} {\bibinfo {author} {\bibfnamefont {H.}~\bibnamefont {Zhao}}, \bibinfo {author} {\bibfnamefont {F.}~\bibnamefont {Mintert}}, \bibinfo {author} {\bibfnamefont {J.}~\bibnamefont {Knolle}},\ and\ \bibinfo {author} {\bibfnamefont {R.}~\bibnamefont {Moessner}},\ }\bibfield  {title} {\bibinfo {title} {Localization persisting under aperiodic driving},\ }\href {https://doi.org/10.1103/PhysRevB.105.L220202} {\bibfield  {journal} {\bibinfo  {journal} {Phys. Rev. B}\ }\textbf {\bibinfo {volume} {105}},\ \bibinfo {pages} {L220202} (\bibinfo {year} {2022})}\BibitemShut {NoStop}%
\bibitem [{\citenamefont {Lazarides}\ \emph {et~al.}(2014)\citenamefont {Lazarides}, \citenamefont {Das},\ and\ \citenamefont {Moessner}}]{lazarides2014equilibrium}%
  \BibitemOpen
  \bibfield  {author} {\bibinfo {author} {\bibfnamefont {A.}~\bibnamefont {Lazarides}}, \bibinfo {author} {\bibfnamefont {A.}~\bibnamefont {Das}},\ and\ \bibinfo {author} {\bibfnamefont {R.}~\bibnamefont {Moessner}},\ }\bibfield  {title} {\bibinfo {title} {Equilibrium states of generic quantum systems subject to periodic driving},\ }\href {https://doi.org/10.1103/PhysRevE.90.012110} {\bibfield  {journal} {\bibinfo  {journal} {Phys. Rev. E}\ }\textbf {\bibinfo {volume} {90}},\ \bibinfo {pages} {012110} (\bibinfo {year} {2014})}\BibitemShut {NoStop}%
\bibitem [{\citenamefont {Kim}\ \emph {et~al.}(2014)\citenamefont {Kim}, \citenamefont {Ikeda},\ and\ \citenamefont {Huse}}]{kim2014testing}%
  \BibitemOpen
  \bibfield  {author} {\bibinfo {author} {\bibfnamefont {H.}~\bibnamefont {Kim}}, \bibinfo {author} {\bibfnamefont {T.~N.}\ \bibnamefont {Ikeda}},\ and\ \bibinfo {author} {\bibfnamefont {D.~A.}\ \bibnamefont {Huse}},\ }\bibfield  {title} {\bibinfo {title} {Testing whether all eigenstates obey the eigenstate thermalization hypothesis},\ }\href {https://doi.org/10.1103/PhysRevE.90.052105} {\bibfield  {journal} {\bibinfo  {journal} {Phys. Rev. E}\ }\textbf {\bibinfo {volume} {90}},\ \bibinfo {pages} {052105} (\bibinfo {year} {2014})}\BibitemShut {NoStop}%
\bibitem [{\citenamefont {D'Alessio}\ and\ \citenamefont {Rigol}(2014)}]{Isolated2014}%
  \BibitemOpen
  \bibfield  {author} {\bibinfo {author} {\bibfnamefont {L.}~\bibnamefont {D'Alessio}}\ and\ \bibinfo {author} {\bibfnamefont {M.}~\bibnamefont {Rigol}},\ }\bibfield  {title} {\bibinfo {title} {Long-time {{Behavior}} of {{Isolated Periodically Driven Interacting Lattice Systems}}},\ }\href {https://doi.org/10.1103/PhysRevX.4.041048} {\bibfield  {journal} {\bibinfo  {journal} {Phys. Rev. X}\ }\textbf {\bibinfo {volume} {4}},\ \bibinfo {pages} {041048} (\bibinfo {year} {2014})}\BibitemShut {NoStop}%
\bibitem [{\citenamefont {Bukov}\ \emph {et~al.}(2015)\citenamefont {Bukov}, \citenamefont {Gopalakrishnan}, \citenamefont {Knap},\ and\ \citenamefont {Demler}}]{bukov2015prethermal}%
  \BibitemOpen
  \bibfield  {author} {\bibinfo {author} {\bibfnamefont {M.}~\bibnamefont {Bukov}}, \bibinfo {author} {\bibfnamefont {S.}~\bibnamefont {Gopalakrishnan}}, \bibinfo {author} {\bibfnamefont {M.}~\bibnamefont {Knap}},\ and\ \bibinfo {author} {\bibfnamefont {E.}~\bibnamefont {Demler}},\ }\bibfield  {title} {\bibinfo {title} {Prethermal floquet steady states and instabilities in the periodically driven, weakly interacting bose-hubbard model},\ }\href {https://doi.org/10.1103/PhysRevLett.115.205301} {\bibfield  {journal} {\bibinfo  {journal} {Phys. Rev. Lett.}\ }\textbf {\bibinfo {volume} {115}},\ \bibinfo {pages} {205301} (\bibinfo {year} {2015})}\BibitemShut {NoStop}%
\bibitem [{\citenamefont {Kuwahara}\ \emph {et~al.}(2016)\citenamefont {Kuwahara}, \citenamefont {Mori},\ and\ \citenamefont {Saito}}]{kuwahara2016floquet}%
  \BibitemOpen
  \bibfield  {author} {\bibinfo {author} {\bibfnamefont {T.}~\bibnamefont {Kuwahara}}, \bibinfo {author} {\bibfnamefont {T.}~\bibnamefont {Mori}},\ and\ \bibinfo {author} {\bibfnamefont {K.}~\bibnamefont {Saito}},\ }\bibfield  {title} {\bibinfo {title} {Floquet--magnus theory and generic transient dynamics in periodically driven many-body quantum systems},\ }\href {https://www.sciencedirect.com/science/article/abs/pii/S0003491616000142} {\bibfield  {journal} {\bibinfo  {journal} {Ann. Phys.}\ }\textbf {\bibinfo {volume} {367}},\ \bibinfo {pages} {96} (\bibinfo {year} {2016})}\BibitemShut {NoStop}%
\bibitem [{\citenamefont {Else}\ \emph {et~al.}(2017)\citenamefont {Else}, \citenamefont {Bauer},\ and\ \citenamefont {Nayak}}]{else2017prethermal}%
  \BibitemOpen
  \bibfield  {author} {\bibinfo {author} {\bibfnamefont {D.~V.}\ \bibnamefont {Else}}, \bibinfo {author} {\bibfnamefont {B.}~\bibnamefont {Bauer}},\ and\ \bibinfo {author} {\bibfnamefont {C.}~\bibnamefont {Nayak}},\ }\bibfield  {title} {\bibinfo {title} {Prethermal phases of matter protected by time-translation symmetry},\ }\href {https://doi.org/10.1103/PhysRevX.7.011026} {\bibfield  {journal} {\bibinfo  {journal} {Phys. Rev. X}\ }\textbf {\bibinfo {volume} {7}},\ \bibinfo {pages} {011026} (\bibinfo {year} {2017})}\BibitemShut {NoStop}%
\bibitem [{\citenamefont {Mori}(2018)}]{mori2018floquet}%
  \BibitemOpen
  \bibfield  {author} {\bibinfo {author} {\bibfnamefont {T.}~\bibnamefont {Mori}},\ }\bibfield  {title} {\bibinfo {title} {Floquet prethermalization in periodically driven classical spin systems},\ }\href {https://doi.org/10.1103/PhysRevB.98.104303} {\bibfield  {journal} {\bibinfo  {journal} {Phys. Rev. B}\ }\textbf {\bibinfo {volume} {98}},\ \bibinfo {pages} {104303} (\bibinfo {year} {2018})}\BibitemShut {NoStop}%
\bibitem [{\citenamefont {Rajak}\ \emph {et~al.}(2019)\citenamefont {Rajak}, \citenamefont {Dana},\ and\ \citenamefont {Dalla~Torre}}]{Josephson-PhysRevB}%
  \BibitemOpen
  \bibfield  {author} {\bibinfo {author} {\bibfnamefont {A.}~\bibnamefont {Rajak}}, \bibinfo {author} {\bibfnamefont {I.}~\bibnamefont {Dana}},\ and\ \bibinfo {author} {\bibfnamefont {E.~G.}\ \bibnamefont {Dalla~Torre}},\ }\bibfield  {title} {\bibinfo {title} {Characterizations of prethermal states in periodically driven many-body systems with unbounded chaotic diffusion},\ }\href {https://doi.org/10.1103/PhysRevB.100.100302} {\bibfield  {journal} {\bibinfo  {journal} {Phys. Rev. B}\ }\textbf {\bibinfo {volume} {100}},\ \bibinfo {pages} {100302} (\bibinfo {year} {2019})}\BibitemShut {NoStop}%
\bibitem [{\citenamefont {Howell}\ \emph {et~al.}(2019)\citenamefont {Howell}, \citenamefont {Weinberg}, \citenamefont {Sels}, \citenamefont {Polkovnikov},\ and\ \citenamefont {Bukov}}]{howell2019asymptotic}%
  \BibitemOpen
  \bibfield  {author} {\bibinfo {author} {\bibfnamefont {O.}~\bibnamefont {Howell}}, \bibinfo {author} {\bibfnamefont {P.}~\bibnamefont {Weinberg}}, \bibinfo {author} {\bibfnamefont {D.}~\bibnamefont {Sels}}, \bibinfo {author} {\bibfnamefont {A.}~\bibnamefont {Polkovnikov}},\ and\ \bibinfo {author} {\bibfnamefont {M.}~\bibnamefont {Bukov}},\ }\bibfield  {title} {\bibinfo {title} {Asymptotic prethermalization in periodically driven classical spin chains},\ }\href {https://doi.org/10.1103/PhysRevLett.122.010602} {\bibfield  {journal} {\bibinfo  {journal} {Phys. Rev. Lett.}\ }\textbf {\bibinfo {volume} {122}},\ \bibinfo {pages} {010602} (\bibinfo {year} {2019})}\BibitemShut {NoStop}%
\bibitem [{\citenamefont {Luitz}\ \emph {et~al.}(2020)\citenamefont {Luitz}, \citenamefont {Moessner}, \citenamefont {Sondhi},\ and\ \citenamefont {Khemani}}]{luitz2020prethermalization}%
  \BibitemOpen
  \bibfield  {author} {\bibinfo {author} {\bibfnamefont {D.~J.}\ \bibnamefont {Luitz}}, \bibinfo {author} {\bibfnamefont {R.}~\bibnamefont {Moessner}}, \bibinfo {author} {\bibfnamefont {S.}~\bibnamefont {Sondhi}},\ and\ \bibinfo {author} {\bibfnamefont {V.}~\bibnamefont {Khemani}},\ }\bibfield  {title} {\bibinfo {title} {Prethermalization without temperature},\ }\href {https://doi.org/10.1103/PhysRevX.10.021046} {\bibfield  {journal} {\bibinfo  {journal} {Phys. Rev. X}\ }\textbf {\bibinfo {volume} {10}},\ \bibinfo {pages} {021046} (\bibinfo {year} {2020})}\BibitemShut {NoStop}%
\bibitem [{\citenamefont {Rubio-Abadal}\ \emph {et~al.}(2020)\citenamefont {Rubio-Abadal}, \citenamefont {Ippoliti}, \citenamefont {Hollerith}, \citenamefont {Wei}, \citenamefont {Rui}, \citenamefont {Sondhi}, \citenamefont {Khemani}, \citenamefont {Gross},\ and\ \citenamefont {Bloch}}]{rubio2020floquet}%
  \BibitemOpen
  \bibfield  {author} {\bibinfo {author} {\bibfnamefont {A.}~\bibnamefont {Rubio-Abadal}}, \bibinfo {author} {\bibfnamefont {M.}~\bibnamefont {Ippoliti}}, \bibinfo {author} {\bibfnamefont {S.}~\bibnamefont {Hollerith}}, \bibinfo {author} {\bibfnamefont {D.}~\bibnamefont {Wei}}, \bibinfo {author} {\bibfnamefont {J.}~\bibnamefont {Rui}}, \bibinfo {author} {\bibfnamefont {S.}~\bibnamefont {Sondhi}}, \bibinfo {author} {\bibfnamefont {V.}~\bibnamefont {Khemani}}, \bibinfo {author} {\bibfnamefont {C.}~\bibnamefont {Gross}},\ and\ \bibinfo {author} {\bibfnamefont {I.}~\bibnamefont {Bloch}},\ }\bibfield  {title} {\bibinfo {title} {Floquet prethermalization in a bose-hubbard system},\ }\href {https://doi.org/10.1103/PhysRevX.10.021044} {\bibfield  {journal} {\bibinfo  {journal} {Phys. Rev. X}\ }\textbf {\bibinfo {volume} {10}},\ \bibinfo {pages} {021044} (\bibinfo {year} {2020})}\BibitemShut {NoStop}%
\bibitem [{\citenamefont {Pizzi}\ \emph {et~al.}(2021)\citenamefont {Pizzi}, \citenamefont {Nunnenkamp},\ and\ \citenamefont {Knolle}}]{pizzi2021classical}%
  \BibitemOpen
  \bibfield  {author} {\bibinfo {author} {\bibfnamefont {A.}~\bibnamefont {Pizzi}}, \bibinfo {author} {\bibfnamefont {A.}~\bibnamefont {Nunnenkamp}},\ and\ \bibinfo {author} {\bibfnamefont {J.}~\bibnamefont {Knolle}},\ }\bibfield  {title} {\bibinfo {title} {Classical prethermal phases of matter},\ }\href {https://doi.org/10.1103/PhysRevLett.127.140602} {\bibfield  {journal} {\bibinfo  {journal} {Phys. Rev. Lett.}\ }\textbf {\bibinfo {volume} {127}},\ \bibinfo {pages} {140602} (\bibinfo {year} {2021})}\BibitemShut {NoStop}%
\bibitem [{\citenamefont {Ye}\ \emph {et~al.}(2021)\citenamefont {Ye}, \citenamefont {Machado},\ and\ \citenamefont {Yao}}]{ye2021floquet}%
  \BibitemOpen
  \bibfield  {author} {\bibinfo {author} {\bibfnamefont {B.}~\bibnamefont {Ye}}, \bibinfo {author} {\bibfnamefont {F.}~\bibnamefont {Machado}},\ and\ \bibinfo {author} {\bibfnamefont {N.~Y.}\ \bibnamefont {Yao}},\ }\bibfield  {title} {\bibinfo {title} {Floquet phases of matter via classical prethermalization},\ }\href {https://doi.org/10.1103/PhysRevLett.127.140603} {\bibfield  {journal} {\bibinfo  {journal} {Phys. Rev. Lett.}\ }\textbf {\bibinfo {volume} {127}},\ \bibinfo {pages} {140603} (\bibinfo {year} {2021})}\BibitemShut {NoStop}%
\bibitem [{\citenamefont {Peng}\ \emph {et~al.}(2021)\citenamefont {Peng}, \citenamefont {Yin}, \citenamefont {Huang}, \citenamefont {Ramanathan},\ and\ \citenamefont {Cappellaro}}]{peng2021floquet}%
  \BibitemOpen
  \bibfield  {author} {\bibinfo {author} {\bibfnamefont {P.}~\bibnamefont {Peng}}, \bibinfo {author} {\bibfnamefont {C.}~\bibnamefont {Yin}}, \bibinfo {author} {\bibfnamefont {X.}~\bibnamefont {Huang}}, \bibinfo {author} {\bibfnamefont {C.}~\bibnamefont {Ramanathan}},\ and\ \bibinfo {author} {\bibfnamefont {P.}~\bibnamefont {Cappellaro}},\ }\bibfield  {title} {\bibinfo {title} {Floquet prethermalization in dipolar spin chains},\ }\href {https://www.nature.com/articles/s41567-020-01120-z} {\bibfield  {journal} {\bibinfo  {journal} {Nat. Phys.}\ }\textbf {\bibinfo {volume} {17}},\ \bibinfo {pages} {444} (\bibinfo {year} {2021})}\BibitemShut {NoStop}%
\bibitem [{\citenamefont {Fleckenstein}\ and\ \citenamefont {Bukov}(2021)}]{fleckenstein2021thermalization}%
  \BibitemOpen
  \bibfield  {author} {\bibinfo {author} {\bibfnamefont {C.}~\bibnamefont {Fleckenstein}}\ and\ \bibinfo {author} {\bibfnamefont {M.}~\bibnamefont {Bukov}},\ }\bibfield  {title} {\bibinfo {title} {Thermalization and prethermalization in periodically kicked quantum spin chains},\ }\href {https://doi.org/10.1103/PhysRevB.103.144307} {\bibfield  {journal} {\bibinfo  {journal} {Phys. Rev. B}\ }\textbf {\bibinfo {volume} {103}},\ \bibinfo {pages} {144307} (\bibinfo {year} {2021})}\BibitemShut {NoStop}%
\bibitem [{\citenamefont {Ikeda}\ and\ \citenamefont {Polkovnikov}(2021)}]{ikeda2021fermi}%
  \BibitemOpen
  \bibfield  {author} {\bibinfo {author} {\bibfnamefont {T.~N.}\ \bibnamefont {Ikeda}}\ and\ \bibinfo {author} {\bibfnamefont {A.}~\bibnamefont {Polkovnikov}},\ }\bibfield  {title} {\bibinfo {title} {Fermi's golden rule for heating in strongly driven floquet systems},\ }\href {https://doi.org/10.1103/PhysRevB.104.134308} {\bibfield  {journal} {\bibinfo  {journal} {Phys. Rev. B}\ }\textbf {\bibinfo {volume} {104}},\ \bibinfo {pages} {134308} (\bibinfo {year} {2021})}\BibitemShut {NoStop}%
\bibitem [{\citenamefont {Mu\~noz Arias}\ \emph {et~al.}(2022)\citenamefont {Mu\~noz Arias}, \citenamefont {Chinni},\ and\ \citenamefont {Poggi}}]{munoz2022floquet}%
  \BibitemOpen
  \bibfield  {author} {\bibinfo {author} {\bibfnamefont {M.~H.}\ \bibnamefont {Mu\~noz Arias}}, \bibinfo {author} {\bibfnamefont {K.}~\bibnamefont {Chinni}},\ and\ \bibinfo {author} {\bibfnamefont {P.~M.}\ \bibnamefont {Poggi}},\ }\bibfield  {title} {\bibinfo {title} {Floquet time crystals in driven spin systems with all-to-all $p$-body interactions},\ }\href {https://doi.org/10.1103/PhysRevResearch.4.023018} {\bibfield  {journal} {\bibinfo  {journal} {Phys. Rev. Res.}\ }\textbf {\bibinfo {volume} {4}},\ \bibinfo {pages} {023018} (\bibinfo {year} {2022})}\BibitemShut {NoStop}%
\bibitem [{\citenamefont {Beatrez}\ \emph {et~al.}(2023)\citenamefont {Beatrez}, \citenamefont {Fleckenstein}, \citenamefont {Pillai}, \citenamefont {de~Leon~Sanchez}, \citenamefont {Akkiraju}, \citenamefont {Diaz~Alcala}, \citenamefont {Conti}, \citenamefont {Reshetikhin}, \citenamefont {Druga}, \citenamefont {Bukov} \emph {et~al.}}]{beatrez2023critical}%
  \BibitemOpen
  \bibfield  {author} {\bibinfo {author} {\bibfnamefont {W.}~\bibnamefont {Beatrez}}, \bibinfo {author} {\bibfnamefont {C.}~\bibnamefont {Fleckenstein}}, \bibinfo {author} {\bibfnamefont {A.}~\bibnamefont {Pillai}}, \bibinfo {author} {\bibfnamefont {E.}~\bibnamefont {de~Leon~Sanchez}}, \bibinfo {author} {\bibfnamefont {A.}~\bibnamefont {Akkiraju}}, \bibinfo {author} {\bibfnamefont {J.}~\bibnamefont {Diaz~Alcala}}, \bibinfo {author} {\bibfnamefont {S.}~\bibnamefont {Conti}}, \bibinfo {author} {\bibfnamefont {P.}~\bibnamefont {Reshetikhin}}, \bibinfo {author} {\bibfnamefont {E.}~\bibnamefont {Druga}}, \bibinfo {author} {\bibfnamefont {M.}~\bibnamefont {Bukov}}, \emph {et~al.},\ }\bibfield  {title} {\bibinfo {title} {Critical prethermal discrete time crystal created by two-frequency driving},\ }\href {https://doi.org/10.1038/s41567-022-01891-7} {\bibfield  {journal} {\bibinfo  {journal} {Nat. Phys.}\ }\textbf {\bibinfo {volume} {19}},\ \bibinfo {pages} {407} (\bibinfo {year} {2023})}\BibitemShut {NoStop}%
\bibitem [{\citenamefont {Jin}\ \emph {et~al.}(2023)\citenamefont {Jin}, \citenamefont {Knolle},\ and\ \citenamefont {Knap}}]{jin2023fractionalized}%
  \BibitemOpen
  \bibfield  {author} {\bibinfo {author} {\bibfnamefont {H.-K.}\ \bibnamefont {Jin}}, \bibinfo {author} {\bibfnamefont {J.}~\bibnamefont {Knolle}},\ and\ \bibinfo {author} {\bibfnamefont {M.}~\bibnamefont {Knap}},\ }\bibfield  {title} {\bibinfo {title} {Fractionalized prethermalization in a driven quantum spin liquid},\ }\href {https://doi.org/10.1103/PhysRevLett.130.226701} {\bibfield  {journal} {\bibinfo  {journal} {Phys. Rev. Lett.}\ }\textbf {\bibinfo {volume} {130}},\ \bibinfo {pages} {226701} (\bibinfo {year} {2023})}\BibitemShut {NoStop}%
\bibitem [{\citenamefont {Ho}\ \emph {et~al.}(2023)\citenamefont {Ho}, \citenamefont {Mori}, \citenamefont {Abanin},\ and\ \citenamefont {Dalla~Torre}}]{ho2023quantum}%
  \BibitemOpen
  \bibfield  {author} {\bibinfo {author} {\bibfnamefont {W.~W.}\ \bibnamefont {Ho}}, \bibinfo {author} {\bibfnamefont {T.}~\bibnamefont {Mori}}, \bibinfo {author} {\bibfnamefont {D.~A.}\ \bibnamefont {Abanin}},\ and\ \bibinfo {author} {\bibfnamefont {E.~G.}\ \bibnamefont {Dalla~Torre}},\ }\bibfield  {title} {\bibinfo {title} {Quantum and classical floquet prethermalization},\ }\href {https://www.sciencedirect.com/science/article/pii/S0003491623000829} {\bibfield  {journal} {\bibinfo  {journal} {Ann. Phys.}\ }\textbf {\bibinfo {volume} {454}},\ \bibinfo {pages} {169297} (\bibinfo {year} {2023})}\BibitemShut {NoStop}%
\bibitem [{\citenamefont {Yue}\ and\ \citenamefont {Cai}(2023)}]{yue2023prethermal}%
  \BibitemOpen
  \bibfield  {author} {\bibinfo {author} {\bibfnamefont {M.}~\bibnamefont {Yue}}\ and\ \bibinfo {author} {\bibfnamefont {Z.}~\bibnamefont {Cai}},\ }\bibfield  {title} {\bibinfo {title} {Prethermal time-crystalline spin ice and monopole confinement in a driven magnet},\ }\href {https://doi.org/10.1103/PhysRevLett.131.056502} {\bibfield  {journal} {\bibinfo  {journal} {Phys. Rev. Lett.}\ }\textbf {\bibinfo {volume} {131}},\ \bibinfo {pages} {056502} (\bibinfo {year} {2023})}\BibitemShut {NoStop}%
\bibitem [{\citenamefont {Hou}\ \emph {et~al.}(2025)\citenamefont {Hou}, \citenamefont {Fu}, \citenamefont {Moessner}, \citenamefont {Bukov},\ and\ \citenamefont {Zhao}}]{hou2024floquet}%
  \BibitemOpen
  \bibfield  {author} {\bibinfo {author} {\bibfnamefont {Y.}~\bibnamefont {Hou}}, \bibinfo {author} {\bibfnamefont {Z.}~\bibnamefont {Fu}}, \bibinfo {author} {\bibfnamefont {R.}~\bibnamefont {Moessner}}, \bibinfo {author} {\bibfnamefont {M.}~\bibnamefont {Bukov}},\ and\ \bibinfo {author} {\bibfnamefont {H.}~\bibnamefont {Zhao}},\ }\bibfield  {title} {\bibinfo {title} {Floquet-engineered emergent massive nambu-goldstone modes},\ }\href {https://doi.org/10.1103/9znd-dsm2} {\bibfield  {journal} {\bibinfo  {journal} {Phys. Rev. B}\ }\textbf {\bibinfo {volume} {112}},\ \bibinfo {pages} {L020305} (\bibinfo {year} {2025})}\BibitemShut {NoStop}%
\bibitem [{\citenamefont {Fu}\ \emph {et~al.}(2024)\citenamefont {Fu}, \citenamefont {Moessner}, \citenamefont {Zhao},\ and\ \citenamefont {Bukov}}]{PhysRevX.14.041070}%
  \BibitemOpen
  \bibfield  {author} {\bibinfo {author} {\bibfnamefont {Z.}~\bibnamefont {Fu}}, \bibinfo {author} {\bibfnamefont {R.}~\bibnamefont {Moessner}}, \bibinfo {author} {\bibfnamefont {H.}~\bibnamefont {Zhao}},\ and\ \bibinfo {author} {\bibfnamefont {M.}~\bibnamefont {Bukov}},\ }\bibfield  {title} {\bibinfo {title} {Engineering hierarchical symmetries},\ }\href {https://doi.org/10.1103/PhysRevX.14.041070} {\bibfield  {journal} {\bibinfo  {journal} {Phys. Rev. X}\ }\textbf {\bibinfo {volume} {14}},\ \bibinfo {pages} {041070} (\bibinfo {year} {2024})}\BibitemShut {NoStop}%
\bibitem [{\citenamefont {Qi}\ \emph {et~al.}(2024)\citenamefont {Qi}, \citenamefont {Wu},\ and\ \citenamefont {Zheng}}]{qi2024topological}%
  \BibitemOpen
  \bibfield  {author} {\bibinfo {author} {\bibfnamefont {H.-Y.}\ \bibnamefont {Qi}}, \bibinfo {author} {\bibfnamefont {Y.}~\bibnamefont {Wu}},\ and\ \bibinfo {author} {\bibfnamefont {W.}~\bibnamefont {Zheng}},\ }\bibfield  {title} {\bibinfo {title} {Topological origin of floquet thermalization in periodically driven many-body systems},\ }\href@noop {} {\bibfield  {journal} {\bibinfo  {journal} {arXiv preprint arXiv:2404.18052}\ } (\bibinfo {year} {2024})}\BibitemShut {NoStop}%
\bibitem [{\citenamefont {Piazza}\ and\ \citenamefont {Ritsch}(2015)}]{piazza2015self}%
  \BibitemOpen
  \bibfield  {author} {\bibinfo {author} {\bibfnamefont {F.}~\bibnamefont {Piazza}}\ and\ \bibinfo {author} {\bibfnamefont {H.}~\bibnamefont {Ritsch}},\ }\bibfield  {title} {\bibinfo {title} {Self-ordered limit cycles, chaos, and phase slippage with a superfluid inside an optical resonator},\ }\href {https://doi.org/10.1103/PhysRevLett.115.163601} {\bibfield  {journal} {\bibinfo  {journal} {Phys. Rev. Lett.}\ }\textbf {\bibinfo {volume} {115}},\ \bibinfo {pages} {163601} (\bibinfo {year} {2015})}\BibitemShut {NoStop}%
\bibitem [{\citenamefont {Bu{\v{c}}a}\ \emph {et~al.}(2019)\citenamefont {Bu{\v{c}}a}, \citenamefont {Tindall},\ and\ \citenamefont {Jaksch}}]{buvca2019non}%
  \BibitemOpen
  \bibfield  {author} {\bibinfo {author} {\bibfnamefont {B.}~\bibnamefont {Bu{\v{c}}a}}, \bibinfo {author} {\bibfnamefont {J.}~\bibnamefont {Tindall}},\ and\ \bibinfo {author} {\bibfnamefont {D.}~\bibnamefont {Jaksch}},\ }\bibfield  {title} {\bibinfo {title} {Non-stationary coherent quantum many-body dynamics through dissipation},\ }\href {https://www.nature.com/articles/s41467-019-09757-y} {\bibfield  {journal} {\bibinfo  {journal} {Nature Communications}\ }\textbf {\bibinfo {volume} {10}},\ \bibinfo {pages} {1730} (\bibinfo {year} {2019})}\BibitemShut {NoStop}%
\bibitem [{\citenamefont {Mori}(2023)}]{mori2023floquet}%
  \BibitemOpen
  \bibfield  {author} {\bibinfo {author} {\bibfnamefont {T.}~\bibnamefont {Mori}},\ }\bibfield  {title} {\bibinfo {title} {Floquet states in open quantum systems},\ }\href@noop {} {\bibfield  {journal} {\bibinfo  {journal} {Annual Review of Condensed Matter Physics}\ }\textbf {\bibinfo {volume} {14}},\ \bibinfo {pages} {35} (\bibinfo {year} {2023})}\BibitemShut {NoStop}%
\bibitem [{\citenamefont {Kawabata}\ \emph {et~al.}(2023)\citenamefont {Kawabata}, \citenamefont {Kulkarni}, \citenamefont {Li}, \citenamefont {Numasawa},\ and\ \citenamefont {Ryu}}]{kawabata2023symmetry}%
  \BibitemOpen
  \bibfield  {author} {\bibinfo {author} {\bibfnamefont {K.}~\bibnamefont {Kawabata}}, \bibinfo {author} {\bibfnamefont {A.}~\bibnamefont {Kulkarni}}, \bibinfo {author} {\bibfnamefont {J.}~\bibnamefont {Li}}, \bibinfo {author} {\bibfnamefont {T.}~\bibnamefont {Numasawa}},\ and\ \bibinfo {author} {\bibfnamefont {S.}~\bibnamefont {Ryu}},\ }\bibfield  {title} {\bibinfo {title} {Symmetry of open quantum systems: Classification of dissipative quantum chaos},\ }\href {https://link.aps.org/doi/10.1103/PRXQuantum.4.030328} {\bibfield  {journal} {\bibinfo  {journal} {PRX Quantum}\ }\textbf {\bibinfo {volume} {4}},\ \bibinfo {pages} {030328} (\bibinfo {year} {2023})}\BibitemShut {NoStop}%
\bibitem [{\citenamefont {Russomanno}(2023)}]{russomanno2023spatiotemporally}%
  \BibitemOpen
  \bibfield  {author} {\bibinfo {author} {\bibfnamefont {A.}~\bibnamefont {Russomanno}},\ }\bibfield  {title} {\bibinfo {title} {Spatiotemporally ordered patterns in a chain of coupled dissipative kicked rotors},\ }\href {https://doi.org/10.1103/PhysRevB.108.094305} {\bibfield  {journal} {\bibinfo  {journal} {Phys. Rev. B}\ }\textbf {\bibinfo {volume} {108}},\ \bibinfo {pages} {094305} (\bibinfo {year} {2023})}\BibitemShut {NoStop}%
\bibitem [{\citenamefont {Yi-Thomas}\ and\ \citenamefont {Sau}(2024)}]{yi2024theory}%
  \BibitemOpen
  \bibfield  {author} {\bibinfo {author} {\bibfnamefont {S.}~\bibnamefont {Yi-Thomas}}\ and\ \bibinfo {author} {\bibfnamefont {J.~D.}\ \bibnamefont {Sau}},\ }\bibfield  {title} {\bibinfo {title} {Theory for dissipative time crystals in coupled parametric oscillators},\ }\href {https://doi.org/10.1103/PhysRevLett.133.266601} {\bibfield  {journal} {\bibinfo  {journal} {Phys. Rev. Lett.}\ }\textbf {\bibinfo {volume} {133}},\ \bibinfo {pages} {266601} (\bibinfo {year} {2024})}\BibitemShut {NoStop}%
\bibitem [{\citenamefont {Gong}\ \emph {et~al.}(2018)\citenamefont {Gong}, \citenamefont {Hamazaki},\ and\ \citenamefont {Ueda}}]{DTC_qed2018}%
  \BibitemOpen
  \bibfield  {author} {\bibinfo {author} {\bibfnamefont {Z.}~\bibnamefont {Gong}}, \bibinfo {author} {\bibfnamefont {R.}~\bibnamefont {Hamazaki}},\ and\ \bibinfo {author} {\bibfnamefont {M.}~\bibnamefont {Ueda}},\ }\bibfield  {title} {\bibinfo {title} {Discrete {{Time-Crystalline Order}} in {{Cavity}} and {{Circuit QED Systems}}},\ }\href {https://doi.org/10.1103/PhysRevLett.120.040404} {\bibfield  {journal} {\bibinfo  {journal} {Phys. Rev. Lett.}\ }\textbf {\bibinfo {volume} {120}},\ \bibinfo {pages} {040404} (\bibinfo {year} {2018})}\BibitemShut {NoStop}%
\bibitem [{\citenamefont {Iemini}\ \emph {et~al.}(2018)\citenamefont {Iemini}, \citenamefont {Russomanno}, \citenamefont {Keeling}, \citenamefont {Schir\`o}, \citenamefont {Dalmonte},\ and\ \citenamefont {Fazio}}]{iemini2018boundary}%
  \BibitemOpen
  \bibfield  {author} {\bibinfo {author} {\bibfnamefont {F.}~\bibnamefont {Iemini}}, \bibinfo {author} {\bibfnamefont {A.}~\bibnamefont {Russomanno}}, \bibinfo {author} {\bibfnamefont {J.}~\bibnamefont {Keeling}}, \bibinfo {author} {\bibfnamefont {M.}~\bibnamefont {Schir\`o}}, \bibinfo {author} {\bibfnamefont {M.}~\bibnamefont {Dalmonte}},\ and\ \bibinfo {author} {\bibfnamefont {R.}~\bibnamefont {Fazio}},\ }\bibfield  {title} {\bibinfo {title} {Boundary time crystals},\ }\href {https://doi.org/10.1103/PhysRevLett.121.035301} {\bibfield  {journal} {\bibinfo  {journal} {Phys. Rev. Lett.}\ }\textbf {\bibinfo {volume} {121}},\ \bibinfo {pages} {035301} (\bibinfo {year} {2018})}\BibitemShut {NoStop}%
\bibitem [{\citenamefont {Ke\ss{}ler}\ \emph {et~al.}(2021)\citenamefont {Ke\ss{}ler}, \citenamefont {Kongkhambut}, \citenamefont {Georges}, \citenamefont {Mathey}, \citenamefont {Cosme},\ and\ \citenamefont {Hemmerich}}]{kessler2021observation}%
  \BibitemOpen
  \bibfield  {author} {\bibinfo {author} {\bibfnamefont {H.}~\bibnamefont {Ke\ss{}ler}}, \bibinfo {author} {\bibfnamefont {P.}~\bibnamefont {Kongkhambut}}, \bibinfo {author} {\bibfnamefont {C.}~\bibnamefont {Georges}}, \bibinfo {author} {\bibfnamefont {L.}~\bibnamefont {Mathey}}, \bibinfo {author} {\bibfnamefont {J.~G.}\ \bibnamefont {Cosme}},\ and\ \bibinfo {author} {\bibfnamefont {A.}~\bibnamefont {Hemmerich}},\ }\bibfield  {title} {\bibinfo {title} {Observation of a dissipative time crystal},\ }\href {https://doi.org/10.1103/PhysRevLett.127.043602} {\bibfield  {journal} {\bibinfo  {journal} {Phys. Rev. Lett.}\ }\textbf {\bibinfo {volume} {127}},\ \bibinfo {pages} {043602} (\bibinfo {year} {2021})}\BibitemShut {NoStop}%
\bibitem [{\citenamefont {Wu}\ \emph {et~al.}(2024)\citenamefont {Wu}, \citenamefont {Wang}, \citenamefont {Yang}, \citenamefont {Gao}, \citenamefont {Liang}, \citenamefont {Tey}, \citenamefont {Li}, \citenamefont {Pohl},\ and\ \citenamefont {You}}]{wu2024dissipative}%
  \BibitemOpen
  \bibfield  {author} {\bibinfo {author} {\bibfnamefont {X.}~\bibnamefont {Wu}}, \bibinfo {author} {\bibfnamefont {Z.}~\bibnamefont {Wang}}, \bibinfo {author} {\bibfnamefont {F.}~\bibnamefont {Yang}}, \bibinfo {author} {\bibfnamefont {R.}~\bibnamefont {Gao}}, \bibinfo {author} {\bibfnamefont {C.}~\bibnamefont {Liang}}, \bibinfo {author} {\bibfnamefont {M.~K.}\ \bibnamefont {Tey}}, \bibinfo {author} {\bibfnamefont {X.}~\bibnamefont {Li}}, \bibinfo {author} {\bibfnamefont {T.}~\bibnamefont {Pohl}},\ and\ \bibinfo {author} {\bibfnamefont {L.}~\bibnamefont {You}},\ }\bibfield  {title} {\bibinfo {title} {Dissipative time crystal in a strongly interacting rydberg gas},\ }\href {https://www.nature.com/articles/s41567-024-02542-9} {\bibfield  {journal} {\bibinfo  {journal} {Nat. Phys.}\ }\textbf {\bibinfo {volume} {20}},\ \bibinfo {pages} {1389} (\bibinfo {year} {2024})}\BibitemShut {NoStop}%
\bibitem [{\citenamefont {Solanki}\ \emph {et~al.}(2024)\citenamefont {Solanki}, \citenamefont {Krishna}, \citenamefont {Hajdu\ifmmode~\check{s}\else \v{s}\fi{}ek}, \citenamefont {Bruder},\ and\ \citenamefont {Vinjanampathy}}]{Solanki2024Exotic}%
  \BibitemOpen
  \bibfield  {author} {\bibinfo {author} {\bibfnamefont {P.}~\bibnamefont {Solanki}}, \bibinfo {author} {\bibfnamefont {M.}~\bibnamefont {Krishna}}, \bibinfo {author} {\bibfnamefont {M.}~\bibnamefont {Hajdu\ifmmode~\check{s}\else \v{s}\fi{}ek}}, \bibinfo {author} {\bibfnamefont {C.}~\bibnamefont {Bruder}},\ and\ \bibinfo {author} {\bibfnamefont {S.}~\bibnamefont {Vinjanampathy}},\ }\bibfield  {title} {\bibinfo {title} {Exotic synchronization in continuous time crystals outside the symmetric subspace},\ }\href {https://doi.org/10.1103/PhysRevLett.133.260403} {\bibfield  {journal} {\bibinfo  {journal} {Phys. Rev. Lett.}\ }\textbf {\bibinfo {volume} {133}},\ \bibinfo {pages} {260403} (\bibinfo {year} {2024})}\BibitemShut {NoStop}%
\bibitem [{\citenamefont {Yao}\ \emph {et~al.}(2020)\citenamefont {Yao}, \citenamefont {Nayak}, \citenamefont {Balents},\ and\ \citenamefont {Zaletel}}]{CDTC2020}%
  \BibitemOpen
  \bibfield  {author} {\bibinfo {author} {\bibfnamefont {N.~Y.}\ \bibnamefont {Yao}}, \bibinfo {author} {\bibfnamefont {C.}~\bibnamefont {Nayak}}, \bibinfo {author} {\bibfnamefont {L.}~\bibnamefont {Balents}},\ and\ \bibinfo {author} {\bibfnamefont {M.~P.}\ \bibnamefont {Zaletel}},\ }\bibfield  {title} {\bibinfo {title} {Classical discrete time crystals},\ }\href {https://doi.org/10.1038/s41567-019-0782-3} {\bibfield  {journal} {\bibinfo  {journal} {Nat. Phys.}\ }\textbf {\bibinfo {volume} {16}},\ \bibinfo {pages} {438} (\bibinfo {year} {2020})}\BibitemShut {NoStop}%
\bibitem [{\citenamefont {Boccaletti}\ \emph {et~al.}(2002)\citenamefont {Boccaletti}, \citenamefont {Kurths}, \citenamefont {Osipov}, \citenamefont {Valladares},\ and\ \citenamefont {Zhou}}]{boccaletti2002synchronization}%
  \BibitemOpen
  \bibfield  {author} {\bibinfo {author} {\bibfnamefont {S.}~\bibnamefont {Boccaletti}}, \bibinfo {author} {\bibfnamefont {J.}~\bibnamefont {Kurths}}, \bibinfo {author} {\bibfnamefont {G.}~\bibnamefont {Osipov}}, \bibinfo {author} {\bibfnamefont {D.}~\bibnamefont {Valladares}},\ and\ \bibinfo {author} {\bibfnamefont {C.}~\bibnamefont {Zhou}},\ }\bibfield  {title} {\bibinfo {title} {The synchronization of chaotic systems},\ }\href {https://www.sciencedirect.com/science/article/pii/S0370157302001370} {\bibfield  {journal} {\bibinfo  {journal} {Phys. Rep.}\ }\textbf {\bibinfo {volume} {366}},\ \bibinfo {pages} {1} (\bibinfo {year} {2002})}\BibitemShut {NoStop}%
\bibitem [{\citenamefont {Pisarchik}\ and\ \citenamefont {Feudel}(2014)}]{pisarchik2014control}%
  \BibitemOpen
  \bibfield  {author} {\bibinfo {author} {\bibfnamefont {A.~N.}\ \bibnamefont {Pisarchik}}\ and\ \bibinfo {author} {\bibfnamefont {U.}~\bibnamefont {Feudel}},\ }\bibfield  {title} {\bibinfo {title} {Control of multistability},\ }\href {https://www.sciencedirect.com/science/article/pii/S0370157314000453} {\bibfield  {journal} {\bibinfo  {journal} {Phys. Rep.}\ }\textbf {\bibinfo {volume} {540}},\ \bibinfo {pages} {167} (\bibinfo {year} {2014})}\BibitemShut {NoStop}%
\bibitem [{\citenamefont {Landa}\ \emph {et~al.}(2020)\citenamefont {Landa}, \citenamefont {Schir\'o},\ and\ \citenamefont {Misguich}}]{landa2020multistability}%
  \BibitemOpen
  \bibfield  {author} {\bibinfo {author} {\bibfnamefont {H.}~\bibnamefont {Landa}}, \bibinfo {author} {\bibfnamefont {M.}~\bibnamefont {Schir\'o}},\ and\ \bibinfo {author} {\bibfnamefont {G.}~\bibnamefont {Misguich}},\ }\bibfield  {title} {\bibinfo {title} {Multistability of driven-dissipative quantum spins},\ }\href {https://doi.org/10.1103/PhysRevLett.124.043601} {\bibfield  {journal} {\bibinfo  {journal} {Phys. Rev. Lett.}\ }\textbf {\bibinfo {volume} {124}},\ \bibinfo {pages} {043601} (\bibinfo {year} {2020})}\BibitemShut {NoStop}%
\bibitem [{\citenamefont {Alaeian}\ and\ \citenamefont {Bu{\v{c}}a}(2022)}]{alaeian2022exact}%
  \BibitemOpen
  \bibfield  {author} {\bibinfo {author} {\bibfnamefont {H.}~\bibnamefont {Alaeian}}\ and\ \bibinfo {author} {\bibfnamefont {B.}~\bibnamefont {Bu{\v{c}}a}},\ }\bibfield  {title} {\bibinfo {title} {Exact multistability and dissipative time crystals in interacting fermionic lattices},\ }\href {https://doi.org/10.1038/s42005-022-01090-z} {\bibfield  {journal} {\bibinfo  {journal} {Commun. Phys.}\ }\textbf {\bibinfo {volume} {5}},\ \bibinfo {pages} {318} (\bibinfo {year} {2022})}\BibitemShut {NoStop}%
\bibitem [{\citenamefont {Zaslavsky}(1978)}]{zaslavsky1978simplest}%
  \BibitemOpen
  \bibfield  {author} {\bibinfo {author} {\bibfnamefont {G.~M.}\ \bibnamefont {Zaslavsky}},\ }\bibfield  {title} {\bibinfo {title} {The simplest case of a strange attractor},\ }\href {https://www.sciencedirect.com/science/article/pii/0375960178901950} {\bibfield  {journal} {\bibinfo  {journal} {Phys. Lett. A}\ }\textbf {\bibinfo {volume} {69}},\ \bibinfo {pages} {145} (\bibinfo {year} {1978})}\BibitemShut {NoStop}%
\bibitem [{\citenamefont {Matsumoto}\ and\ \citenamefont {Tsuda}(1983)}]{matsumoto1983noise}%
  \BibitemOpen
  \bibfield  {author} {\bibinfo {author} {\bibfnamefont {K.}~\bibnamefont {Matsumoto}}\ and\ \bibinfo {author} {\bibfnamefont {I.}~\bibnamefont {Tsuda}},\ }\bibfield  {title} {\bibinfo {title} {Noise-induced order},\ }\href {https://link.springer.com/article/10.1007/BF01010923} {\bibfield  {journal} {\bibinfo  {journal} {J. Stat. Phys.}\ }\textbf {\bibinfo {volume} {31}},\ \bibinfo {pages} {87} (\bibinfo {year} {1983})}\BibitemShut {NoStop}%
\bibitem [{\citenamefont {Van~den Broeck}\ \emph {et~al.}(1994)\citenamefont {Van~den Broeck}, \citenamefont {Parrondo},\ and\ \citenamefont {Toral}}]{van1994noise}%
  \BibitemOpen
  \bibfield  {author} {\bibinfo {author} {\bibfnamefont {C.}~\bibnamefont {Van~den Broeck}}, \bibinfo {author} {\bibfnamefont {J.~M.~R.}\ \bibnamefont {Parrondo}},\ and\ \bibinfo {author} {\bibfnamefont {R.}~\bibnamefont {Toral}},\ }\bibfield  {title} {\bibinfo {title} {Noise-induced nonequilibrium phase transition},\ }\href {https://doi.org/10.1103/PhysRevLett.73.3395} {\bibfield  {journal} {\bibinfo  {journal} {Phys. Rev. Lett.}\ }\textbf {\bibinfo {volume} {73}},\ \bibinfo {pages} {3395} (\bibinfo {year} {1994})}\BibitemShut {NoStop}%
\bibitem [{\citenamefont {Zhou}\ and\ \citenamefont {Kurths}(2002)}]{zhou2002noise}%
  \BibitemOpen
  \bibfield  {author} {\bibinfo {author} {\bibfnamefont {C.}~\bibnamefont {Zhou}}\ and\ \bibinfo {author} {\bibfnamefont {J.}~\bibnamefont {Kurths}},\ }\bibfield  {title} {\bibinfo {title} {Noise-induced phase synchronization and synchronization transitions in chaotic oscillators},\ }\href {https://doi.org/10.1103/PhysRevLett.88.230602} {\bibfield  {journal} {\bibinfo  {journal} {Phys. Rev. Lett.}\ }\textbf {\bibinfo {volume} {88}},\ \bibinfo {pages} {230602} (\bibinfo {year} {2002})}\BibitemShut {NoStop}%
\bibitem [{\citenamefont {Teramae}\ and\ \citenamefont {Tanaka}(2004)}]{teramae2004robustness}%
  \BibitemOpen
  \bibfield  {author} {\bibinfo {author} {\bibfnamefont {J.-n.}\ \bibnamefont {Teramae}}\ and\ \bibinfo {author} {\bibfnamefont {D.}~\bibnamefont {Tanaka}},\ }\bibfield  {title} {\bibinfo {title} {Robustness of the noise-induced phase synchronization in a general class of limit cycle oscillators},\ }\href {https://doi.org/10.1103/PhysRevLett.93.204103} {\bibfield  {journal} {\bibinfo  {journal} {Phys. Rev. Lett.}\ }\textbf {\bibinfo {volume} {93}},\ \bibinfo {pages} {204103} (\bibinfo {year} {2004})}\BibitemShut {NoStop}%
\bibitem [{\citenamefont {Goldobin}\ and\ \citenamefont {Pikovsky}(2005)}]{goldobin2005synchronization}%
  \BibitemOpen
  \bibfield  {author} {\bibinfo {author} {\bibfnamefont {D.~S.}\ \bibnamefont {Goldobin}}\ and\ \bibinfo {author} {\bibfnamefont {A.}~\bibnamefont {Pikovsky}},\ }\bibfield  {title} {\bibinfo {title} {Synchronization and desynchronization of self-sustained oscillators by common noise},\ }\href {https://doi.org/10.1103/PhysRevE.71.045201} {\bibfield  {journal} {\bibinfo  {journal} {Phys. Rev. E}\ }\textbf {\bibinfo {volume} {71}},\ \bibinfo {pages} {045201} (\bibinfo {year} {2005})}\BibitemShut {NoStop}%
\bibitem [{\citenamefont {Goychuk}\ \emph {et~al.}(2006)\citenamefont {Goychuk}, \citenamefont {Casado-Pascual}, \citenamefont {Morillo}, \citenamefont {Lehmann},\ and\ \citenamefont {H\"anggi}}]{goychuk2006quantum}%
  \BibitemOpen
  \bibfield  {author} {\bibinfo {author} {\bibfnamefont {I.}~\bibnamefont {Goychuk}}, \bibinfo {author} {\bibfnamefont {J.}~\bibnamefont {Casado-Pascual}}, \bibinfo {author} {\bibfnamefont {M.}~\bibnamefont {Morillo}}, \bibinfo {author} {\bibfnamefont {J.}~\bibnamefont {Lehmann}},\ and\ \bibinfo {author} {\bibfnamefont {P.}~\bibnamefont {H\"anggi}},\ }\bibfield  {title} {\bibinfo {title} {Quantum stochastic synchronization},\ }\href {https://doi.org/10.1103/PhysRevLett.97.210601} {\bibfield  {journal} {\bibinfo  {journal} {Phys. Rev. Lett.}\ }\textbf {\bibinfo {volume} {97}},\ \bibinfo {pages} {210601} (\bibinfo {year} {2006})}\BibitemShut {NoStop}%
\bibitem [{\citenamefont {Huang}\ \emph {et~al.}(2020)\citenamefont {Huang}, \citenamefont {Qian}, \citenamefont {Wang}, \citenamefont {Ye},\ and\ \citenamefont {Yi}}]{huang2020synchronization}%
  \BibitemOpen
  \bibfield  {author} {\bibinfo {author} {\bibfnamefont {W.}~\bibnamefont {Huang}}, \bibinfo {author} {\bibfnamefont {H.}~\bibnamefont {Qian}}, \bibinfo {author} {\bibfnamefont {S.}~\bibnamefont {Wang}}, \bibinfo {author} {\bibfnamefont {F.~X.-F.}\ \bibnamefont {Ye}},\ and\ \bibinfo {author} {\bibfnamefont {Y.}~\bibnamefont {Yi}},\ }\bibfield  {title} {\bibinfo {title} {Synchronization in discrete-time, discrete-state random dynamical systems},\ }\href {https://doi.org/10.1137/19M1244883} {\bibfield  {journal} {\bibinfo  {journal} {SIAM J. Appl. Dyn. Syst.}\ }\textbf {\bibinfo {volume} {19}},\ \bibinfo {pages} {233} (\bibinfo {year} {2020})}\BibitemShut {NoStop}%
\bibitem [{\citenamefont {Schmolke}\ and\ \citenamefont {Lutz}(2022)}]{schmolke2022noise}%
  \BibitemOpen
  \bibfield  {author} {\bibinfo {author} {\bibfnamefont {F.}~\bibnamefont {Schmolke}}\ and\ \bibinfo {author} {\bibfnamefont {E.}~\bibnamefont {Lutz}},\ }\bibfield  {title} {\bibinfo {title} {Noise-induced quantum synchronization},\ }\href {https://doi.org/10.1103/PhysRevLett.129.250601} {\bibfield  {journal} {\bibinfo  {journal} {Phys. Rev. Lett.}\ }\textbf {\bibinfo {volume} {129}},\ \bibinfo {pages} {250601} (\bibinfo {year} {2022})}\BibitemShut {NoStop}%
\bibitem [{\citenamefont {Yan}(2025)}]{yan2025bifurcationsintermittencycoupleddissipative}%
  \BibitemOpen
  \bibfield  {author} {\bibinfo {author} {\bibfnamefont {J.}~\bibnamefont {Yan}},\ }\bibfield  {title} {\bibinfo {title} {Bifurcations and intermittency in coupled dissipative kicked rotors},\ }\href {https://doi.org/10.1063/5.0271877} {\bibfield  {journal} {\bibinfo  {journal} {Chaos: An Interdisciplinary Journal of Nonlinear Science}\ }\textbf {\bibinfo {volume} {35}} (\bibinfo {year} {2025})}\BibitemShut {NoStop}%
\bibitem [{Note1()}]{Note1}%
  \BibitemOpen
  \bibinfo {note} {We discuss different parameter regimes and the stability of the transition rule in the Sec.~SM~1 in the Supplementary Material}\BibitemShut {NoStop}%
\bibitem [{\citenamefont {Meyn}\ and\ \citenamefont {Tweedie}(2012)}]{meyn2012markov}%
  \BibitemOpen
  \bibfield  {author} {\bibinfo {author} {\bibfnamefont {S.~P.}\ \bibnamefont {Meyn}}\ and\ \bibinfo {author} {\bibfnamefont {R.~L.}\ \bibnamefont {Tweedie}},\ }\href {https://link.springer.com/book/10.1007/978-1-4471-3267-7} {\emph {\bibinfo {title} {Markov chains and stochastic stability}}}\ (\bibinfo  {publisher} {Springer Science \& Business Media},\ \bibinfo {year} {2012})\BibitemShut {NoStop}%
\bibitem [{sup()}]{supplement}%
  \BibitemOpen
  \href@noop {} {}\bibinfo {note} {See the Supplemental Material at \url{http://link.aps.org/supplemental/10.1103/2yzx-jfky} for discussions on the finite size effect, experimental proposals and the initial state dependence. The Supplemental Material also contains Refs.~\cite{PhysRevX.11.011010,iskhakovFastControlCurrent2012a,yan2020engineeringframeworkoptimizingsuperconducting}}\BibitemShut {NoStop}%
\bibitem [{\citenamefont {Berger}\ \emph {et~al.}(2021)\citenamefont {Berger}, \citenamefont {Qian}, \citenamefont {Wang},\ and\ \citenamefont {Yi}}]{berger2021intermittent}%
  \BibitemOpen
  \bibfield  {author} {\bibinfo {author} {\bibfnamefont {A.}~\bibnamefont {Berger}}, \bibinfo {author} {\bibfnamefont {H.}~\bibnamefont {Qian}}, \bibinfo {author} {\bibfnamefont {S.}~\bibnamefont {Wang}},\ and\ \bibinfo {author} {\bibfnamefont {Y.}~\bibnamefont {Yi}},\ }\bibfield  {title} {\bibinfo {title} {Intermittent synchronization in finite-state random networks under markov perturbations},\ }\href {https://link.springer.com/article/10.1007/s00220-021-04104-z} {\bibfield  {journal} {\bibinfo  {journal} {Commun. Math. Phys.}\ }\textbf {\bibinfo {volume} {384}},\ \bibinfo {pages} {1945} (\bibinfo {year} {2021})}\BibitemShut {NoStop}%
\bibitem [{Note2()}]{Note2}%
  \BibitemOpen
  \bibinfo {note} {Long-time saturation values are obtained by averaging order parameters(system size $L=500$) at 100 stroboscopic times, starting from $t=10^5$.}\BibitemShut {Stop}%
\bibitem [{\citenamefont {Wolf}\ \emph {et~al.}(1985)\citenamefont {Wolf}, \citenamefont {Swift}, \citenamefont {Swinney},\ and\ \citenamefont {Vastano}}]{wolf1985determining}%
  \BibitemOpen
  \bibfield  {author} {\bibinfo {author} {\bibfnamefont {A.}~\bibnamefont {Wolf}}, \bibinfo {author} {\bibfnamefont {J.~B.}\ \bibnamefont {Swift}}, \bibinfo {author} {\bibfnamefont {H.~L.}\ \bibnamefont {Swinney}},\ and\ \bibinfo {author} {\bibfnamefont {J.~A.}\ \bibnamefont {Vastano}},\ }\bibfield  {title} {\bibinfo {title} {Determining lyapunov exponents from a time series},\ }\href {https://www.sciencedirect.com/science/article/pii/0167278985900119} {\bibfield  {journal} {\bibinfo  {journal} {Phys. D: Nonlinear Phenom.}\ }\textbf {\bibinfo {volume} {16}},\ \bibinfo {pages} {285} (\bibinfo {year} {1985})}\BibitemShut {NoStop}%
\bibitem [{\citenamefont {Xu}\ \emph {et~al.}(2024)\citenamefont {Xu}, \citenamefont {Sun}, \citenamefont {Zhang}, \citenamefont {He},\ and\ \citenamefont {Pu}}]{xu2024phase}%
  \BibitemOpen
  \bibfield  {author} {\bibinfo {author} {\bibfnamefont {Y.}~\bibnamefont {Xu}}, \bibinfo {author} {\bibfnamefont {F.-X.}\ \bibnamefont {Sun}}, \bibinfo {author} {\bibfnamefont {W.}~\bibnamefont {Zhang}}, \bibinfo {author} {\bibfnamefont {Q.}~\bibnamefont {He}},\ and\ \bibinfo {author} {\bibfnamefont {H.}~\bibnamefont {Pu}},\ }\bibfield  {title} {\bibinfo {title} {Phase transition and multistability in dicke dimer},\ }\href {https://doi.org/10.1103/PhysRevLett.133.233604} {\bibfield  {journal} {\bibinfo  {journal} {Phys. Rev. Lett.}\ }\textbf {\bibinfo {volume} {133}},\ \bibinfo {pages} {233604} (\bibinfo {year} {2024})}\BibitemShut {NoStop}%
\bibitem [{\citenamefont {Xu}\ \emph {et~al.}(2014)\citenamefont {Xu}, \citenamefont {Tieri}, \citenamefont {Fine}, \citenamefont {Thompson},\ and\ \citenamefont {Holland}}]{xu2014synchronization}%
  \BibitemOpen
  \bibfield  {author} {\bibinfo {author} {\bibfnamefont {M.}~\bibnamefont {Xu}}, \bibinfo {author} {\bibfnamefont {D.~A.}\ \bibnamefont {Tieri}}, \bibinfo {author} {\bibfnamefont {E.~C.}\ \bibnamefont {Fine}}, \bibinfo {author} {\bibfnamefont {J.~K.}\ \bibnamefont {Thompson}},\ and\ \bibinfo {author} {\bibfnamefont {M.~J.}\ \bibnamefont {Holland}},\ }\bibfield  {title} {\bibinfo {title} {Synchronization of two ensembles of atoms},\ }\href {https://doi.org/10.1103/PhysRevLett.113.154101} {\bibfield  {journal} {\bibinfo  {journal} {Phys. Rev. Lett.}\ }\textbf {\bibinfo {volume} {113}},\ \bibinfo {pages} {154101} (\bibinfo {year} {2014})}\BibitemShut {NoStop}%
\bibitem [{\citenamefont {Cataliotti}\ \emph {et~al.}(2001)\citenamefont {Cataliotti}, \citenamefont {Burger}, \citenamefont {Fort}, \citenamefont {Maddaloni}, \citenamefont {Minardi}, \citenamefont {Trombettoni}, \citenamefont {Smerzi},\ and\ \citenamefont {Inguscio}}]{cataliottiJosephsonJunctionArrays2001}%
  \BibitemOpen
  \bibfield  {author} {\bibinfo {author} {\bibfnamefont {F.~S.}\ \bibnamefont {Cataliotti}}, \bibinfo {author} {\bibfnamefont {S.}~\bibnamefont {Burger}}, \bibinfo {author} {\bibfnamefont {C.}~\bibnamefont {Fort}}, \bibinfo {author} {\bibfnamefont {P.}~\bibnamefont {Maddaloni}}, \bibinfo {author} {\bibfnamefont {F.}~\bibnamefont {Minardi}}, \bibinfo {author} {\bibfnamefont {A.}~\bibnamefont {Trombettoni}}, \bibinfo {author} {\bibfnamefont {A.}~\bibnamefont {Smerzi}},\ and\ \bibinfo {author} {\bibfnamefont {M.}~\bibnamefont {Inguscio}},\ }\bibfield  {title} {\bibinfo {title} {Josephson {{Junction Arrays}} with {{Bose-Einstein Condensates}}},\ }\href {https://doi.org/10.1126/science.1062612} {\bibfield  {journal} {\bibinfo  {journal} {Science}\ }\textbf {\bibinfo {volume} {293}},\ \bibinfo {pages} {843} (\bibinfo {year} {2001})}\BibitemShut {NoStop}%
\bibitem [{\citenamefont {Zhang}\ \emph {et~al.}(2021)\citenamefont {Zhang}, \citenamefont {Chakram}, \citenamefont {Roy}, \citenamefont {Earnest}, \citenamefont {Lu}, \citenamefont {Huang}, \citenamefont {Weiss}, \citenamefont {Koch},\ and\ \citenamefont {Schuster}}]{PhysRevX.11.011010}%
  \BibitemOpen
  \bibfield  {author} {\bibinfo {author} {\bibfnamefont {H.}~\bibnamefont {Zhang}}, \bibinfo {author} {\bibfnamefont {S.}~\bibnamefont {Chakram}}, \bibinfo {author} {\bibfnamefont {T.}~\bibnamefont {Roy}}, \bibinfo {author} {\bibfnamefont {N.}~\bibnamefont {Earnest}}, \bibinfo {author} {\bibfnamefont {Y.}~\bibnamefont {Lu}}, \bibinfo {author} {\bibfnamefont {Z.}~\bibnamefont {Huang}}, \bibinfo {author} {\bibfnamefont {D.~K.}\ \bibnamefont {Weiss}}, \bibinfo {author} {\bibfnamefont {J.}~\bibnamefont {Koch}},\ and\ \bibinfo {author} {\bibfnamefont {D.~I.}\ \bibnamefont {Schuster}},\ }\bibfield  {title} {\bibinfo {title} {Universal fast-flux control of a coherent, low-frequency qubit},\ }\href {https://doi.org/10.1103/PhysRevX.11.011010} {\bibfield  {journal} {\bibinfo  {journal} {Phys. Rev. X}\ }\textbf {\bibinfo {volume} {11}},\ \bibinfo {pages} {011010} (\bibinfo {year} {2021})}\BibitemShut {NoStop}%
\bibitem [{\citenamefont {Iskhakov}\ \emph {et~al.}(2012)\citenamefont {Iskhakov}, \citenamefont {Balakshina}, \citenamefont {Pospelov},\ and\ \citenamefont {Skovpen'}}]{iskhakovFastControlCurrent2012a}%
  \BibitemOpen
  \bibfield  {author} {\bibinfo {author} {\bibfnamefont {A.~S.}\ \bibnamefont {Iskhakov}}, \bibinfo {author} {\bibfnamefont {L.~V.}\ \bibnamefont {Balakshina}}, \bibinfo {author} {\bibfnamefont {V.~{\relax Ya}.}\ \bibnamefont {Pospelov}},\ and\ \bibinfo {author} {\bibfnamefont {S.~M.}\ \bibnamefont {Skovpen'}},\ }\bibfield  {title} {\bibinfo {title} {Fast control of current of a pulse-width modulator},\ }\href {https://doi.org/10.3103/S1068371212060065} {\bibfield  {journal} {\bibinfo  {journal} {Russian Electrical Engineering}\ }\textbf {\bibinfo {volume} {83}},\ \bibinfo {pages} {302} (\bibinfo {year} {2012})}\BibitemShut {NoStop}%
\bibitem [{\citenamefont {Yan}\ \emph {et~al.}(2020)\citenamefont {Yan}, \citenamefont {Sung}, \citenamefont {Krantz}, \citenamefont {Kamal}, \citenamefont {Kim}, \citenamefont {Yoder}, \citenamefont {Orlando}, \citenamefont {Gustavsson},\ and\ \citenamefont {Oliver}}]{yan2020engineeringframeworkoptimizingsuperconducting}%
  \BibitemOpen
  \bibfield  {author} {\bibinfo {author} {\bibfnamefont {F.}~\bibnamefont {Yan}}, \bibinfo {author} {\bibfnamefont {Y.}~\bibnamefont {Sung}}, \bibinfo {author} {\bibfnamefont {P.}~\bibnamefont {Krantz}}, \bibinfo {author} {\bibfnamefont {A.}~\bibnamefont {Kamal}}, \bibinfo {author} {\bibfnamefont {D.~K.}\ \bibnamefont {Kim}}, \bibinfo {author} {\bibfnamefont {J.~L.}\ \bibnamefont {Yoder}}, \bibinfo {author} {\bibfnamefont {T.~P.}\ \bibnamefont {Orlando}}, \bibinfo {author} {\bibfnamefont {S.}~\bibnamefont {Gustavsson}},\ and\ \bibinfo {author} {\bibfnamefont {W.~D.}\ \bibnamefont {Oliver}},\ }\href {https://arxiv.org/abs/2006.04130} {\bibinfo {title} {Engineering framework for optimizing superconducting qubit designs}} (\bibinfo {year} {2020}),\ \Eprint {https://arxiv.org/abs/2006.04130} {arXiv:2006.04130 [quant-ph]} \BibitemShut {NoStop}%
\end{thebibliography}%

\newpage

% \author{Zhuocheng Ma} 
% \affiliation{\affA} 
% \author{Jin Yan} 
% \affiliation{\affB} 	
% \author{Hongzheng Zhao}	
% \email{hzhao@pku.edu.cn} 
% \affiliation{\affA} 
% \author{Liang-You Peng} 
% \email{liangyou.peng@pku.edu.cn} 
% \affiliation{\affA } 
% \affiliation{ \affC } 
% \affiliation{ \affD}
\let\addcontentsline\oldaddcontentsline
	\cleardoublepage
	\onecolumngrid

\begin{center}
\textbf{\large{\textit{Supplementary Material} \\ \smallskip
	Stable time rondeau crystals in dissipative many-body systems }}\\
		% \hfill \break
		% \smallskip
\end{center}

% \maketitle
% \let\oldaddcontentsline\addcontentsline
% \renewcommand{\addcontentsline}[3]{}

\onecolumngrid
	\renewcommand{\thefigure}{S\arabic{figure}}
        \setcounter{figure}{0}
    \renewcommand{\thesection}{SM\;\arabic{section}}
	\setcounter{section}{0}
	\renewcommand{\theequation}{S.\arabic{equation}}
        \setcounter{equation}{0}
    \renewcommand{\thesubsection}{\arabic{subsection}}
	\setcounter{section}{0}
    \tableofcontents

\section{Various types of dynamics for single rotor}

\begin{figure}[h]
    \centering
\includegraphics[width=.7\linewidth]{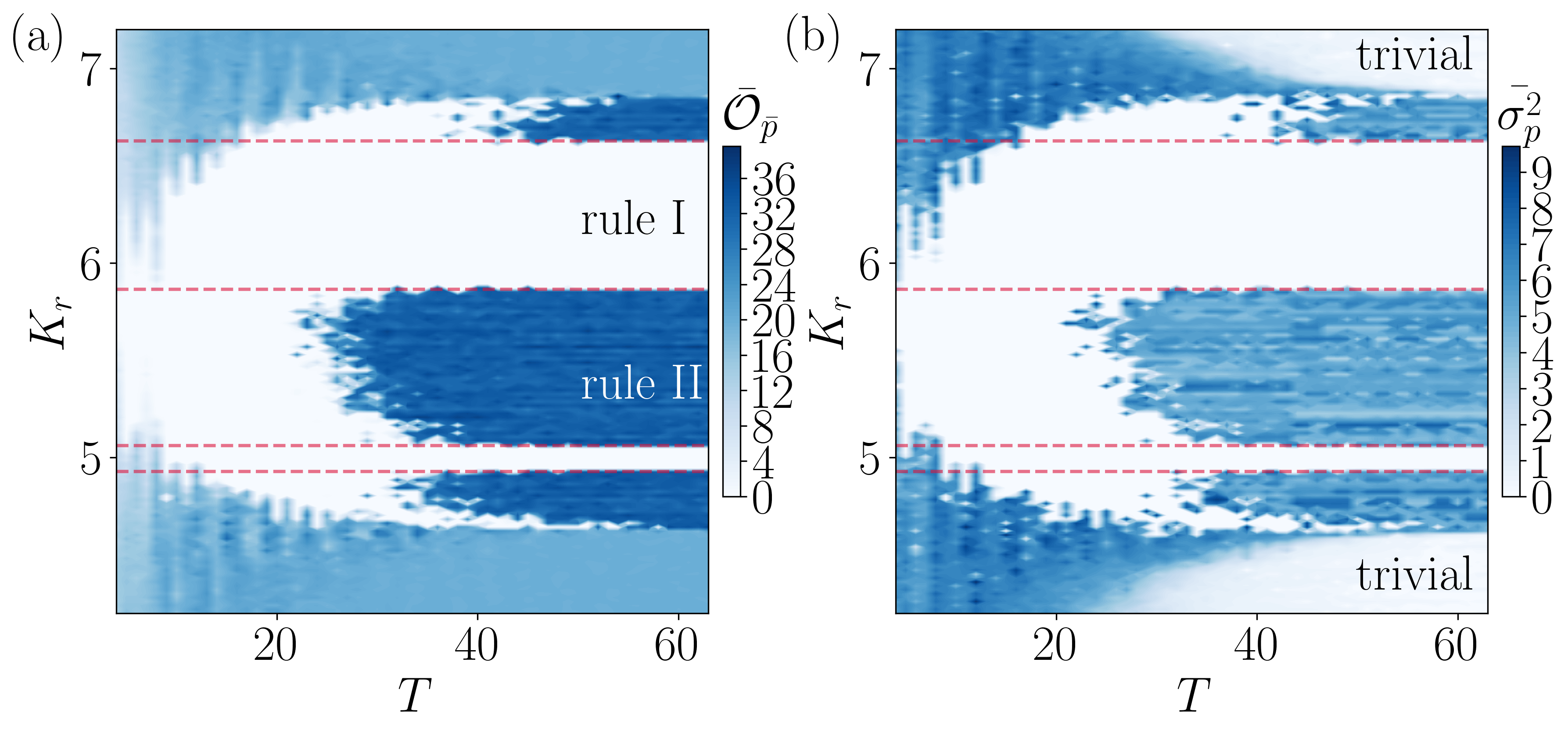}
    \caption{Phase diagram for the single-rotor system for different waiting time $T$ and random drive strength $K_r$. The order parameter $\bar{\mathcal{O}}_{p}$ and the variance of momentum $\bar{\sigma}^2_p$ are plotted in (a) and (b), respectively. We average over 10 stroboscopic times around $t=2\times 10^4$ to obtain their long-time saturation values. ``Rule I" corresponds to the synchronized time rondeau order. ``Rule II" and other trivial phases are discussed in the text. Rules I and II are separated by red lines when $T$ is sufficiently large. These phase boundaries are obtained by analyzing the basin structure of the periodically kicked rotors when $T\to\infty$, but it can still capture the phase boundary for a large range of $T$. When $T$ is further reduced, e.g. $T\approx 20$, Rule I remains remarkably stable, but Rule II no longer persists. The initial conditions of momenta have Gaussian distributions with zero average and standard deviation $\sigma_p=6$.}
    \label{fig:phase2T}
\end{figure}

In the absence of many-body interactions, an ensemble of kicked rotors can exhibit various types of dynamics that depend on both the waiting time $T$ and the random kick strength $K_r$. For $T\to\infty$, as elaborated in the main text, the system exhibits the time rondeau order and synchronization protects it against initial state perturbations. Here, we show that this phenomenon indeed persists for finite $T$, and we depict the entire phase diagram in Fig.~\ref{fig:phase2T}.

We numerically obtain two order parameters at long times: $\bar{\mathcal{O}}_{p}$, the deviation from the perfect rondeau evolution for $T\to\infty$ and the fluctuation of the momentum of the entire rotor ensemble, $\bar{\sigma}^2_p$. There are four possible phases as shown in Fig.~\ref{fig:phase2T}:
\begin{itemize}
    \item The synchronized time rondeau, induced by Transition Rule I (Fig.~1(b) in the main text in the main text) appears when both order parameters vanish (white region). 
    \item  The light blue region in Fig.~\ref{fig:phase2T}(a) corresponds to the random motion that exhibits neither the rondeau order nor the synchronized behavior.
    \item The dark blue regions in Fig.~\ref{fig:phase2T}(a) correspond to dynamics induced by Transition Rule II, which will be explained in detail later.
    \item In the ``trivial" region (white in Fig.~\ref{fig:phase2T}(b)), rotors rapidly synchronize since they relax to the fixed point $P_0$. Hence, no stroboscopic transition between fixed points can exist, and $\bar{\mathcal{O}}_{p}$ remains finite.

\end{itemize}

\textit{Transition rule I.---}For large $T$, the stroboscopic transition rule can be precisely determined by analyzing the structure of the basins of fixed points as shown in Fig.~1(a) in the main text. To realize the Transition Rule I, we numerically find that the strength $K_r$ should be in the region $(4.93,5.06)\cup(5.87,6.63)$ so that a positive kick can drive points A and B to the blue region (the basin of the fixed point $P_+$) simultaneously. We plot the boundaries of $K_r$ as red dashed lines in Fig.~\ref{fig:phase2T}, which precisely capture the phase boundary of Rule I when $T$ is sufficiently large.

\textit{Transition rule II.---} When $K_r$ lies in the parameter region $(4.63,4.93)\cup (5.06, 5.86)\cup (6.63,6.85)$, different stroboscopic dynamics can appear in the limit of $T\to\infty$
\begin{equation}
    \begin{matrix}
 +K_r: & P_+ \to P_0   & P_0 \to P_+  & P_- \to P_0 \\
 -K_r: & P_+ \to P_0  & P_0 \to P_-  & P_- \to P_0, 
\end{matrix}
\label{rule2}
\end{equation}
which we dub as Rule II. This type of dynamics appears in the dark blue regions in Fig.~\ref{fig:phase2T}(a). The only difference from Rule I is $P_+\to P_0$ for a positive kick and $P_-\to P_0$ for a negative kick.

The dynamics after a dipolar kick can also be analyzed via a Markovian process. The corresponding stochastic matrix reads
\begin{equation}
    \boldsymbol{A}{=}
\begin{pmatrix}
  1/2 &  &1/2 \\
  & 1 & \\
  1/2&  & 1/2
\end{pmatrix},
\end{equation}
and by diagonalizing this matrix, we find that it has two invariant subspaces spanned by $\vec{e}_2$ and $\{\vec{e}_1+\vec{e}_3, \vec{e}_1-\vec{e}_3\}$, where, for example, $\vec{e}_1=(1,0,0)^T$ is a unit base vector. These two subspaces correspond to two possible motions. $\vec{e}_2$ corresponds to Motion I, where the rotor always stays at the fixed point $P_0$ at time $2mT$,
\begin{equation}
    \begin{matrix}
 (+K_r,-K_r): & P_0 \to P_+ \to P_0,\\
 (-K_r,+K_r): & P_0 \to P_- \to P_0,
\end{matrix}
\label{rule2}
\end{equation}
which also reproduces the synchronized steady state as in the main text. In contrast, in the subspace spanned by $\{\vec{e}_1+\vec{e}_3, \vec{e}_1-\vec{e}_3\}$, Motion II appears 
\begin{equation}
    \begin{matrix}
 (+K_r,-K_r): & P_+ \to P_0 \to P_-, & P_- \to P_0 \to P_-,\\
 (-K_r,+K_r): & P_+ \to P_0 \to P_+, & P_- \to P_0 \to P_+,
\end{matrix}
\label{rule2}
\end{equation}
and the rotor jumps randomly between $P_+$ and $P_-$ at time $2mT$. Therefore, rotors starting from different initial conditions will not synchronize and exhibit one of these two motions at long times.  

Rotors starting from widely distributed initial conditions prefer Motion II in the large $T$ limit. To see this, we first notice that the momentum difference at stroboscopic times between motion II and motion I is exact $2\pi$. Also, the order parameter in the dark blue region in Fig.~\ref{fig:phase2T}(a) is 
 $\bar{\mathcal{O}}_{p}\approx 4\pi^2$, suggesting that most rotors exhibit Motion II. Further, $\bar{\mathcal{O}}_{p}$ is large enough to distinguish Motion II from the light blue region, where rotors exhibit a random motion. We can understand this phenomenon again by analyzing the basin structure shown in Fig.~1(a) in the main text: Clearly, the area of the white region is larger than the blue or the red region. Hence, after the first stroboscopic kick, rotors with widely distributed initial conditions quickly converge to the fixed point $P_0$ within maximum probability and evolve according to Motion II afterward. 
 
When $T$ is finite, analysis becomes complicated since one cannot construct the stochastic matrix according to the basin structure. Instead, we perform extensive numerical simulations to map out the phase diagram. We note that Rule I is remarkably stable and survives in a wide parameter space, while Rule II becomes unstable at a finite $T$.
It remains an interesting open question to further justify the stability of Rule I, and we leave this to future work. In the current study, we fix $T=10$, which is indeed far from the $T\to\infty$ limit, and focus on the stability of time rondeau crystals in the presence of many-body interactions.

\section{Details of many-body kicked rotors}

\subsection{System size dependence of order parameters and intermittent synchronization}
In the main text, we demonstrate the existence of the synchronization phase in the presence of weak many-body interactions and a de-synchronization phase for large $J$. Here, we provide more numerical evidence to show these two phases are thermodynamically stable.

\begin{figure}[h]
    \centering
\includegraphics[width=0.5\linewidth]{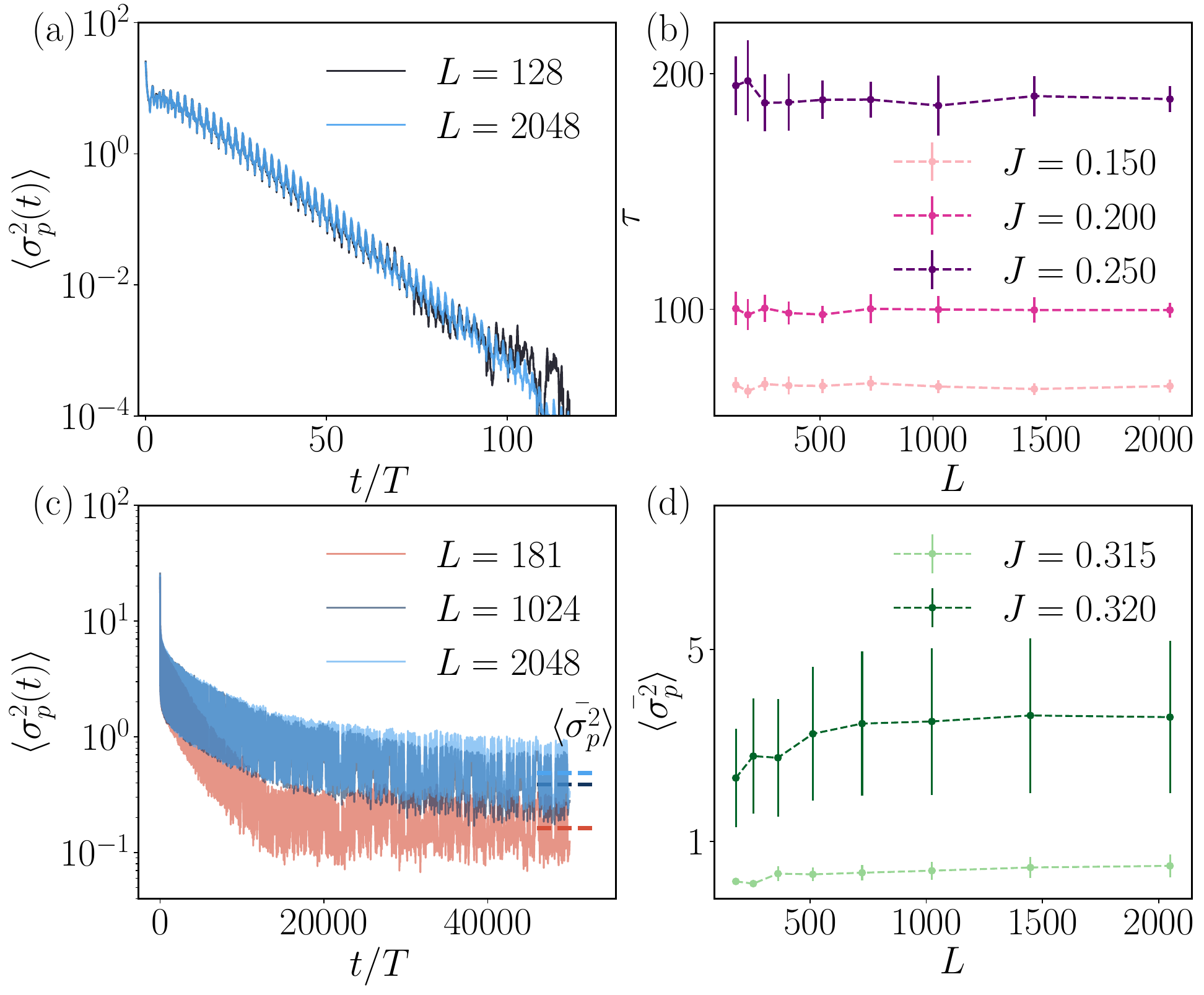}
    \caption{(a) Synchronization occurs for weak interaction strength, where the momentum variance decays to zero exponentially fast in time. Finite-size effects only become visible at late times, e.g., $t\approx80T$. Here we use $J=0.2$. (b) The time scale $\tau$ before synchronization is independent of the system size. Longer time is needed for a larger interaction strength. (c) For $J>J_c$, de-synchronization occurs and the spatial fluctuation does not vanish. $J_c$ is about $0.31$. (d) The long-time average $\langle\bar{\sigma_p^2}\rangle$ (dashed line in (c)) converges for larger system sizes. The increase of $J$ generally induces a larger spatial fluctuation. The initial momentum is sampled from a Gaussian distribution with
a standard deviation of $6$ and zero average. We use $K_r=5.5$ for these numerical simulations.}
    \label{fig:var-t_syn}
\end{figure}

As shown in Fig.~\ref{fig:var-t_syn}(a), for weak interaction $J=0.2$, the ensemble-averaged order parameter $\langle\sigma_p^2\rangle$ decays to zero, suggesting that the synchronization phase is quickly established. By comparing the numerical simulation performed for different system sizes (black and blue lines), we conclude that such a decaying process is largely independent of $L$, and finite-size effects only appear at longer times, e.g.,  $t/T>80$. Also, it occurs exponentially fast in time, $\langle\sigma_p^2\rangle \propto e^{-t/\tau}$, and the corresponding synchronization time scale $\tau$ can be obtained by performing a linear fit in panel (a), where a log scale is used. In Fig.~\ref{fig:var-t_syn}(b), we can see $\tau$ becomes independent of the system sizes for large $L$. 

As we increase the interaction strength, $\tau$ also grows and a phase transition to de-synchronization occurs for large $J$. We extract the saturation value of the $\langle\sigma_p^2(t)\rangle$ at sufficiently long times ( dashed lines in Fig.~\ref{fig:var-t_syn}(c)). As shown in Fig.~\ref{fig:var-t_syn}(d), we illustrate its dependence on the system size, and clearly it converges to a non-vanishing value in the thermodynamic limit for $J>J_c$. 

\begin{figure}[h]
    \centering
    \includegraphics[width=0.5\linewidth]{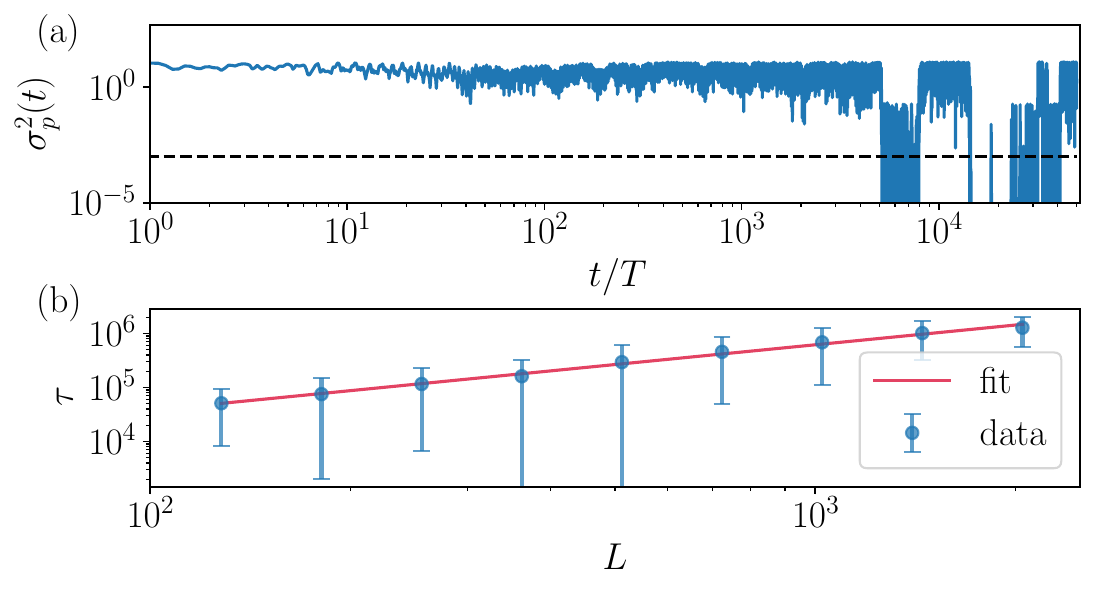}
    \caption{(a) Intermittent synchronization when $J=0.32,L=128$. The dashed black line denotes $\sigma^2_p(t)=10^{-3}$. (b) $\tau$ denotes the first time where the spatial fluctuation drops below the threshold value (black dashed line in (a)).
It follows a power-law dependence on the system size. We use $J=0.318$ for numerical simulation. For each system size, we simulate 112 trajectories to get the average value of $\tau$ and the error bar corresponds to the standard deviation.}
    \label{fig:int_syn}
\end{figure}

Interestingly, as shown in Fig.~\ref{fig:int_syn}(a), we notice that for a finite-size system, intermittent synchronization can occur near the phase transition point, where full synchronization and the non-synchronized dynamics alternate irregularly in time. 
However, this phenomenon is thermodynamically unstable. To show this, we use the threshold value $10^{-3}$ (dashed black line) and extract the synchronization time scale $\tau$, after which the system's spatial fluctuation first drops below this threshold. We consider many different realizations and plot the mean value of $\tau$ versus different system sizes in Fig.~\ref{fig:int_syn}(b), where the error bar denotes the standard deviation. We use a log-log scale in Fig.~\ref{fig:int_syn}(b). Numerical results fit well with a straight line (red), suggesting a power law dependence of $\tau$ on size $L$. Therefore, we conclude that for large interaction strength $J$, intermittent synchronization will not occur in the thermodynamic limit for $L\to\infty$.

\subsection{Phase diagram for different periodic driving strength $K_0$}

\par Realization of TRC requires a proper design of the driving parameters. In the main text and the End Matter, we have extensively discussed the robustness of TRC in a suitably chosen range of the random kick strength $K_r$ and the dissipation rate. Here, we further present its phase diagram for different values of the periodic kick strength $K_0$.

\begin{figure}[H]
	\centering
        \begin{minipage}{0.32\linewidth}
		\centering
		\includegraphics[width=0.9\linewidth]{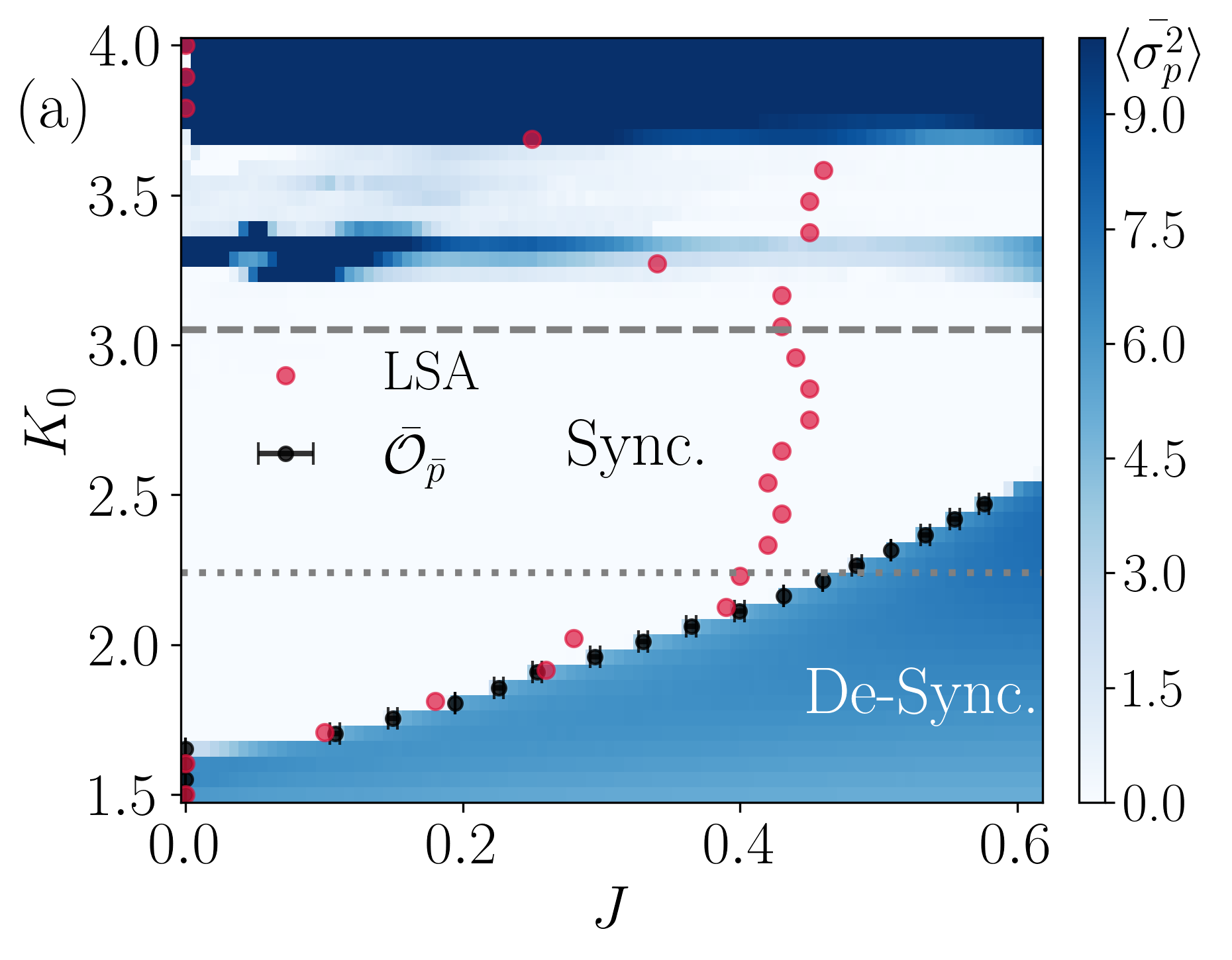}
        \end{minipage}
	\begin{minipage}{0.32\linewidth}
		\centering
		\includegraphics[width=0.9\linewidth]{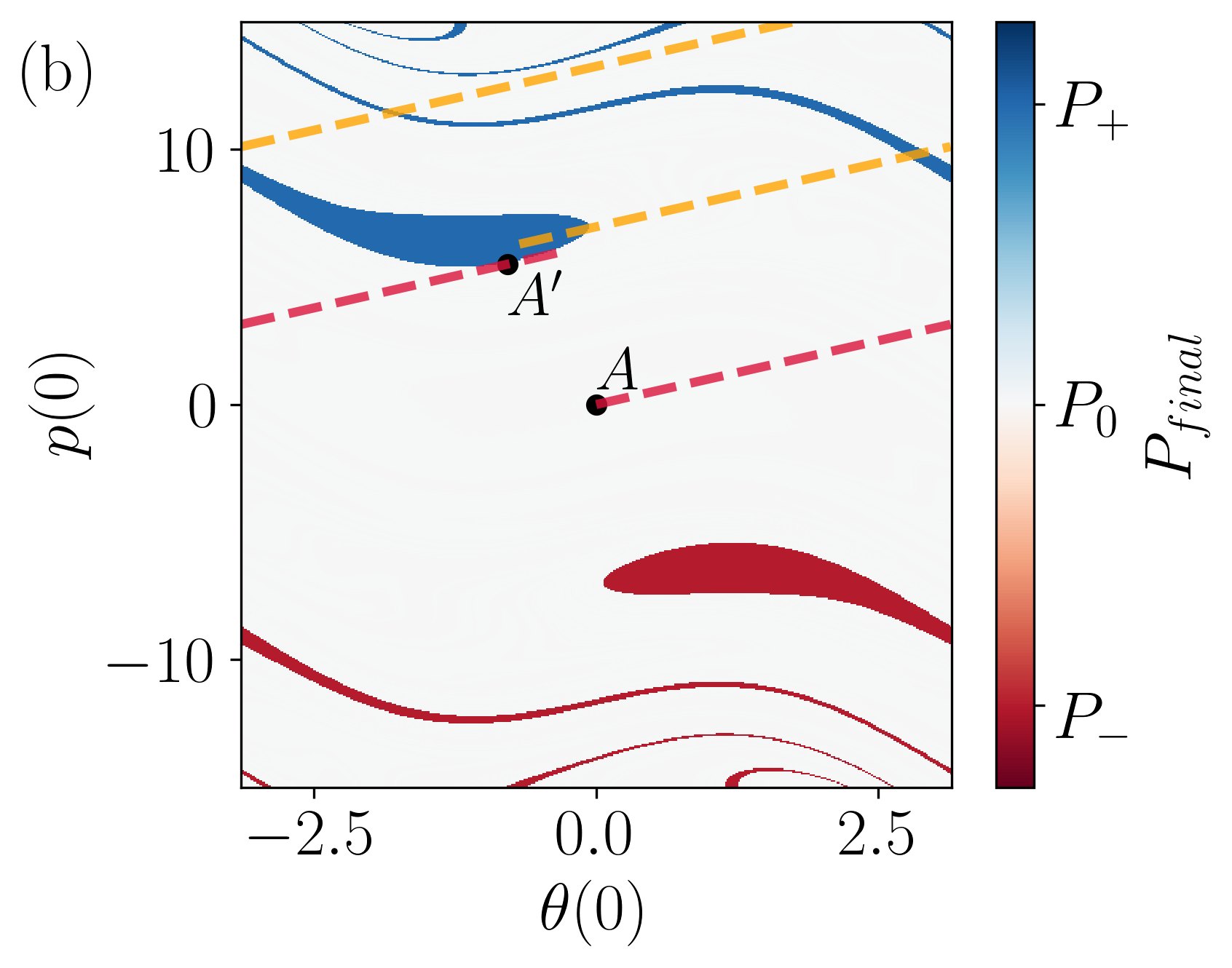}
	\end{minipage}
        \begin{minipage}{0.32\linewidth}
		\centering
		\includegraphics[width=0.9\linewidth]{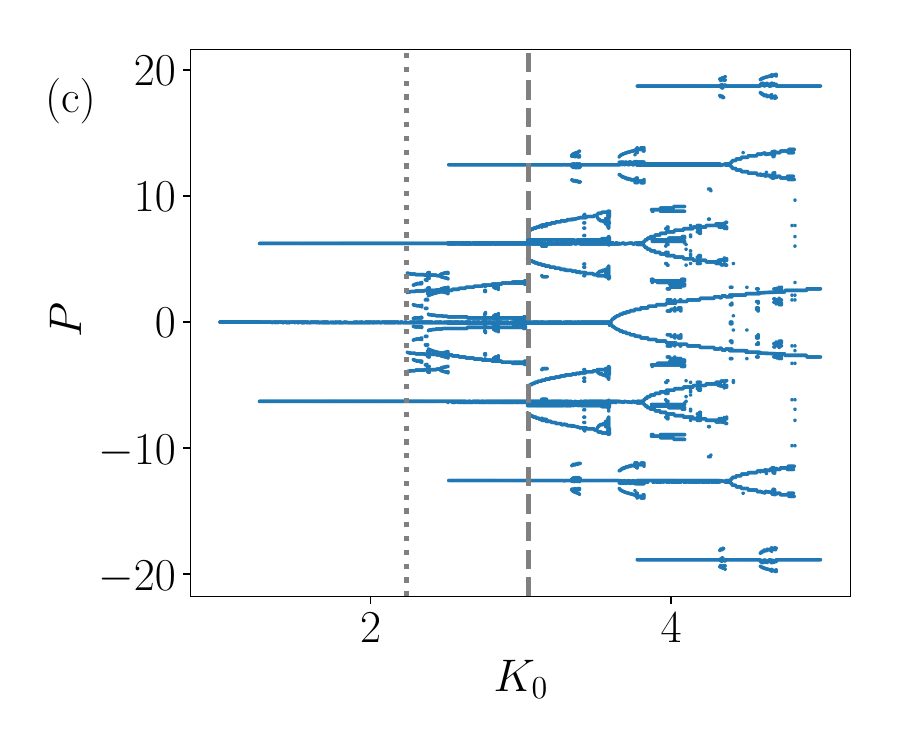}
	\end{minipage}
	\caption{(a) Phase diagram for different $K_0$ and $J$ while we fix $K_r=5.5,\gamma=0.8,T=10$. White and blue regions correspond to the synchronization and de-synchronization phases, respectively. A linear stability analysis (LSA) captures the phase transition (red). The black dots denote the phase boundary obtained by $\bar{\mathcal{O}}_{\bar{p}}$. (b) Basins of fixed points in the dissipative kicked rotor model at $K_0=1.5,\gamma=0.8$. $A^\prime$ is not in the blue region, so the transition rules designed in the main text cannot be obeyed. (c) Bifurcations with varying $K_0$ in the single dissipative rotor system for $\gamma=0.8$. The fixed point $P_0$ begins to bifurcate at about $K_0=3.6$. Both (a) and (c) share the same gray dotted lines at $K_0=2.24$ and gray dashed lines at $K_0=3.05$. 
    % In $2.24<K_0<3.05$, period-4 state emerges with momentum near $P_0$. In $K_0>3.05$, period-3 state occurs near $P_{\pm}$ \cite{yan2025bifurcationsintermittencycoupleddissipative}. 
    }
    \label{Bifurcations}
\end{figure}

\par The phase diagram is shown in Fig.~\ref{Bifurcations}(a). For simplicity, let us first focus on the parameter region below the gray dotted line in the non-interacting system ($J=0$). As shown in the bifurcation diagram in Fig.~\ref{Bifurcations}(c), this regime generally has three stable fixed points. However, for different values of $K_0$, these fixed points have different basin structures. In general, when $K_0$ decreases, the basins of fixed points $P_+$ and $P_-$ shrink, as shown in Fig.~\ref{Bifurcations}(b). For the $K_r$ value being used here, the single rotor may not enter the basin of the fixed point $P_+$ after a positive kick from $P_0$, see red lines in Fig.~\ref{Bifurcations}(b). Therefore, our transition rule, designed in Fig.~1 in the main text, fails at the single rotor level. This sets the lower bound of the white region (synchronization phase) in Fig.~\ref{Bifurcations}(a), and a finite interaction strength $J$ generally reduces its support.

\par Above the gray dotted line, periodic orbits appear around $P_0$, see Fig.~\ref{Bifurcations}(c), while $P_{\pm}$ remains largely unaffected. This ends at the gray dashed line where more periodic orbits appear around $P_{\pm}$. {The appearance of those complicated orbits can induce significant changes in the basin structure. We examine the stability of synchronized TRCs in many body systems numerically. As shown in Fig.~\ref{Bifurcations}(a), synchronized TRC indeed remains stable between the gray dashed and dotted lines.}

\par We can use linear stability analysis (red dots) to investigate the phase boundaries. It accurately captures the phase boundary below the gray dotted line in Fig.~\ref{Bifurcations}(a). This once again demonstrates the effectiveness of our method. However, above the gray dotted line, it generally fails, as seen in Fig.~\ref{Bifurcations}(c), where the mean-field trajectory becomes less predictable and unstable due to the appearance of more periodic orbits and bifurcations.

\subsection{ Independence of initial conditions }
\begin{figure}[h]
    \centering
    \includegraphics[width=0.5\linewidth]{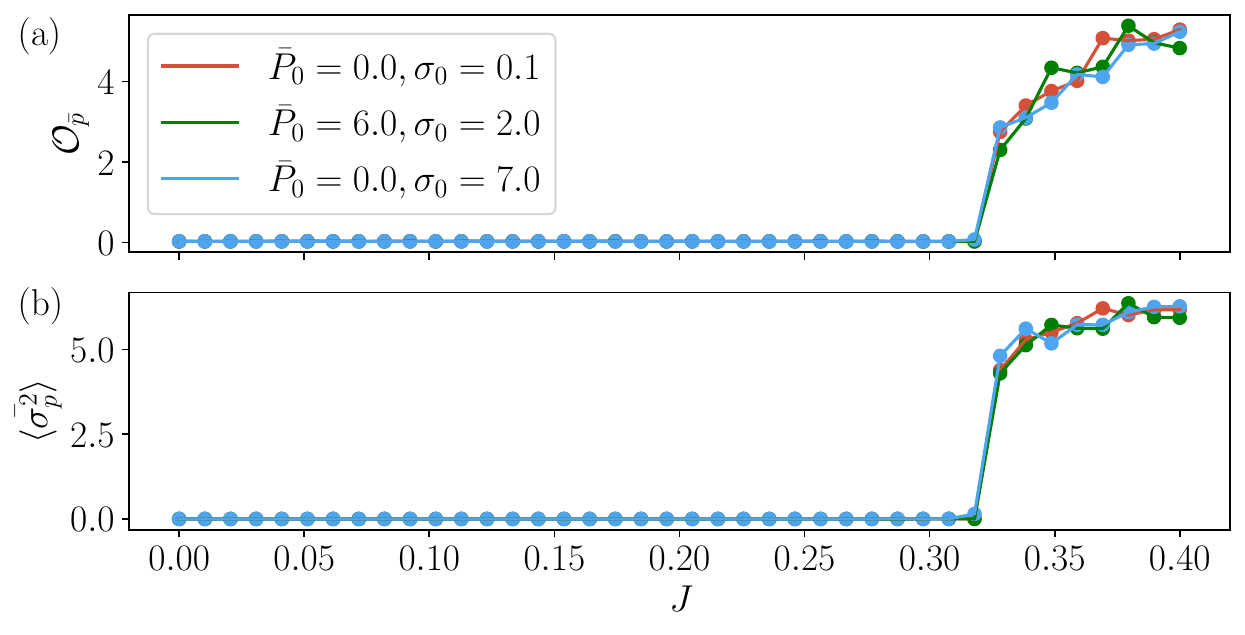}
    \caption{The order parameters $\mathcal{O}_{\bar{p}}$ and $\langle\bar{\sigma_p^2} \rangle$ are independent of initial conditions. Here the evolution of the many-body system begins at a random Gaussian distribution of both momentum and angle for each site, exhibited in different colors. $\bar{P}_0$ is the average and $\sigma_0$ is the standard deviation of the Gaussian distribution {for the initial momenta}. The standard deviation of the initial angles is $\sigma=3$. We use  $K_r=5.5,T=10,$ and $ L=100$ for numerical simulations and extract order parameters at $t/2T=10^4$.}
    \label{fig:ini}
\end{figure}

The phase diagram Fig.~4 in the main text does not depend on the specific choices of initial conditions. In Fig.~\ref{fig:ini}, we plot the order parameters for different $J$, calculated with different initial conditions. Small differences are observed in the de-synchronization phase, whereas the phase boundary remains unchanged. 

\subsection{Convergence of the distribution of the largest Lyapunov exponents}

\begin{figure}[h]
	\centering
	\begin{minipage}{0.49\linewidth}
		\centering
		\includegraphics[width=0.8\linewidth]{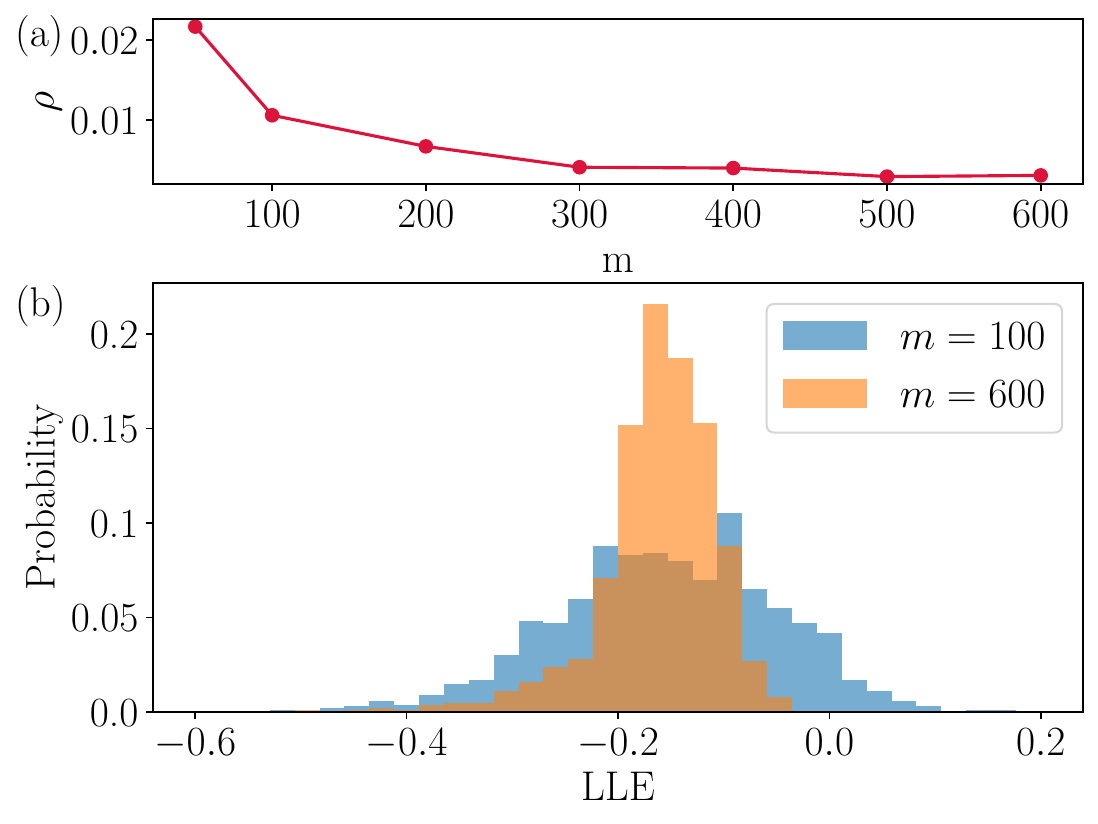}
		
	\end{minipage}
	\begin{minipage}{0.49\linewidth}
		\centering
		\includegraphics[width=0.8\linewidth]{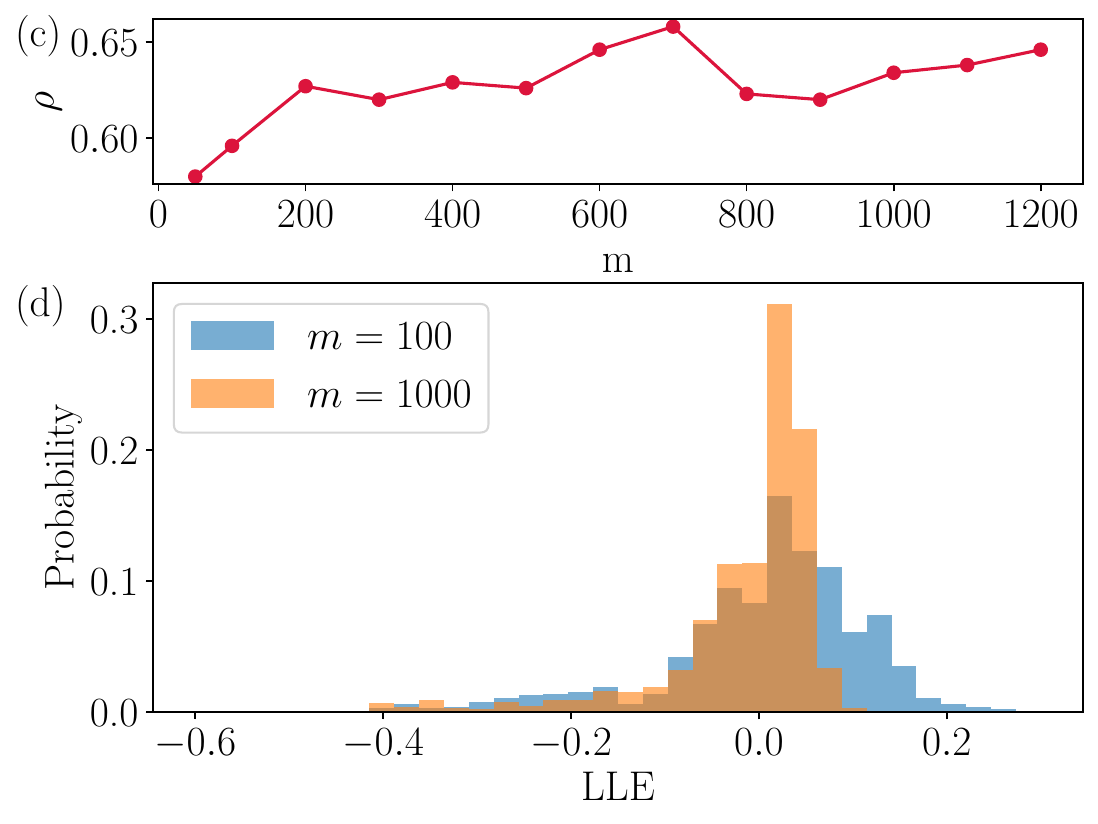}
		
	\end{minipage}
	\caption{(a)(c) $\rho$ denotes the fraction of positive LLEs. As the number of dipolar kicks $m$ increases, $\rho$ converges to zero in panel (a), while it remains finite in panel (c), corresponding to the synchronization and de-synchronization phases, respectively. (b)(d) The corresponding normalized distribution of LLEs for different realizations of the mean-field trajectories shows that larger $m$ leads to sharper distributions. Numerically, we use interaction strength $J=0.28$ for panels (a) and (b), and $J=0.35$ for (c) and (d).
    }
    \label{fig:LLE4}
\end{figure}

The distribution of the largest Lyapunov exponents(LLE) converges when the number of dipolar kicks, $m$, is sufficiently large. To see this, we evolve 1000 single-rotor trajectories during $m$ dipolar drives with a random Gaussian-distributed initial momentum, $\sigma^2_p=0.01,\bar{p}=0$, and all the initial angles fixed at zero. For each mean-field trajectory $\{\bar{\theta}(t)\}$, we can compute a series of matrix products $\boldsymbol{D}_{k}{=}\prod_{i=0}^{2mT-1} \boldsymbol{A}_k(\bar{\theta}(i))$. Then the LLE of each trajectory can be calculated from its maximum eigenvalue. We present the histogram of all the LLEs
of the synchronization and de-synchronization phases in Fig.~\ref{fig:LLE4}(b) and Fig.~\ref{fig:LLE4}(d), respectively.  We notice that the distribution of LLEs sharpens, and crucially, the fraction of positive LLEs converges for larger $m$. To quantify this,
we define $\rho$ as the fraction of LLEs that become positive. Clearly, it quickly converges to zero in Fig.~\ref{fig:LLE4}(a) but remains finite in (c), corresponding to 
the synchronization and de-synchronization phase. In the main text, we use $m=300$ that is sufficiently large to capture the phase boundary.

\section{Quantum system in Rydberg atoms} 
While we have demonstrated stable TRCs in classical many-body systems in the main text, our construction is readily generalizable to quantum systems. For concreteness, here we provide one example using dissipative Rydberg atoms. The Hamiltonian reads
\begin{equation}\nonumber
    H(t){=}\sum_{k=1}^{N}\left[\Omega_{x}(t) \sigma_{k}^{x}+\Omega_{y}(t) \sigma_{k}^{y}+\Delta(t) n_{k}\right]+\sum_{k \neq p}^{N} V_{k p} n_{k} n_{p},
\end{equation}
where $\sigma^x_k,\sigma^y_k$ are the Pauli matrices at the $k^{th}$ site, 
$n_k{=}(1{+}\sigma^z_k)/2$ is the Rydberg number operator. Spontaneous emission induces dissipation, leading to the time evolution of the density matrix $\rho$, $\partial_t \rho{=}-i[H,\rho]{+}\mathcal{L}[\rho]$, with $\mathcal{L}[\rho]{=}\Gamma \sum_{k=1}^{N}\left[\sigma_{k}^{-} \rho \sigma_{k}^{+}{-}\frac{1}{2}\left\{\sigma_{k}^{+} \sigma_{k}^{-}, \rho\right\}\right]$. 
For $\Omega_{x}=0.7\Gamma,\Omega_{y}=0,\Delta=-3.5\Gamma,V=6\Gamma$, at the mean-field level, the system has two fixed points, each of which exhibits distinctive values of total magnetization~\cite{PhysRevLett.122.015701}. We use this bistability to construct TRC in quantum systems, by considering the three-stage driving protocol $(\Omega_x(t),\Omega_y(t),\Delta(t))$ as follows,
\begin{equation}
    \left\{\begin{matrix}
 (\Omega_x,\Omega_y,\Delta), & t\in [mT,mT+T_1)\\
  (\Omega_x^{(1)},\Omega_y^{(1)},\Delta^{(1)}),& t\in [mT+T_1,mT+T_1+\delta t)\\
 (\Omega_x,\Omega_y,\Delta) ,& t\in [mT+T_1+\delta T,(m+1)T)
\end{matrix}\right.
\end{equation}
where 
$T_1$ is randomly chosen from $0.4T$ or $0.6T$ with the same probability and $m{\in} \mathbb{R}$. We consider $(\Omega_x^{(1)},\Omega_y^{(1)},\Delta^{(1)}){=}(0,94,-314)$, such that the system transits from one stable fixed point to the other during a short time $\delta t{=}10^{-2} \Gamma^{-1}$~\cite{PhysRevLett.122.015701}, where $\Gamma^{-1}$ is used as the unit of time. The random selection of $T_1$ suggests that such a transition happens either at $mT{+}0.4T$ or $mT{+}0.6T$, inducing the short-time disordered micro-motion. In contrast, at stroboscopic times, e.g., $t{=}mT$, the system exhibits deterministic period-doubling evolution: after every $2T$ duration, the system returns to the original fixed point. The coexistence of both long-time order and short-time disorder is the characteristic feature of TRC.

\begin{figure}[h]
    \centering
\includegraphics[width=.6\linewidth]{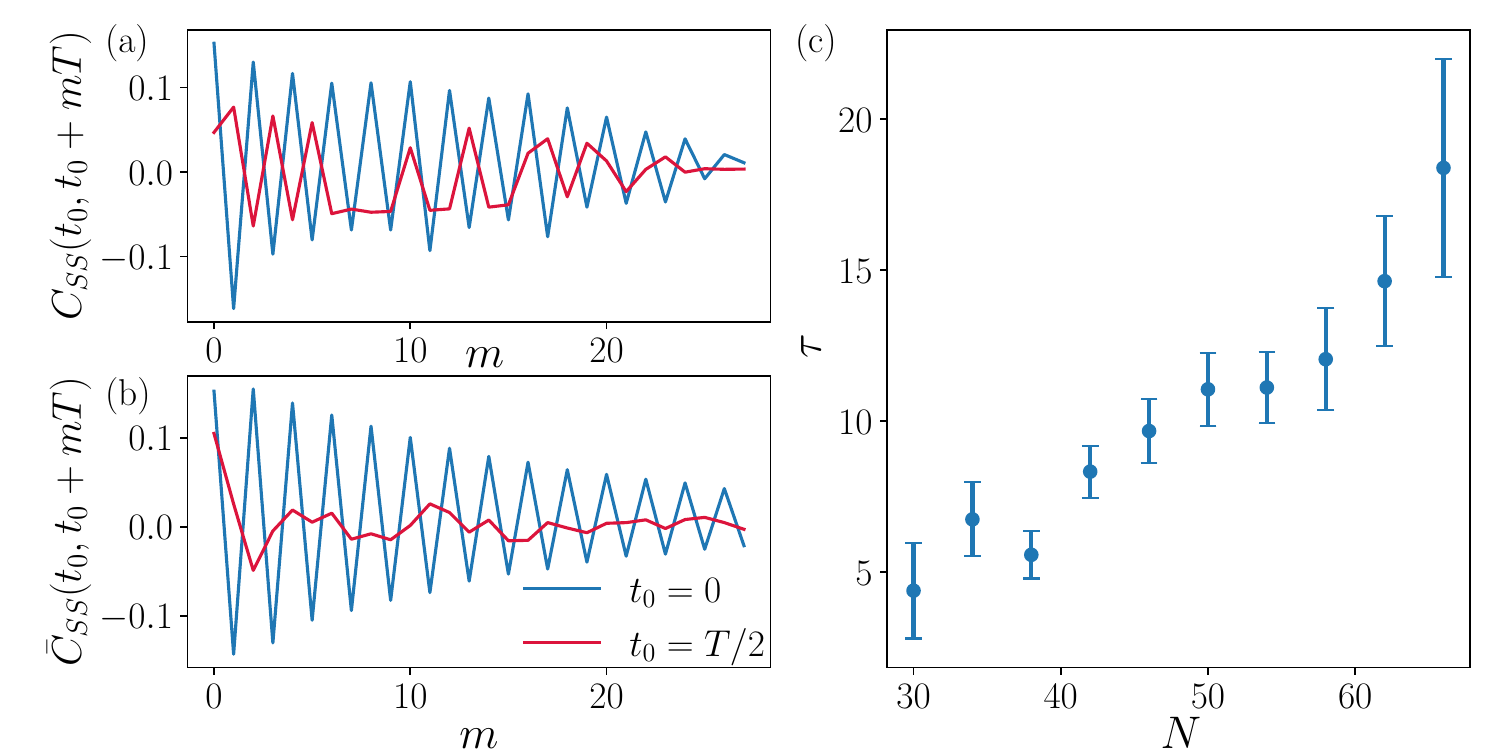}
    \caption{(a) Correlation function of the Rydberg system. The blue line shows the regular oscillations, while the red line is short-time disordered. The system size is $N=66$. (b) Ensemble-averaged correlation function. (c) The lifetime increases with the system size, suggesting that TRC is stable in the thermodynamic limit.}
    \label{fig:Ryd}
\end{figure}
We consider the fully connected quantum model with $V_{k,p}{=}V/N,k{\ne} p$ with permutation symmetry, for which the mean-field solution becomes reliable in the thermodynamic limit. For a finite system size $N$, there is only one stable fixed point (the other fixed point becomes metastable), which we use as the initial state. To diagnose the TRC, we use the correlation function $C_{SS}(t_0,t_0+mT)= S^x(t_0)S^x(t_0+mT)$, where $ S^x(t){=}1/N\sum^N_i \mathrm{Tr}[\sigma^x_i\rho(t)]$ is the expectation of the total $x$-magnetization. As shown in Fig.~\ref{fig:Ryd}(a), when $t_0{=}0$ (blue), the stroboscopic dynamics have long-range order and oscillate regularly. When $t_0=T/2$ (red), the evolution is temporally disordered, confirming the key feature of the TRC. For $N{\to}\infty$, TRC should be stable, while the correlation function decays in time due to the finite size effects. 
To extract the dependence of the lifetime on the system size, we first perform the ensemble-averaged correlation function $\bar{C}_{SS}(t_0,t_0+mT)$ over $20$ trajectories. As shown in Fig.~\ref{fig:Ryd}(b), it exhibits an exponential decay.
In contrast, the micro-motion quickly averages to zero. We numerically fit the envelop of the blue curve according to the function $e^{-mT/\tau}$, where $\tau$ is used to quantify the lifetime, and present the results in Fig.~\ref{fig:Ryd}(c) over many numerical simulations.
Clearly, the lifetime grows with $N$, suggesting that TRC is a stable quantum phase of matter for $N{\to}\infty$.

 Both the Rydberg and rotor systems employ multistability to construct stable TRC. However, there are two key differences between these models. First, in the kicked rotors we obtain synchronized TRC while here synchronization is not necessary in Rydberg atoms.
 Second, interaction plays different roles in the two systems. For kicked rotors, interaction tends to spread defect which destabilizes synchronization. In contrast, interaction in the Rydberg model is essential for creating bistability in the mean-field approximation. We leave the systematic characterization of the phase diagram in Rydberg systems to future works.

\section{Experimental realization of kicked rotors}
The kicked protocol discussed in the main text can be experimentally realized in superconducting quantum simulation platforms.

\begin{figure}[h]
    \centering
    \includegraphics[width=0.8\linewidth]{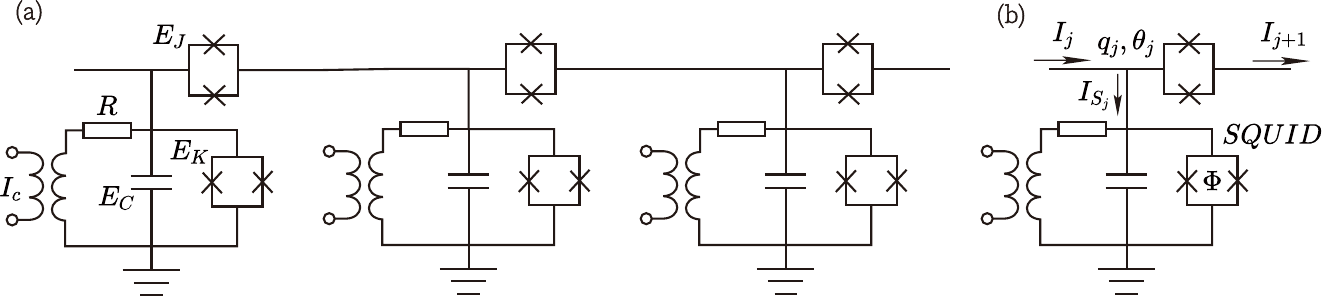}
    \caption{(a) The experimental realization of our random kicked rotor model. Here we just show three of the rotors in the periodic boundary. (b) The circuit for the $j^{th}$ kicked rotor.}
    \label{fig:super}
\end{figure}

\begin{figure}[h]
    \centering
    \includegraphics[width=0.5\linewidth]{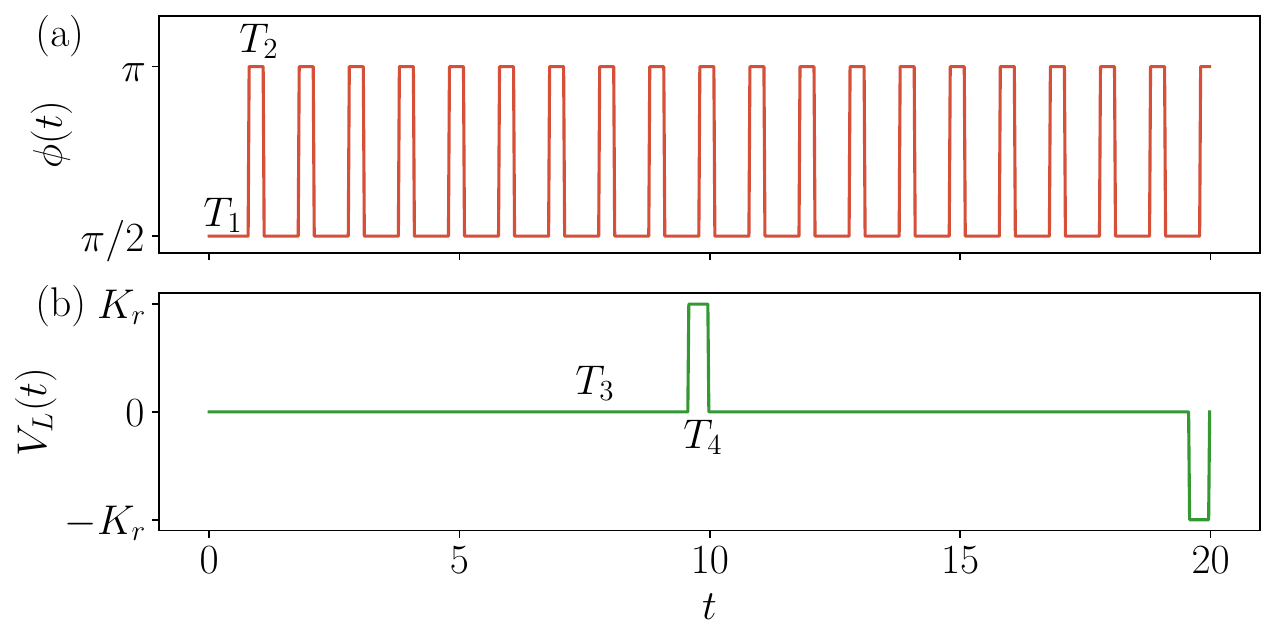}
    \caption{Time-dependent protocol of $\phi(t)$ and $V_L(t)=MI_c(t)/R$ in the  experimental setting to mimic the kicked rotor system. $\phi(t)$ generates periodic kicks and $V_L(t)$ generates random dipolar kicks to induce the time rondeau order. $T_1$ and $T_3$ are the driving periods. $T_2$ and $T_4$ are width of pulses.}
    \label{fig:exp_T}
\end{figure}

We first consider the systems without dissipation and mutual inductance.
As shown in Fig.~\ref{fig:super}(b), each rotor can be simulated by a resonator composed of one capacitor and superconducting quantum interference devices (SQUIDs). Each SQUID has two Josephson junctions to form a loop. The Hamiltonian of the resonator on the $j^{th}$ site is $\hat{H}_j=\hat{q}^2_j/2C+ E_K \cos(2\pi \Phi_K/\Phi_0)\cos(\hat{\theta}_j)$, where the first and the second term correspond to the capacitor and the SQUID, respectively. In the Hamiltonian, $\hat{q}_j$ is the charge on the $j^{th}$ capacitor,  $\Phi_K$ is the total magnetic flux of two Josephson junctions, $\Phi_0=hc/(2e)$ is the Cooper-pair flux quantum, and $\hat{\theta}_j$ is their average phase. As shown in Fig.~\ref{fig:super}(a), by introducing one SQUID linking the $j^{th}$ resonator with the $(j+1)^{th}$ resonator, $\hat{H}_{j,j+1}=E_J \cos(2\pi \Phi_J/\Phi_0)\cos(\hat{\theta}_j-\hat{\theta}_{j+1})$, one can now simulate the many-body interacting system. The phase $\Phi_J$ and $\Phi_K$ are tunable time-dependent parameters, which can control the strength of interaction and the on-site potential energy. Therefore, the Hamiltonian of the entire many-body system reads
\begin{equation}
    \hat{H}=\sum_j \left\{\frac{\hat{q}^2_j}{2C} +E_K \cos(\phi(t))\cos(\hat{\theta}_j)+E_J \cos(\phi(t))[\cos(\hat{\theta}_j-\hat{\theta}_{j+1})+\cos(\hat{\theta}_j-\hat{\theta}_{j-1})]\right\},
\label{exp}
\end{equation}
where we set two magnetic fluxes the same, $\phi(t)=2\pi\Phi_{J}(t)/\Phi_0=2\pi\Phi_{K}(t)/\Phi_0$. We can apply the driving pulses shown in Fig.~\ref{fig:exp_T}, and if $T_1 \gg T_2$, $\cos(\phi(t))$ can approximately generate dynamical effects induced by periodic kicks $\delta(t)$ as considered in the main text. 

Then we introduce dissipation and mutual inductance by connecting a branch composed of a resistor and an instrument transformer, as shown in Fig.~\ref{fig:super}(a). To simulate our model, we discuss EOMs under the classical approximation. In this approximation, we change $\hat{q},\hat{\theta}$ to $q,\theta$ and obtain their equation of motion via the Poisson bracket. We denote the current moving from the $(j-1)^{th}$ resonator to the $j^{th}$ resonator as $I_{j}$ and the $j^{th}$ resonator to the $(j+1)^{th}$ resonator as $I_{j+1}$ shown in Fig.~\ref{fig:super}(b). These currents go through SQUIDs following the EOM,
\begin{equation}
\begin{aligned}
 I_j  & = \frac{2e}{\hbar}E_J\cos(\phi)\sin(\theta_{j}-\theta_{j-1}),  \\
 I_{j+1}  & = \frac{2e}{\hbar}E_J\cos(\phi)\sin(\theta_{j+1}-\theta_{j}).
\end{aligned}
\end{equation}
We apply Kirchhoff’s law and get $I_{S_j}=I_j-I_{j+1}$. Then the current splits into the capacitor with current $\dot{q}_j$, resistor with $(q_j/C-MI_c)/R$, and SQUIDs with $-\frac{2e}{\hbar}E_J\cos(\phi)\sin(\theta_{j})$. One can also assume that the self-inductance of resonators is negligible. Then we get the EOM that mimics our kicked rotor model,
\begin{equation}
    \dot{q_j}=-q_j/CR+\frac{2e}{\hbar}E_K\cos(\phi(t))\sin(\theta_{j})+\frac{2e}{\hbar}E_J\cos(\phi(t))[\sin(\theta_{j}-\theta_{j-1})+\sin(\theta_{j}-\theta_{j+1})]+MI_c(t)/R,
    \label{Exp:qd}
\end{equation}

where $I_c(t)$ denotes the control current that induces dipolar kicks. Its driving protocol is depicted in Fig.~\ref{fig:exp_T}(b) and we also require $T_3\gg T_4$ to approximate the delta kick. Note that the realization of the dipolar kicks can indeed be quite flexible in practice, for instance, the transformer can be replaced by any other voltage source. 

By using the relation between the phase $\theta_j$ and voltage $U_j=q_j/C$, we obtain the other set of EOMs
\begin{equation}
    \dot{\theta_j}=\frac{2e U_j}{\hbar}= \frac{2e q_j}{C\hbar},
    \label{Exp:th}
\end{equation}
where $U_j$ is the voltage between the two sides of SQUIDs and it is equal to the voltage of the capacitor. In conclusion, this circuit with EOMs Eqs.~\eqref{Exp:qd}
and \eqref{Exp:th} can simulate our dipolar kicked many-rotor system.

We rewrite the superconducting Hamiltonian (Eq.~\eqref{exp}) in a dimensionless form 
\begin{equation}
    H=\sum_i[\frac{P^2_i}{2}-\tilde{K}_0\cos\theta_i \sum_n \delta(\tilde{t}{-}n\tilde{T}_1)]-\tilde{J}\sum_i[\cos(\theta_i-\theta_{i+1}){+}\cos(\theta_i{-}\theta_{i-1})]\sum_n \delta(\tilde{t}{-}n\tilde{T}_1), 
\end{equation}
where $\tilde{t}=tE_K/\hbar$, $\tilde{T}_1=T_1E_K/\hbar$ (time is defined using the Josephson energy $E_K$), and $\tilde{K}_0=E_KT_2/\hbar$, $\tilde{K}_r=\frac{MI_c T_4}{2eR}$, $\tilde{J}=E_J T_2/\hbar$ and $P_i=q_i/\sqrt{CE_K}$. Based on $C,E_K$, we can determine the initial charges of capacitance (for classical limit) or the initial energy level (for quantum limit) to set the initial momentum of rotors. $T_1(T_3)$ is the driving period(waiting time of random driving), and $T_2(T_4)$ is the pulse width of periodic kicks(random kicks) in FIG.~\ref{fig:exp_T}. If $\cos(\phi (t))$ in Eq.~\ref{exp} follows the pulse shape in FIG.~\ref{fig:exp_T}, we can approximate it as $T_2\delta(t-nT_1)$ which further determines $\tilde{K}_0,\tilde{J}$. The dissipation rate reads $1-\gamma=\exp(-\frac{ T_1}{CR})$. In superconductors, the energy of a Josephson junction is typically $10^{9}\sim 10^{11}\hbar\text{Hz}$. Capacitance is normally $C=50\text{fF}$. If we choose $\tilde{T}_1=10\tilde{T}_2=10\tilde{T}_4=1$, $\gamma=0.2$ and $\tilde{K}_0=2$ as considered in the main text, we need the energy of SQUID $E_K/\hbar \in [0,10^9]\text{Hz}$, $1/T_2=5\times 10^{8}\text{Hz}$ and resistance $R\approx 24.9\text{k}\Omega$. Parameters $\tilde{J}$ can be justified by the phase $2\pi \Phi_J/\Phi_0$ of SQUIDs with a range $E_J/\hbar \in [0,5\times10^8]\text{Hz}$. These can be achieved in experiments to simulate the time-dependent many-body rotors with dissipation~\cite{PhysRevX.11.011010,iskhakovFastControlCurrent2012a,yan2020engineeringframeworkoptimizingsuperconducting}.
 
\end{document}